\documentclass[11pt]{article}
\usepackage{arxiv}
\usepackage{import}
\usepackage[utf8]{inputenc}
\usepackage{fullpage}
\usepackage{graphicx}
\usepackage[export]{adjustbox}
\usepackage{hyperref}
\usepackage{float}
\usepackage{url}
\usepackage{mathtools} 
\usepackage{tabto}
\usepackage{todonotes}
\usepackage{comment}
\usepackage{caption}
\usepackage{subcaption}

\usepackage{amsmath} 
\usepackage{amsfonts}
\usepackage{wrapfig} 

\usepackage{algcompatible}
\usepackage[ruled]{algorithm}
\usepackage[noend]{algpseudocode}
\usepackage{systeme}
\usepackage{multirow}
\usepackage[group-separator={,},group-minimum-digits=4]{siunitx}
\usepackage{arydshln}

\usepackage[normalem]{ulem}
\newcommand{\msout}[1]{\text{\sout{\ensuremath{#1}}}}

\usepackage[maxbibnames=99,backend=bibtex]{biblatex}
\graphicspath{ {graphics/} } 

\numberwithin{equation}{section}

\algdef{SE}[EVENT]{Event}{EndEvent}[1]{\textbf{upon event}\ #1\ \algorithmicdo}{\algorithmicend\ \textbf{event}}
\algtext*{EndEvent}

\usepackage{eqparbox}
\newdimen{\algindent}
\algnewcommand\LeftComment[2]{%
\hspace{#1\algindent}$\triangleright$ \eqparbox{COMMENT}{#2} \hfill %
}

\algnewcommand{\Defs}[2]{%
  \State \textbf{#1:}
  \Statex \hspace*{\algorithmicindent}\parbox[t]{.8\linewidth}{\raggedright #2}
}

\DeclarePairedDelimiter\ceil{\lceil}{\rceil}
\DeclarePairedDelimiter\floor{\lfloor}{\rfloor}

\newcommand{\bs}[1]{\lceil\log_2{#1}\rceil}

\usepackage{pifont}
\usepackage{amssymb}

\pdfoutput=1 

\title{Distributed Recoverable Sketches}

\author{
    \href{https://orcid.org/0009-0003-6061-7635}{\includegraphics[scale=0.06]{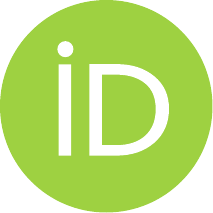}\hspace{1mm}
    Diana Cohen} \\
    Computer Science Department\\
    Technion, Haifa, Israel\\
    \texttt{diana.cohen@cs.technion.ac.il} \\
    \And
    \href{https://orcid.org/0000-0001-6460-9665}{\includegraphics[scale=0.06]{orcid.pdf}\hspace{1mm}
    Roy Friedman} \\
    Computer Science Department\\
    Technion, Haifa, Israel\\
    \texttt{roy@cs.technion.ac.il} \\
    \And
    \href{https://orcid.org/0000-0002-9254-8529}{\includegraphics[scale=0.06]{orcid.pdf}\hspace{1mm}
    Rana Shahout} \\
    Computer Science Department\\
    Harvard University, Cambridge, United States\\
    \texttt{rana@seas.harvard.edu} \\
}

\begin{document}
\maketitle

\begin{abstract}
Sketches are commonly used in computer systems and network monitoring tools to provide efficient query executions while maintaining a compact data representation. 
Switches and routers maintain sketches to track statistical characteristics of network traffic. 
The availability of such data is essential for the network analysis as a whole. 
Consequently, being able to recover sketches is critical after a switch~crash. 

In this work, we explore how nodes in a network environment can cooperate to recover sketch data whenever any subset of them crashes. 
In particular, we focus on frequency estimation linear sketches, such as the Count-Min Sketch. 
We consider various approaches to ensure data reliability and explore the trade-offs between space consumption, runtime overheads, and traffic during recovery, which we point out as design guidelines. 
Besides different aspects of efficacy, we design a modular system for ease of maintenance and further~scaling. 

A key aspect we examine is how the nodes update each other regarding their sketch content as it evolves over time. 
In particular, we compare periodic full updates vs.\ incremental updates. 
We also examine several data structures to economically represent and encode a batch of latest changes. 
Our framework is generic, and other data structures can be plugged-in via an abstract API as long as they implement the corresponding API~methods. 
\end{abstract}


\section{Introduction}
\label{chap:intro}
Network monitoring is a crucial part of network management.
Effective routing, load balancing, and DDoS, anomaly, and intrusion detection are examples of applications that rely on monitoring network flows frequencies~\cite{MicroTE_Routing, IBM_LoadBalancing, Afek_MitigatingDDoS, IntrusionDetection}.
Internet flows are often identified by the source and destination IP addresses.
However, the usual IPv4 5-tuple 
($src.ip$, $src.port$, $dest.ip$, $dest.port$, $protocol$) 
can be used, uniquely identifying each connection or session~\cite{TCPIPbookStevens}.

Monitoring tools track a significant amount of flows, updating the frequency count of every flow upon each of its packet arrivals~\cite{RAP}.
Hence, maintaining exact counter-per-flow is often not feasible due to the high line rates of modern networks as well as lack of space of SRAM memory.
To overcome these limitations, it is common to trade the accuracy for memory space with sketching algorithms that utilize hashing to summarize traffic data using fewer counters and without explicitly storing flows' identifiers (e.g., \emph{Count Sketch}~\cite{CountSketch}, \emph{Spectral Bloom Filter}~\cite{SpectralBloomFilter}, and \emph{Count-Min Sketch}~\cite{CountMinSketch}).

{Count-Min Sketch} (CMS) is a popular \emph{sublinear space} data structure for estimating a flow size.
It uses pairwise independent hash functions to map flows to frequencies, while sublinear space is achieved at the expense of overestimating some flows' frequencies due to hash collisions.
Given a stream of $N$ items and accuracy parameters $(\epsilon, \delta)$, 
CMS guarantees that the estimation error for any \emph{flow} (distinct item) is bounded by $\epsilon N$, with probability at least $(1-\delta)$. 

Switches and routers constantly produce statistics of tracked traffic.
The availability of such data is essential for the network analysis as a whole and to enable continuous and long-term survivability of its applications. 
The ability to recover the statistics after a crash failure is therefore highly desirable.
Obviously, ensuring such information durability requires a certain level of redundancy and replication.
However, tracing and processing information at the high line rate of modern networks might consume significant networking, computation, and storage resources.
On the other hand, the fact that we specifically target sketches may offer opportunity to reduce these costs compared to direct utilization of common techniques used for general data availability.
To that end, in this work we explore various approaches to ensure data survivability of fast updating sketches in terms of their computational costs, communication cost under normal operation, recovery communication cost after a failure, and storage overhead.

\begin{figure}[ht]
    \centering
    \includegraphics
    [width=0.35\textwidth]
    {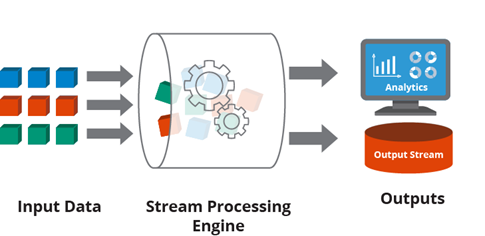}
    \caption{Stream processing diagram of a switch}
    \label{fig:stream-processing}
\end{figure}

\subsection{Contributions}
\label{sec:contributions}
Our goal is to design a flexible distributed software system consisting of fully functional switches, each maintaining its own {local} sketch (referred to as \emph{data node}), while sharing different partitions of its sketch with other switches (referred to as \emph{redundant nodes}) for backup.
In case of a switch's crash failure, the others will contribute to the recovery process, restoring the latest backup of a failed~sketch.

One of our goals is to provide a modular system having a simple design which is easy to maintain and to further scale.
We believe that simplicity is crucial for any system, especially in the long run.
Modularity is inherent in our design, which relies on swappable building blocks, providing a variety of configurations and settings, as well as extendability through API.
During the pre-processing stage, the global~configuration is generated based on user-specified parameters.
Further, we assume that all nodes are globally configured with the same parameters and same hash functions as well, resulting in identical sketch structure.
Additionally, we assume that all the nodes comply with the~protocol.

Given the physic parameters associated with system topology, such as the number of switches, our framework enables tuning the necessary parameters for the preferred performance goals. 
To enable {self-configuration} we are looking for algorithms that optimize the trade-offs between space redundancy overhead, communication overhead during redundancy maintenance, and the amount of recovery traffic after the node's failure.
To that end, we distinguish between the \emph{redundant space} required to store the redundant sketches and the \emph{extra local space} required to represent a batch of delayed items, used for incremental~updates.

Redundant information can be constructed by summing the sketches into a \emph{sum-sketch}, having the same structure, but with longer counters to count up to the sum of items in the data streams it covers. 
Inspired by RAID~\cite{RAID} we introduce a simple, yet scalable method to generate redundant sums to tolerate any number of concurrent data failures, as described in \autoref{sec:RAID4-ext}. 
Our generation method has small coefficients, which imply fast multiply~operations. 

To optimize communication overhead, batching and periodic incremental updates provide some highly desired benefits. 
As we point out in \autoref{sec:RAID}, since our work is sketch-aware, we are able to use less space than required with RAID while buffering the changes into a batch. 
Having a batch next to the original sketch, we might \emph{boost the sketch} in terms of its running time, when managing ongoing arrivals using a batch and updating the original sketch periodically. 
Additionally, when a batch is represented by a hash table, this technique provides us with a more accurate estimation to answer point queries, since it partially depends on the exact frequency accumulated during the latest~batch. 

We evaluated the usage of compact hash tables to economically represent a batch of latest changes, as well as the impact of a batch size on the estimation error of CMS after recovery. 
In our experiments, we used CAIDA Internet traces for measurements and consider the results as the range for any smaller Internet trace (e.g., within a campus). 
The traces were processed using the \verb|Go| programming language, while the plots were generated using \verb|Python|.
The source code is available on GitHub~\cite{Cohen_Measuring_average_frequency_2024}.

\subsection{Organization}
\label{sec:organization}
The rest of this paper is organized as follows.
\autoref{chap:related} shortly covers the related work and the associated properties on which we rely.
In \autoref{chap:prelim} we formulate the model and its key entities, while in \autoref{chap:design} we highlight the key aspects of efficacy that we focus on during the system design.
\autoref{chap:redundancy} describes the two redundancy strategies and a trade-off between redundant space and recovery traffic.
To enable a fine grain redundant information sharing, the concept of \emph{sketch partitioning} is introduced in \autoref{sec:sketch-partitioning}.
In \autoref{chap:updates} we introduce several data structures to economically represent a batch of the latest changes. 
While \autoref{tab:batch-summary} lists the main aspects of representation's complexity (e.g., space, communication, and computation), \autoref{chap:eval} contains a more detailed discussion, which extends our previous version~\cite{opodis24}.
In \autoref{chap:experiment} we evaluate the usage of compact hash tables to economically represent a batch of the latest changes, as well as the impact of a batch size on the estimation error of CMS after recovery. 
Finally, in \autoref{chap:conclusions} we conclude and discuss some future work.
Technical details on core parameters and completion of missing proofs
are provided in~\autoref{chap:appendix}.
\section{Background and Related Work}
\label{chap:related}

\subsection{Count-Min Sketch}
\label{sec:CMS}
Count-Min Sketch (CMS), introduced by Cormode and Muthukrishnan in~\cite{CountMinSketch}, 
is a probabilistic data structure for summarizing data streams, which can be used to identify \emph{heavy hitters} (frequent items). 
Using several hash functions and the corresponding arrays of counters, it responds to a point query with a probably approximately correct (PAC) answer by \emph{counting} first and computing the \emph{minimum} next.
This ensures that the overestimation caused by hash collisions is~minimal. 

\begin{definition}[Pairwise independent family]
A family of hash functions is called 2-independent (or pairwise independent) if it satisfies \emph{distribution} and \emph{independence} properties, meaning that any pair of elements is unlikely to~collide.
\end{definition}

\begin{definition}[Distribution property]
Each element should have an equal probability of being placed in each~slot. 
\end{definition}

\begin{definition}[Independence property]
The place where one element is placed should not affect where the second~goes.
\end{definition}

Given user-specified parameters $(\epsilon, \delta)$, a CMS is represented by a matrix of initially zeroed counters, $\mathit{Count}_{d \times w}$, 
where $d=\ceil*{\ln{\frac{1}{\delta}}}$ and $w=\ceil*{\frac{e}{\epsilon}}$.
Furthermore, $d$ hash functions
$h_1 \dots h_d: \{1 \dots n\} \rightarrow \{1 \dots w\}$ 
are chosen uniformly at random from a pairwise independent family, mapping an item to one cell, for each row of~$\mathit{Count}$.

\begin{figure}[ht]
    \centering
    \includegraphics
    [width=0.35\textwidth]
    {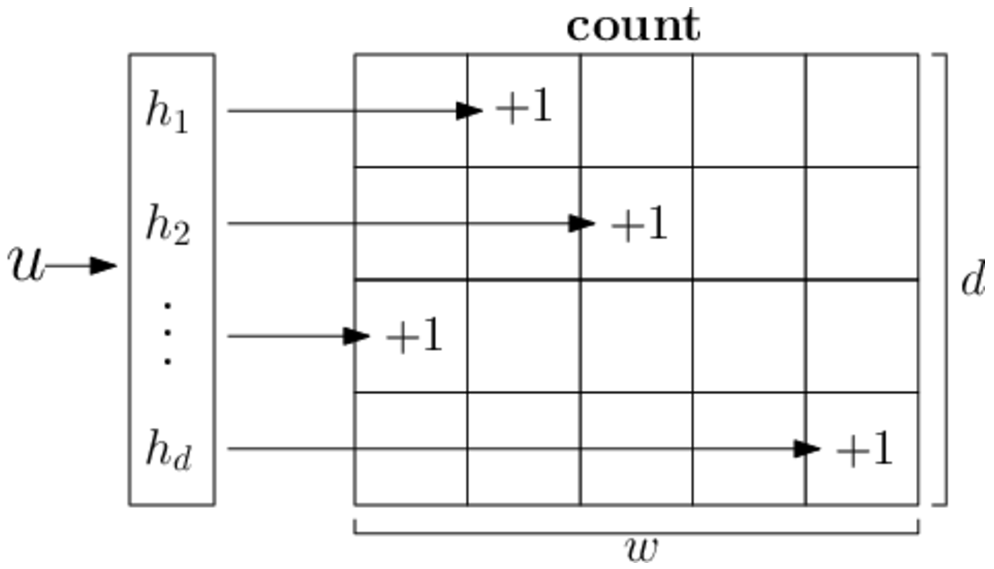}
    \caption{CMS $Update(u,1)$}
    \label{fig:CMS-update}
\end{figure}

When an update $(x, c_t)$ arrives, meaning that item $x$ is updated by a quantity of $c_t$, then $c_t$ is added to the corresponding counters, determined by $h_i(x)$ for each row $i$.
Formally, 
$\forall \ 1 \leq i \leq d$ set
$\mathit{Count}[i, h_i(x)] \mathrel{+}= c_t$.

The answer to a point query $Q(x)$ is given by 
$\hat{c_x}=\min_{i} \mathit{Count}[i, h_i(x)]$.
The estimate $\hat{c_x}$ has the following guarantees: 
($i$) ${c_x} \leq \hat{c_x}$, where ${c_x}$ is the true frequency of $x$ within the stream,
and ($ii$) with probability at least $(1-\delta)$, 
$\hat{c_x} \leq {c_x} + \epsilon N$, where $N = \sum_{x=1}^{n} |{c_x}|$ is the total number of items in the stream.
Alternatively, the probability of exceeding the additive error $\epsilon N$ is bounded by $\delta$, i.e., 
$\Pr[\hat{c_x} > {c_x} + \epsilon N] < e^{-d} \leq \delta$.

\subsubsection*{Linear sketches}
We note that CMS is a \emph{linear sketch}, i.e., given two streams, constructing a sketch for each stream and then summing the sketches yields the same result as concatenating the streams first and constructing a sketch next.
With linearity preserved, estimating the frequency of a system-wide item (e.g., within a grid or cluster) becomes an easy task, simply executing a point query against \emph{sum-sketch}.
However, the estimated answer might drift away from the true frequency due to additive~error.

\begin{observation}[No show]
When a sum-sketch answers a point query $Q(x)$ with zero, we are certain that $x$ did not flow through any of the data streams covered by that particular sum-sketch. 
\end{observation}

\begin{conjecture} 
Sharding the items of a given stream into several sub-streams can both load balance across the nodes and reduce estimation error of the original~sketch.
\end{conjecture}

\subsection{RAID}
\label{sec:RAID}
Chen, Lee, Gibson, Katz, and Patterson described in 1994 the seven disk array architectures called Redundant Arrays of Inexpensive Disks, aka RAID levels 0-6~\cite{RAID}.
These schemes were developed based on the two architectural techniques used in disk arrays: \emph{striping} data across multiple disks to improve performance and \emph{redundancy} to improve reliability.
Data striping results in uniform load balancing across all the disks, while redundancy in the form of error-correcting codes (like parity, Hamming or Reed-Solomon codes) tolerates disk failure.
Parity information can be used to recover from any single disk~failure. 

\begin{wrapfigure}{r}{0.25\textwidth}
    \centering
    \includegraphics
    [width=0.8\linewidth]
    {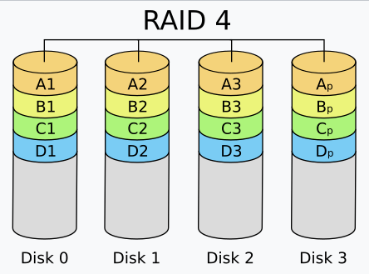}
    \label{fig:RAID4}
\end{wrapfigure}

RAID 4 concentrates the parity on the dedicated redundant disk, while RAID 5 distributes the parity uniformly across the disk array.
Thinking of a redundant disk as holding the sum of all the contents in data disks, when a data disk fails, all the data on non-failed disks can be subtracted from that sum, and the remaining information must be the missing one.
RAID 6, also called $P+Q$ redundancy, uses Reed-Solomon codes to tolerate up to two disk failures. 
RAID 1, also called mirroring, provides the highest data availability due to~replicas.

Our work is inspired by RAID, but optimized for the realm of sketches, which in networking applications must cope with extremely fast update rates across the network.
In RAID, parity can be either reconstructed from scratch or incrementally updated by applying the differences in parity between the \emph{new} and \emph{old} data.
This general technique can be applied to sketches that process data streams and send incremental updates periodically.
However, two data sketches would be required to compute the differences in parity: the original sketch, whose content is updated during normal execution, and the latest backup from the prior update cycle, aka the new and old~data.

As our work is sketch-aware, we use simplified design, focusing on integers instead of bits.
With linear sketches, we are able to update the redundant nodes in an incremental manner, producing the sums as if these were calculated from scratch.
To that end, next to its original sketch, each data node holds only the ongoing changes of a cycle.
The sending of updates can be further optimized with compact encoding to reduce the communication cost when only a few changes~occurred.

\subsection{Peer-to-Peer Backup}
\label{sec:p2p-storage}
In the distributed setting, the nodes form a network of peers, taking a decentralized approach. 
Each node can contribute a portion of its disk free space, while the overall aggregated storage can be viewed as a global storage space.
This architecture is referred to as the Peer-to-Peer (P2P) storage architecture~\cite{p2p-arc}.
Redundancy is essential for the recoverability of failed data. 
Common redundancy schemes include \emph{replication}, where multiple copies of data are shared with different nodes, and a more complex \emph{erasure code redundancy}, which requires much less storage space than replicas~\cite{p2p-backup}.

In our work, erasure-like redundancy is used for compact backups.  
Addressing both centralized and decentralized approaches to maintain redundancy,
\autoref{chap:redundancy} covers the two main strategies, whether to add dedicated redundant nodes or to distribute redundancy among existing nodes (peers), next to their normal data storage.
In other words, in a distributed setting, we adopt a hybrid scheme of P2P storage, where each peer holds a full replica of its own data sketch, while redundancy is held in a sum-sketch, backing up other~peers.

\subsection{Batching and Buffering}
\label{sec:micro-batching}
We consider the general message-passing model with $send$ and $receive$ operations\footnote{Alternatively, one can think of RDMA-like $write$ and $read$ model.},  
and assume that messages are sent over reliable communication links, similar to TCP/IP or QUIC, 
that provide an end-to-end service to applications running on nodes.
Data nodes could update the redundant nodes on each item arrival by simply forwarding, as they process their data streams.
An IP packet is the smallest message entity exchanged via the Internet Protocol across IP networks.
As each message consists of a header and data (payload), 
it is reasonable to send updates whenever there are enough changes to fill an IP packet.
This would \emph{increase network throughput} by reducing the communication overhead associated with the transition of many small frames.
Batching and buffering are commonly used techniques, where events are buffered first and sent as batch operation next.
With the batch publishing strategy, \verb|RabbitMQ| publisher~\cite{RabbitMQ-batch-publishing} batches the messages before sending them to consumers, reducing the overhead of processing each message individually.

To efficiently maintain redundant information, we consider periodic updates of incremental changes and examine several data structures to economically represent and encode a batch.
A batch may contain only a small portion of overall flows, making it feasible to maintain exact counter-per-flow (e.g., using a compact hash table~\cite{CompactHash}).
We begin with traditional buffering of items as they arrive, and then switch to flows, grouping the items by their identifiers while maintaining their accumulated frequencies using \emph{key-value}~pairs.

\subsection{Compact Hash Tables}
\label{sec:hash-tables}
Dictionaries are commonly used to implement sets, using an associative array of slots or buckets to store {key-value} pairs, with the constraint of unique keys.
The most efficient dictionaries are based on hashing techniques~\cite{CuckooHash}, but hash collisions might occur when more than one key is mapped to the same slot of a hash table. 

\begin{definition}
A \emph{load factor} of a hash table is the ratio between the number of elements occupied in a hash table and its capacity, signifying how full a hash table~is.
\end{definition}

Different implementations exist that aim to provide good performance at very high table load factors. 
A hash table has to re-hash the values once its load factor reaches its \emph{load threshold}, i.e., the maximum load factor, which is an implementation-specific parameter.
To avoid collisions, some extra space is required in the hash table, which can be achieved by having a capacity greater than the number of elements at all~times.

We consider an abstract compact hash table with $\alpha$ load threshold. 
As we express the extra local space in terms of~$\alpha$, any hash table implementation can be integrated into our work.
High performance hash tables often rely on the bucketized cuckoo hash table, since it features an excellent read performance by guaranteeing that the \emph{value} associated with some \emph{key} can be found in less than three memory accesses~\cite{Cuckoo++}.
Although the original Cuckoo hashing~\cite{CuckooHash} requires the average {load factor} to be kept less than $0.5$, 
with Blocked Cuckoo hashing~\cite{BlockedCuckoo}, using four keys per bucket increases the threshold to~$0.98$.
\section{Modeling Problem Statement}
\label{chap:prelim}
We assume a collection of $k$ nodes, each with a unique identifier known to all others, whose goal is (possibly, among other things) to monitor \emph{streams} of the data items flowing through them.
For example, these nodes could be routers in a network or servers in a decentralized scalable web application.
We also assume that the nodes are equipped with reliable memory and can communicate with each other by exchanging messages over reliable communication links.
That is, each message sent by a non-failed node to another non-failed node is eventually delivered to its intended recipient using an end-to-end communication service.
Otherwise, we detect it as a failed node, and can initiate the recovery procedure to reconstruct its sketch up to the latest restoring point.
We focus on \emph{erasures} only, due to reliable memory and communication links, as well as error corrections and retransmissions provided by the lower layer protocols, which are out of our~scope.

Every stream $S_{i \in \{1 \dots k\}}$ is modeled as a sequence of items 
$S_i = (x^{(i)}_1, \dots, x^{(i)}_{N})$, 
each carrying an identifier and possibly additional data, which is irrelevant to our work and is therefore ignored from this point on.
At any given point in time $t$, part of each stream 
$S_i = (x^{(i)}_1, \dots, x^{(i)}_{N_t})$ 
has already arrived, while the rest of the stream is still unknown.
To avoid clatter, whenever $i$ is clear from the context, we omit it.
Within these streams, we refer to all items having the same identifier $x$ 
as instances of a particular \emph{flow}, denoted as $f_x$, 
or simply referred to by its identifier $x$.
The goal of each node $i$ is to track the \emph{frequency} of every flow $x$ within its stream $S_i$, i.e., to count the number of items having the identifier $x$ that have flowed through it so far.
To reduce space overhead, nodes only maintain an estimation of each flow's frequency through Count-Min Sketches, which are configured with the same global parameters $(\epsilon, \delta)$ 
and same hash functions as well.
Upon a point query for a given flow $x$, whose true frequency in a stream after arrivals of $N_t$ items is ${c_x}$, the CMS maintained by a node $i$ would return the estimate $\hat{c_x}$, 
such that 
${c_x} \leq \hat{c_x} + \epsilon N_t$ 
with probability at least $(1-\delta)$.

For ease of reference, \autoref{tab:notations} summarizes the notations presented here as well as later in this~work.
{
\renewcommand{\arraystretch}{1.2}
\begin{table}[ht]
\footnotesize
    {
    \begin{tabular}
    {|p{0.06\textwidth}||p{0.44\textwidth}|p{0.4\textwidth}|}
    \hline
    Symbol & Meaning & Usage
    \\
    \hline
    \hline
    $(\epsilon, \delta)$ 
    & accuracy guarantees of CMS: $\epsilon \in (0,1]$ and $\delta \in (0,1]$
    & the error in answering a query is within an additive factor of $\epsilon$ with probability $(1-\delta)$
    \\
    \hline
    $c_x$, $\hat{c_x}$ 
    & true and estimated frequencies of a flow $x$ 
    & $c_x \leq \hat{c_x}$ and
    $\Pr[\hat{c_x} - c_x > \epsilon N] < \delta$
    \\
    \hline
    $n$, $N$ 
    & number of flows (distinct items) in data stream having 
    $N=\sum_{x=1}^{n}|c_x|$ items
    & stream counters are $\ceil*{\log_2{(N+1)}}$ bits long, we omit (+1) from now on
    \\
    \hline
    $w$ 
    & number of counters-per-row in CMS
    & $w \leftarrow \ceil*{\frac{e}{\epsilon}}$
    \\
    \hline
    $d$ 
    & number of CMS hash functions and corresponding rows in $\mathit{Count}$
    & $d \leftarrow \ceil*{\ln{\frac{1}{\delta}}}$, 
    hash functions
    $h_1 \dots h_d$ map flows' ids $[n]$
    to counters' indices $[w]$
    \\
    \hline
    \hline
    $k$ 
    & number of data nodes 
    & routers, switches or servers
    \\
    \hline
    $f$ 
    & number of concurrent failures, 
    $f \leq k$ 
    & set $f \leftarrow 1$ to tolerate any single failure
    \\
    \hline
    $r$ 
    & number of redundant nodes, 
    $f \leq r \leq f \cdot k$ 
    & with distributed redundancy 
    $f \leq r \leq k$
    \\
    \hline
    $p$
    & number of sketch's partitions, each of which is shared with $f$ redundant nodes
    & balance the communication and space overhead across the redundant nodes
    \\
    \hline
    $B$ 
    & batch size -- maximum number of items to be delayed between consecutive shares 
    & batch counters are shorter, 
    $1 \leq B \leq N$
    \\
    \hline
    $\alpha$ 
    & load threshold of data structure,
    $0< \alpha \leq 1$;
    for array, $\alpha=1$
    & need to resize whenever its load factor reaches $\alpha$
    \\
    \hline
    \end{tabular}
    \caption{Notations.}
    \label{tab:notations}
    }
\end{table}
}

\subsection{The Role-based Approach}
Our goal is to share redundant information regarding each node's sketch elsewhere, so if that node fails, it would be possible to recover the missing data.
To that end, we refer to the nodes holding sketches capturing data streams flowing through them as \emph{data nodes}.
In particular, we have $k$ data nodes.
We also introduce into the model \emph{redundant nodes}, whose purpose is to store redundant information needed for the recovery of failed~sketches.

We consider two main use cases:
($i$) redundant nodes only store redundant information, in which case we refer to them as \emph{dedicated redundant nodes}, and assume that they can communicate reliably with each other as well as with all the data nodes; 
alternatively, ($ii$) each data node may also serve as a redundant node, in which case we have \emph{distributed redundancy}, where a set of redundant nodes is a subset of data nodes.
In case ($i$), the roles are exclusive, i.e., a node can be either \emph{data} or \emph{redundant}, but not both. 
Hence, a pure redundant node can be realized as a centralized controller or backup server.
In case ($ii$), in contrast to RAID 5, every node holds its own original sketch (no data striping), while it may (or may not) provide backup services to some others. 
With this approach, the nodes perform as a distributed backup service, storing redundant sum-sketches next to their original~sketches.

\subsection{System Model in Linear Vector Space}
In our model, each data node holds its own original sketch. 
Hence, the original sketch of a node $i$ can be represented as a row-vector in $\mathbb{R}^{k}$, containing $1$ in its $i^\mathrm{th}$ entry, while all other entries contain $0$.
In other words, in a system with data nodes 
$\vec{D}=(D_1, \dots, D_{k})$, 
their original sketches are represented as vectors of the standard basis of $\mathbb{R}^{k}$, forming the identity matrix ${I}_{k}$.
Similarly, redundant nodes are represented as redundant vectors, indicating the linear combination of data nodes' sketches that they~cover.

\begin{observation}
To recover a failed data node $j$, a redundant vector with non-zero $j^\mathrm{th}$ entry is~essential.
\end{observation}
\section{Design Guidelines}
\label{chap:design}
During the design of the system, we explore some trade-offs that might reduce the total environmental impact in terms of resource consumption. 
Our key~guidelines:

\begin{description}
\item[Storage.] 
Trade recovery traffic for redundant space, using \emph{sum-sketches} instead of full replicas.
Following the approach from \autoref{sec:contributions}, redundant information can be constructed by summing the original (data) sketches into redundant sum-sketches, requiring less memory than storing full~replicas.

\item[Communication.] 
Reduce maintenance traffic, using \emph{incremental updates} organized in batches.
We generally prioritize reducing unnecessary communication traffic during the runtime.
Thus, we also embrace compact message encoding and early-transmission~avoidance. 

\item[Runtime.]
Reduce the number of rehash operations and improve performance, providing a data structure with \emph{initial capacity} upon memory allocation.
Don't forget to take into account the load threshold of a data structure, 1.0 by default.
Various data structures can be used to represent a batch and can be plugged into our framework via an abstract API.  
A dynamic data structure starts with some default capacity and might need to be resized (e.g., the capacity of \verb|Java| $\mathit{HashMap}$ is doubled each time it reaches the load threshold, $0.75$ by default~\cite{Java-HashMap}).
In case of hash tables, resizing also implies rehashing the elements, which can be a computationally costly operation in stream~processing.

\item[Memory.] 
Avoid memory re-allocation.
This can be done by overallocating the memory in advance, but we also need to take into account the space~limitations.

\item[Computation.]
Reduce an overall computation and total resource consumption, 
using the strategy in which the data nodes perform \emph{local computations} once and encode the share of redundant content in an easy-to-consume manner, implying fast updates at the redundant~nodes. 
\begin{note}
To tolerate up to $f$ concurrent failures, 
each data node must be covered by at least $f$ redundant~nodes.
\end{note}
\end{description}

\section{Redundancy Strategies}
\label{chap:redundancy}
Redundancy is essential for recoverability of a failed sketch.
Upon adding some redundancy into the system, various solutions may differ in: ($i$) extra space they require to support redundancy (beyond the space required to store the original sketches at data nodes), 
($ii$) communication overhead they impose in order to maintain this redundant information, and ($iii$) recovery traffic that is needed to restore failed data node~sketch.

Denote $k$ the total number of data nodes in the system, $f$ the number of concurrent failures that we wish to be able to recover from, and $r$ the number of redundant nodes holding redundant information.
To recover from any $f$ concurrent failures, each data node must back up at least $f$ copies of its sketch at different redundant nodes.
Thus, at least\footnote{We do not use compression due to its computational overhead, which is too high for in-network stream processing.} $f$ redundant sketches are required to hold redundant information of the entire system (e.g., using \emph{erasure code redundancy}). 
Considering the case of a fully \emph{replication redundancy} as a sufficient upper bound, we derive the range $f \leq space \leq f \cdot k$, where $space$ is equivalent to the number of redundant sketches within a~system.

Optimal erasure coding implies the lowest redundant \emph{space}, at the expense of high \emph{recovery traffic}, requiring $k$ nodes to recover failed data first and all $k$ data nodes to reconstruct every failed redundant node next.
In contrast, full replicas imply the lowest recovery traffic, since only a single sketch is sent during the recovery of a failed node.
In \autoref{sec:RAID4-ext} we introduce an erasure-like method to generate $f$ redundant sum-sketches for compact backups, which require less recovery traffic in some~cases.

Recall that summing CMSs results in sum-sketch, which is also a CMS.
Positive integer coefficients of a sum imply that redundant counters fit the non-negative case of the CMS, and never underestimate the counts of each component CMS within a sum. 
We assume that the counters of a sum-sketch are wide enough to avoid overflow associated with regular arithmetic. 
Thus, one of our goals is to use small integer~coefficients.

We start with $f=1$, (i.e., our initial goal is to recover from any single failure), and then extend to a general case. 
Inspired by RAID, we consider two main strategies to maintain the redundant sum-sketches -- dedicated vs.\ distributed~redundancy.

\subsection{Dedicated Redundancy}
\label{sec:dedicated-strategy}
Using this strategy, we are able to tolerate $f \leq k$ concurrent data failures. 
For this purpose, in addition to the $k$ data nodes, $r \geq f$ dedicated redundant nodes are allocated for backups, resulting in the overall $(k+r)$ nodes in the~system.

For $f=1$, similar to RAID~4, a single dedicated redundant node is sufficient to hold the sum of all the data nodes' sketches.
Denote $R=\sum_{i=1}^{k} D_{i}$ the sum-sketch held by the dedicated redundant node.
To recover the sketch of a failed data node $i$, we simply compute 
$D_{i}={R}-\sum_{j \neq i} D_{j}$,
where the plus and minus operations are performed as matrix~operations.

For any $f \leq k$, our goal is to be able to recover the sketches held by the data nodes, despite the failure of any $f$ nodes or less.
Using erasure code redundancy, once the failed data nodes are recovered, the failed redundancy can be reconstructed.
In terms of linear algebra, this can be realized by generating a system of $(k+r)$ linear equations with $k$ variables.
For a unique solution to exist, a set of vectors that {span $\mathbb{R}^{k}$} is~required. 

\begin{corollary}
Any $k$ linearly independent vectors must be available at all~times.
\end{corollary}

\subsubsection{Extending RAID for \emph{f} Erasures}
\label{sec:RAID4-ext}
We now introduce our simple method for producing exactly $r=f$ linearly independent redundant vectors in $\mathbb{R}^{k}$.
These row-vectors construct an elastic redundant matrix ${MR}_{f} \in \mathcal{M}_{f \times k}(\mathbb{R})$, where any subset of $f$ out of its $k$ columns spans $\mathbb{R}^{f}$.
Refer to \autoref{lemma:f-span} and further discussion for more~details.

\begin{algorithm}[ht]
    \caption{Generate redundant matrix ${MR} \in \mathcal{M}_{f \times k}(\mathbb{R})$}
    \label{alg:RedundantMatrix}
    \small
    \begin{algorithmic}
        \Defs{Inputs}{
        $k$ -- number of data nodes\\
        $f$ -- number of concurrent failures, $f \leq k$\\
        } 
        
        \Procedure{Init}{$f, k$}
            \State ${MR} \gets new\ \mathit{Matrix}(f, k)$ 
            \Comment{$f$ rows, $k$ columns, each cell is $k$ bits long}
            \State ${MR}[1] \gets 1_{k}$
            \Comment{First row is full of 1's}
            
            \For{$row \gets 2$ to $f$}
                \State ${MR}[row][1] \gets 1$
                \Comment{First column is full of 1's}
                \For{$col \gets 2$ to $k$}
                \Comment{Fill cells inductively}
                    \State $MR[row][col] \gets MR[row-1][col-1]+MR[row][col-1]$
                \EndFor
            \EndFor
        \EndProcedure
    \end{algorithmic}
\end{algorithm}

The full ${MR}_k$ can be generated during the pre-processing stage, and then its $f$ leading rows can be used for any given $f$.
The first row and column are filled with $1$'s, holding the base case $f=1$.
Then, all other entries are calculated inductively, depending on the preceding column's values of the current row and the row above it.
Using one-based indices, we produce ${MR}_5$ as an example for $k=5$ and compare it to a Vandermonde matrix ${V}$, where each entry~${[V]}_{i,j}=i^{j-1}$. 

\begin{center}
\footnotesize
$MR_5 = 
\begin{bmatrix}
1 & 1 & 1 & 1 & 1\\
1 & 2 & 3 & 4 & 5\\
1 & 2 & 4 & 7 & 11\\
1 & 2 & 4 & 8 & 15\\
1 & 2 & 4 & 8 & 16
\end{bmatrix}$ 
vs
${V}=\begin{bmatrix}
1^0 & \ 1^1 & \ 1^2 & \ 1^3 & \ 1^4\\
2^0 & 2 & 4 & 8 & 16\\
3^0 & 3 & 9 & 27 & 81\\
4^0 & 4 & 16 & 64 & 256\\
5^0 & 5 & 25 & 125 & 625
\end{bmatrix}$
\end{center}

We note that ${[MR]}_{i,j} \leq {[V]}_{i,j}$ and ${[MR]}_{i,j} \leq {[V]}_{j,i}={[V^\intercal]_{i,j}}$, 
due to its small coefficients, which are bounded by $2^{k-1}$, while with a Vandermonde matrix the bound is $k^{k-1}$.
Small integer coefficients are sufficient for our purpose, since we deal with CMS erasures.
As CMS counts are also integers, we expect quicker execution associated with hardware~operations.

\begin{definition}
Denote 
${G}_{(k+f) \times k}
= 
\begin{bsmallmatrix}
{I}_k \\
{MR}_f
\end{bsmallmatrix}$ 
the generation matrix of a system with data nodes 
$\vec{D}=(D_1, \dots, D_k)$ 
and dedicated redundant nodes, 
denoting their sum-sketches 
$\vec{R}=(R_1, \dots, R_f)$. 
Hence, the 
overall nodes and their corresponding 
sketches are represented by 
${G} \vec{D}^\intercal = 
\begin{bsmallmatrix}
{I}_k\\
{MR}_f
\end{bsmallmatrix}
\vec{D}^\intercal
=
(D_1, \dots, D_k, R_1, \dots, R_f)^\intercal$.
\end{definition}

\begin{observation}
$MR_f$'s construction using shifted sums ensures that its $f$ redundant row-vectors are linearly independent of any subset of $(k-f)$ data vectors, meaning that even after $f$ \emph{data} erasures, we are left with $k$ linearly independent vectors, revealing a solution to recover all failed~data.
\end{observation}
Being able to recover from \emph{any two} concurrent failures is usually considered~satisfactory. 

\begin{observation}[Optimal two-erasures code]
For $f \leq 2$, ${G}$ has the property of \emph{optimal erasure codes}, 
where \emph{any} $k$ out of the $(k+f)$ nodes are sufficient to recover all failed data nodes first and reconstruct redundancy~next.
\end{observation}

\paragraph*{Elastic matrix.}
Easy-scale adapting to changes in parameters $k$ or $f$.
For example, we visualize the transition from $k=4$ and $f=2$ to $k'=5$ and $f'=3$, where $MR_{2}$ is nested within~$MR_{3}$.
\begin{center}
\footnotesize
$MR_{2 \times 4}=
\begin{bmatrix}
1 & 1 & 1 & 1\\
1 & 2 & 3 & 4\\
\end{bmatrix}$ 
and 
$MR_{3 \times 5}=
\begin{bmatrix}
1 & 1 & 1 & 1 & | & 1\\
1 & 2 & 3 & 4 & | & 5\\
\hline
1 & 2 & 4 & 7 & | & 11\\
\end{bmatrix}$
\end{center}
\subsubsection{Data Recovery in Dedicated Setting}
\label{sec:dedicated-recovery}
Recall that the generation matrix of a system with dedicated redundancy consists of $I_k$, representing the data, and $MR_f$ representing the redundancy. 
When a redundant node fails, its sum-sketch is reconstructed according to a linear combination of data nodes, defined by the corresponding redundant vector. 
However, when a data node fails, its row-vector becomes $0_k$, i.e., erased.
As shown for $f=1$, 
when a redundant vector consists of all non-zero coefficients, it can be used together with the other $(k-1)$ non-failed data vectors to recover the failed one.
Similarly, $MR_f$ consists of such redundant vectors, covering all the~data.

To recover the failed data nodes, we replace each erased data vector with the first redundant vector still available, 
and then we need to solve the equations left (e.g., by Gaussian elimination).
Notice that replacing the erased data vectors in such an order results in $k$ leading independent vectors, which form an invertible matrix~$k \times k$.

\begin{example}
\label{ex:dedicated-d1-d2-d3}
Let $(D_1, \dots, D_5)=I_5 \vec{D}^\intercal$ the five data nodes.
For $f=3$, we construct the dedicated redundant sum-sketches, according to the coefficients in $MR_{3}$.
Any 3 columns of $MR_{3}$ are linearly independent in $\mathbb{R}^{3}$, and hence, for any 3 erased variables (data nodes), there exists a \emph{unique solution} to a corresponding system of linear equations.
Suppose that $D_1, D_2$ and $D_3$ failed, i.e., their data vectors were~erased. 
\begin{center}
\footnotesize
$
{G} \vec{D}^\intercal =
\begin{bmatrix}
{I}_5 \\
{MR}_3
\end{bmatrix}
\vec{D}^\intercal=
\begin{matrix}
\begin{pmatrix}
\msout{D_1}\\ 
\msout{D_2}\\ 
\msout{D_3}\\ 
D_4\\ 
D_5
\end{pmatrix}\\*
\begin{pmatrix}
R_1\\ 
R_2\\ 
R_3
\end{pmatrix}
\end{matrix}
=
\begin{matrix}
\begin{bmatrix}
\msout{1} & \msout{0} & \msout{0} & \msout{0} & \ \msout{0}\\
\msout{0} & \msout{1} & \msout{0} & \msout{0} & \ \msout{0}\\
\msout{0} & \msout{0} & \msout{1} & \msout{0} & \ \msout{0}\\
0 & 0 & 0 & 1 & \ 0\\
0 & 0 & 0 & 0 & \ 1
\end{bmatrix}\\*
\begin{bmatrix}
1 & 1 & 1 & 1 & 1\\
1 & 2 & 3 & 4 & 5\\
1 & 2 & 4 & 7 & 11
\end{bmatrix}
\end{matrix}
\begin{pmatrix}
D_1\\ 
D_2\\ 
D_3\\ 
D_4\\ 
D_5
\end{pmatrix}
$
\end{center}
To recover these nodes, simply replace the erased rows of ${I}_5$ with the available redundant rows from $MR_{3}$ and then, invert the resulting matrix revealing the solution for data recovery, e.g., $D_1=(2R_1-2R_2+R_3-D_4-3D_5)$.
\begin{center}
\footnotesize
$
\begin{pmatrix}
R_{1}\\ 
R_{2}\\ 
R_{3}\\ 
D_4\\ 
D_5
\end{pmatrix} 
=
\begin{bmatrix}
1 & 1 & 1 & 1 & 1\\
1 & 2 & 3 & 4 & 5\\
1 & 2 & 4 & 7 & 11\\
0 & 0 & 0 & 1 & 0\\
0 & 0 & 0 & 0 & 1
\end{bmatrix}
\begin{pmatrix}
D_1\\ 
D_2\\ 
D_3\\ 
D_4\\ 
D_5
\end{pmatrix}
\Rightarrow
\begin{pmatrix}
D_1\\ 
D_2\\ 
D_3\\ 
D_4\\ 
D_5
\end{pmatrix}=
\begin{bmatrix}
\ \ 2 & -2 & \ \ 1 & -1 & -3\\
-1 & \ \ 3 & -2 & \ \ 3 & \ \ 8\\
\ \ 0 & -1 & \ \ 1 & -3 & -6\\
\ \ 0 & \ \ 0 & \ \ 0 & \ \ 1 & \ \ 0\\
\ \ 0 & \ \ 0 & \ \ 0 & \ \ 0 & \ \ 1
\end{bmatrix}
\begin{pmatrix}
R_{1}\\ 
R_{2}\\ 
R_{3}\\ 
D_4\\ 
D_5
\end{pmatrix}$
\end{center}
\end{example}

\paragraph{Case \emph{f} = \emph{k}.}
In this special case, the full ${MR}_k$ can be generated by \autoref{alg:RedundantMatrix} during the pre-processing stage, and its inverse matrix can also be~calculated. 

\begin{example}
Suppose that all data nodes failed.
We use pre-calculated $MR_5^{-1}$ to reveal the solution for data recovery of every original sketch, which in some cases requires less than $k$ nodes to recover, e.g., $D_5=(-R_4+R_5)$.
\begin{center}
\footnotesize
$
\begin{pmatrix}
R_1\\ 
R_2\\ 
R_3\\ 
R_4\\ 
R_5
\end{pmatrix} 
=
\begin{bmatrix}
1 & 1 & 1 & 1 & 1\\
1 & 2 & 3 & 4 & 5\\
1 & 2 & 4 & 7 & 11\\
1 & 2 & 4 & 8 & 15\\
1 & 2 & 4 & 8 & 16
\end{bmatrix}
\begin{pmatrix}
D_1\\ 
D_2\\ 
D_3\\ 
D_4\\ 
D_5
\end{pmatrix}
\Rightarrow
\begin{pmatrix}
D_1\\ 
D_2\\ 
D_3\\ 
D_4\\ 
D_5
\end{pmatrix}=
\begin{bmatrix}
\ \ 2 & -2 & \ \ 2 & -2 & \ \ 1\\
-1 & \ \ 3 & -5 & \ \ 7 & -4\\
\ \ 0 & -1 & \ \ 4 & -9 & \ \ 6\\
\ \ 0 & \ \ 0 & -1 & \ \ 5 & -4\\
\ \ 0 & \ \ 0 & \ \ 0 & -1 & \ \ 1
\end{bmatrix}
\begin{pmatrix}
R_1\\ 
R_2\\ 
R_3\\ 
R_4\\ 
R_5
\end{pmatrix}$
\end{center}
\end{example}

\subsection{Distributed Redundancy}
\label{sec:distributed-strategy}
With this strategy, redundancy is distributed across existing $k$ data nodes, each of which can also serve as a redundant node.
This time, when a node goes down, both its original sketch (aka data) and the redundant information that it held for others fail. 
Hence, we are able to tolerate no more than $\floor*{\frac{k}{2}}$ concurrent failures. 
For more details, refer to \autoref{lemma:2-tolerance-distributed}, which shows a peer-to-peer backup scheme with $k=4$ peers, which is resilient up to \emph{two} peer~failures.

\begin{lemma}
When redundancy is distributed across existing data nodes, 
more redundant sketches are required than with dedicated redundant nodes, i.e., $space > f$.
\end{lemma}
\begin{proof}
We show that it holds for the base case $f=1$.
A node $i$ can cover up to the rest of $(k-1)$ data nodes.
However, to be able to recover from $i$'s failure as well, 
some other node $j$ must cover $i$'s data (within its redundant sketch).
Hence, using this strategy, even for a single failure resilience, at least two redundant sketches are~required.
\end{proof}
\subsubsection{Data Recovery in Distributed Setting} 
\label{sec:distributed-recovery}
In order to distinguish between the roles of a node $i$, denote $D_i$ the original (data) sketch and $R_i$ the redundant sum-sketch held by $i$. 
For any $f \leq \floor*{\frac{k}{2}}$, 
the full ${MR}_k$ is generated by \autoref{alg:RedundantMatrix}, during the pre-processing stage.
Then, $MR_k'$ is produced by circular displacement of rows from $MR_k$, such that its $((i+\ceil*{\frac{k}{2}})\mod{k})^\mathrm{th}$ row-vector is associated with $R_i$. 
Therefore, the generation matrix of such a system is 
${G'}_{k \times 2k} = 
\begin{bmatrix}
{I}_k | {MR}_k'
\end{bmatrix}$.

When a node $i$ fails, its data vector is erased and its redundant vector becomes unavailable.
As before, we first recover the data and then reconstruct the redundancy.
Similarly to the recovery procedure described in~\autoref{sec:dedicated-recovery}, we replace each erased data vector with the first redundant vector still~available.

\begin{example}
Let $k=5$ and suppose that nodes 1 and 3~failed.
\begin{center}
\footnotesize
$
\begin{pmatrix}
\msout{D_1}\\ 
{D_2}\\ 
\msout{D_3}\\ 
D_4\\ 
D_5
\end{pmatrix},
\begin{pmatrix}
\msout{R_1}\\ 
{R_2}\\ 
\msout{R_3}\\ 
R_4\\ 
R_5
\end{pmatrix}=
\begin{matrix}
\begin{bmatrix}
\msout{1} & \msout{0} & \msout{0} & \msout{0} & \msout{0}\\
{0} & {1} & {0} & {0} & {0}\\
\msout{0} & \msout{0} & \msout{1} & \msout{0} & \msout{0}\\
0 & 0 & 0 & 1 & 0\\
0 & 0 & 0 & 0 & 1
\end{bmatrix}, 
\begin{bmatrix}
\msout{1} & \msout{2} & \msout{4} & \msout{8} & \msout{15}\\
1 & 2 & 4 & 8 & 16\\
\msout{1} & \msout{1} & \msout{1} & \msout{1} & \msout{1}\\
1 & 2 & 3 & 4 & 5\\
1 & 2 & 4 & 7 & 11
\end{bmatrix}
\end{matrix}
$
\end{center}

After replacement, we get $(R_2, D_2, R_4, D_4, D_5)$. 
The inverse of the resulting matrix reveals the solution for data recovery, e.g., $D_3=(R_2-R_4-4D_4-11D_5)$.

\begin{center}
\footnotesize
$
\begin{pmatrix}
R_2\\ 
D_2\\ 
R_4\\ 
D_4\\ 
D_5
\end{pmatrix} 
=
\begin{bmatrix}
1 & 2 & 4 & 8 & 16\\
0 & 1 & 0 & 0 & 0\\
1 & 2 & 3 & 4 & 5\\
0 & 0 & 0 & 1 & 0\\
0 & 0 & 0 & 0 & 1
\end{bmatrix}
\begin{pmatrix}
D_1\\ 
D_2\\ 
D_3\\ 
D_4\\ 
D_5
\end{pmatrix}
\Rightarrow
\begin{pmatrix}
D_1\\ 
D_2\\ 
D_3\\ 
D_4\\ 
D_5
\end{pmatrix}=
\begin{bmatrix}
-3 & -2 & \ \ 4 & \ \ 8 & \ \ 28\\
\ \ 0 & \ \ 1 & \ \ 0 & \ \ 0 & \ \ 0\\
\ \ 1 & \ \ 0 & -1 & -4 & -11\\
\ \ 0 & \ \ 0 & \ \ 0 & \ \ 1 & \ \ 0\\
\ \ 0 & \ \ 0 & \ \ 0 & \ \ 0 & \ \ 1
\end{bmatrix}
\begin{pmatrix}
R_2\\ 
D_2\\ 
R_4\\ 
D_4\\ 
D_5
\end{pmatrix}$
\end{center}
\end{example}

\begin{note}
In a distributed setting, a redundant node has access to its original sketch due to its data node role. 
As there is no need for a redundant node to cover itself, a modified redundancy matrix $MR_k''$ can be used, which differs from $MR_k'$ by having a zero~diagonal.
\end{note}
\subsubsection{Sketch Partitioning}
\label{sec:sketch-partitioning}
To enable fine grain redundant information sharing, we introduce the concept of \emph{sketch partitioning}, where data nodes divide their sketches into $p$
non-overlapping partitions, another global parameter derived during the pre-processing stage.
Every data node is mapped\footnote{We introduce coverage mapping in \autoref{sec:coverage-mapping}} to $r_c \leq r$ redundant nodes, 
each of which covers some of its partitions, and together they cover $f$ full copies of its sketch.
Each redundant node holds at least one partition and may serve up to the $k$ data nodes.
Also, each data node knows the partition every sketch counter is mapped to, and which $f$ redundant nodes cover it, while each redundant node knows which partitions of which data nodes it~covers.

\begin{figure}[ht]
    \begin{subfigure}[b]{0.3\textwidth}
        \includegraphics[width=\textwidth]{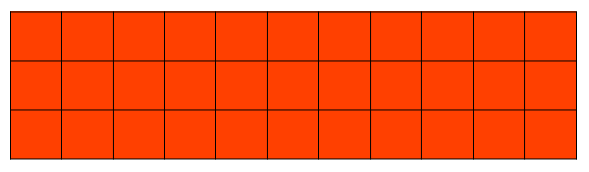}
        \caption{single partition}
        \label{fig:part-table}
    \end{subfigure}
    \hfill
    \begin{subfigure}[b]{0.3\textwidth}
        \includegraphics[width=\textwidth]{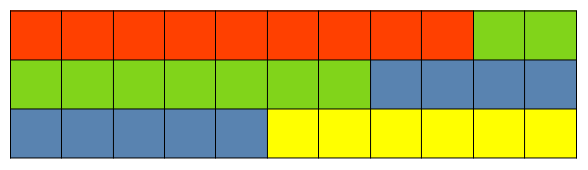}
        \caption{partitioning by cells}
        \label{fig:part-cells}
    \end{subfigure}
    \hfill
    \begin{subfigure}[b]{0.3\textwidth}
        \includegraphics[width=\textwidth]{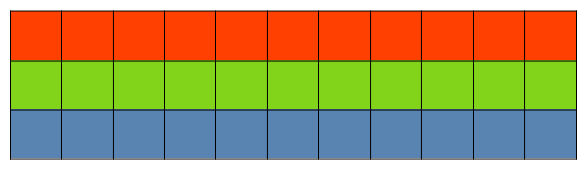}
        \caption{partitioning by rows}
        \label{fig:part-rows}
    \end{subfigure}
    \caption{Sketch partitioning types}
    \label{fig:partitioning}
\end{figure}

Denote $cpp$ the upper bound for consecutive cells-per-partition.
Practically, the sketch can be viewed as a one-dimensional array of $d$ concatenated rows, where its partitions contain exactly $cpp$ cells each, except for the last partition containing the leftover cells.
When $p=1$, the entire sketch is mapped to a $single$ partition, as shown in \autoref{fig:part-table}, and hence there are only $r_c=f$ covering nodes.
With partitioning by $cells$ (\autoref{fig:part-cells}) a data node is able to spread its sketch uniformly across its covering nodes.
Thus, a sketch row may appear on more than one partition. 
With partitioning by $rows$ (\autoref{fig:part-rows})
the entire row (its first counter through the last one) 
is a member of a particular partition, implying that $p \leq \min \{d, r_c\}$.

\begin{example}
\autoref{fig:partitioning} illustrates the three types of sketch partitioning and their impact on the ability to engage as many other nodes as possible.
It reflects a system with $k=5$ data nodes, using distributed redundancy to tolerate any single failure, where the redundant nodes cover other data nodes, $p \leq (k-f)=4$.
Given the parameters 
$(\epsilon, \delta)=(\frac{1}{4}, \frac{1}{10})$, 
the sketch capacity is defined by $d \leftarrow \ceil*{\ln{\frac{1}{\delta}}}$ rows and $w \leftarrow \ceil*{\frac{e}{\epsilon}}$ columns,
i.e., CMS generates a matrix of $3 \times 11$ counters.
With partitioning by $cells$, the sketch can be divided into $p=4$ partitions, where the first 3 partitions contain 9 cells each, and the $4^\mathrm{th}$ contains the remaining 6 cells.
However, with partitioning by $rows$, no more than 3 partitions can~coexist. 
\end{example}

\begin{note}
When CMS is partitioned by rows, each of the resulting partitions is also a CMS, but with lower confidence.
For example, a CMS with 10 rows has a confidence of five nines ($99.999\%$), while a CMS with 5 rows has only two~($99\%$).
\end{note}
\subsubsection{Coverage Mapping}
\label{sec:coverage-mapping}
We now combine the sketch partitioning into distributed redundancy, and briefly describe three coverage mapping types, which slightly differ in their preferences regarding the space-recovery trade-off.
The mapping, or its generation algorithm, is given to the system, such that all the nodes are globally configured with the same mapping, defining the relations between the nodes, which form a peer-to-peer backup~system.

\begin{figure}[ht]
    \centering
    \begin{subfigure}[b]{0.32\textwidth}
        \centering
        \includegraphics[width=\textwidth]{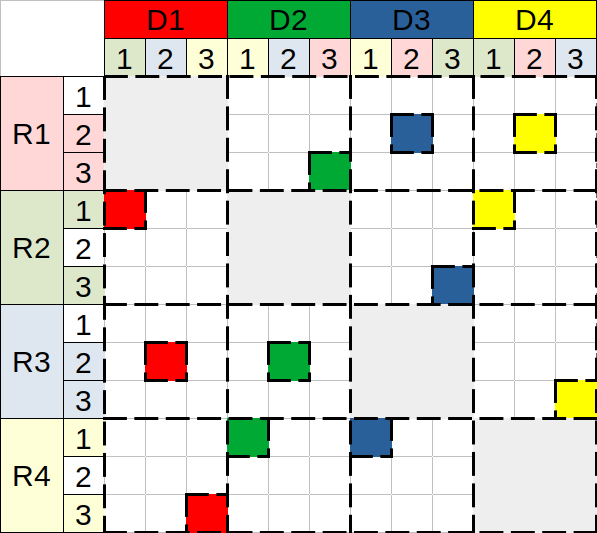}
        \caption{balanced traffic-oriented}
        \label{fig:map-clique}
    \end{subfigure}
    \hfill
    \begin{subfigure}[b]{0.32\textwidth}
        \centering
        \includegraphics[width=\textwidth]{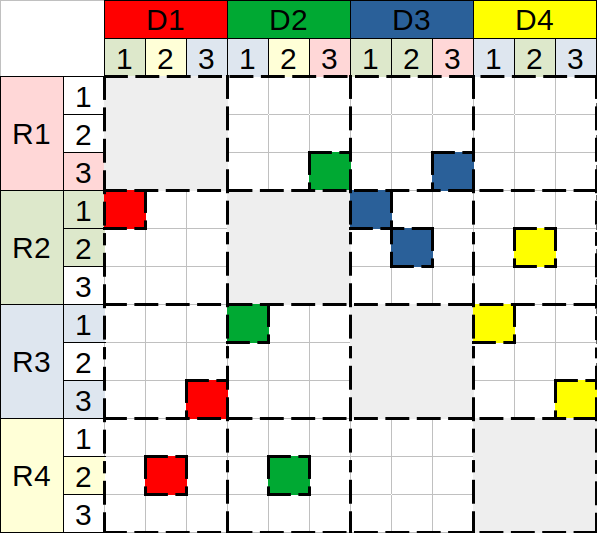}
        \caption{imbalanced space-oriented}
        \label{fig:map-imbalanced-space}
    \end{subfigure}
    \hfill
    \begin{subfigure}[b]{0.32\textwidth}
        \centering
        \includegraphics[width=0.73\textwidth]{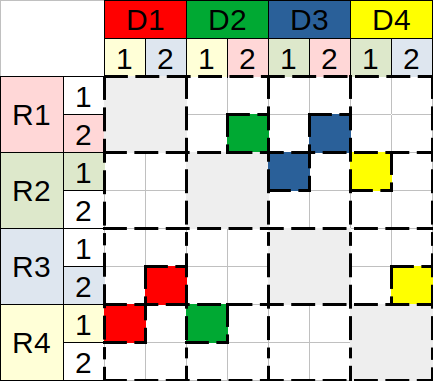}
        \caption{optimized}
        \label{fig:map-sweet-spot}
    \end{subfigure}
    \caption{Coverage mapping types}
    \label{fig:mapping}
\end{figure}

\autoref{fig:mapping} reflects a system with $k=4$ data nodes, 
using distributed redundancy to tolerate any single failure, $f=1$, 
where the redundant nodes cover other data nodes. 
Hence, for such a system, $p \leq 3$ and $space \geq 2$.
The mapping is illustrated using a single coverage matrix, which can be divided into a series of partition-specific mappings with respective matrices.
The columns represent the partitioned data, denoted $D_i$ for a node~$i$, while the rows represent the redundant partitions that a node covers.
The redundant space of a node is therefore proportional to the number of partitions it actually covers, which are colored next to each $R_i$.
When a node fails, all of its partition-rows become invalid.
To validate them back again and fully recover a node, we need to recover its data and reconstruct the redundancy it held prior to~failure.

\paragraph{Evolution of Mapping Types.}
We begin with maximal $p=3$ and aim to achieve a balanced redundant space. 
Using a clique topology, a node $i$ covers some partitions of all other nodes, each of which covers a different partition of $D_i$. 
The redundant vectors of the first mapping, shown in \autoref{fig:map-clique}, vary in the number of their non-zero coefficients (e.g., 
$R_1^{(2)}=D_3^{(2)}+D_4^{(2)}$ but $R_1^{(3)}=D_2^{(3)}$). 
This implies that the overall redundant space is not minimal. 
In fact, we get $\frac{8}{3}>2$ as each of the 4 nodes covers 2 out of 3 partitions.
However, this type is slightly better for recovery traffic than space-oriented types, since it requires only $\frac{8}{3}<3$ sketches to fully recover a failed~node.

Towards the space-oriented mapping, 
we define all (non-zero) redundant vectors to contain the same amount of non-zero coefficients.
The minimal $space=2$ is achieved in \autoref{fig:map-imbalanced-space}, 
using six redundant partitions in total. 
However, they cannot be divided equally among the four nodes.
As a result, the redundant space per node is imbalanced, implying a slight variance in recovery traffic, $3 \pm \frac{1}{3}$.
Such a mapping can be useful in a heterogeneous setting, where the nodes differ in their capabilities (e.g., due to an overlap period during storage upgrade).
Hence, delegating more work to the capable ones suits~\emph{fairness design}. 

Aiming to achieve an optimized mapping which is both balanced and space-oriented, we tune $p$ by reducing its prior value, and propose a final mapping, described in \autoref{fig:map-sweet-spot} for $p=2$, which consists of four redundant vectors (one sum-partition for each node), having exactly two non-zero coefficients each. 
In addition to its space properties, this mapping requires exactly $3$ sketches to fully recover any of the failed~nodes.

\begin{example}
Suppose node 1 failed. 
To recover its partitioned data, we need to compute 
$D_1^{(1)}=R_4^{(1)}-D_2^{(1)}$ and 
$D_1^{(2)}=R_3^{(2)}-D_4^{(2)}$, 
while for redundancy reconstruction, we require 
$R_1^{(2)}=D_2^{(2)}+D_3^{(2)}$, i.e., 
each of the other nodes sends two~partitions.
\end{example}

\section{Periodic Updates}
\label{chap:updates}
We examine the sharing of redundant information upon regular time intervals, and assume that the maximum number of arrivals per time \emph{interval} is bounded by a batch size $B \leq N$. 
The nodes operate in cycles, running the same protocol while exchanging messages over a reliable end-to-end communication service,
i.e., the time interval encapsulates the retransmissions performed by the lower-level protocols we rely~on.

We assume timely delivery of messages, as well as that each node is equipped with an incoming message queue having large enough memory to buffer all the messages that arrive during the cycle.
In case a message from a certain data node fails to arrive on time, we consider this to be a crash failure, and initiate the \emph{recovery} procedure of its~sketch.
If a data node has nothing to share, it sends the minimum message signalling that it is still \emph{alive}.
On each cycle, a redundant node reads all the incoming messages and updates its sum-sketch accordingly.
Note that in practice, when a message containing a sketch update is delivered at a redundant node, handling this message temporarily takes a corresponding extra space until the sketch data is consumed, and its memory is~freed.

\subsection{Update Policies}
\label{sec:share-policies}
When $B=1$, no delay is activated, and we update the redundant nodes on each item arrival. 
Otherwise, to maintain redundancy, we consider the following policies.

\paragraph{Full Share policy.} 
With this policy, the data nodes send their entire sketches causing the redundant nodes to reconstruct their sum-sketches from scratch, i.e., to sum all the data sketches, which were delivered by the start of a cycle.
Recall that each sketch counter is associated with a particular partition.
Given a coverage mapping from \autoref{sec:coverage-mapping}, a data node encodes each of its partitions by concatenating the relevant range of its $d \cdot w$ sketch counters. 
To do so, the counters are tested for membership.
However, with partitioning by $rows$, all counters in a particular row are mapped into the same partition. 
Hence, it is enough to test only one of them, reducing the number of membership tests.
Moreover, with the $single$ partition, no membership testing is required since all counters are mapped together.
When done with encoding of partitions, a data node constructs a message for each of its covering redundant nodes, consisting of only the partitions it covers.
In such a manner, a data node is able to send the bare minimum of $f$ copies for each counter. 

{
\renewcommand{\arraystretch}{1.2}
\begin{table}[ht]
\footnotesize
{
    \centering
    \begin{tabular}
    {|l|l|c||l|l|}
    \hline
    Type
    & Local space
    & $p_{type}$
    & Payload of a share
    & Extra computation
    \\
    \hline
    \multirow{3}{*}{$\mathit{Count}$}
    & \multirow{3}{*}{$d \cdot w \cdot \bs{N}$}
    & single
    & \multirow{3}{*}{$f \cdot d \cdot w \cdot \bs{N}$}
    &\\
    \cline{3-3}\cline{5-5}
    && rows && local membership: $d$ 
    \\
    \cline{3-3}\cline{5-5}
    && cells && local membership: $d \cdot w$ 
    \\
    \hline
    \end{tabular}
    \caption{Full share of the original sketch}
    \label{tab:tbl-full-share}
}
\end{table}
}

We note that full share is essential for the \emph{recovery} procedure of a failed sketch, which subtracts the non-failed counts from the overall sum. 
However, sending the entire sketch every time interval during \emph{maintenance}, over and over again, implies unnecessary communication overhead, especially when only few changes occurred.
Moreover, a redundant node must wait to receive the updates from all data nodes that under its cover, before an older sum can be safely freed.
Therefore, as a communication-oriented solution, we consider sharing in an incremental manner only the changes caused by the items that arrived during the cycle.

\paragraph{Incremental Share policy.}
With this policy, the data nodes need to encode the changes that occurred during the cycle. 
This can be done by either ($i$) recording each item arrival, ($ii$) capturing the identifiers of the items that arrived and the number of times each of them arrived, or ($iii$) recording the counters that have been changed and by how much.
In options ($i$) and ($ii$), to reduce space requirements, we consider capturing only the identifiers\footnote{In case of long identifiers, one can apply a cryptographic hash function to produce shorter fixed-length fingerprints, which are unlikely to collide.} of the items arrived.
In options ($ii$) and ($iii$) we need counters, but these counters might be much smaller than the original sketch counters, as they only need to count up to a batch size. 
In a general case, and in particular when data rates vary significantly, we may have a different local $1 \leq B_l \leq B$ parameter for each data node, with the corresponding counter sizes. 
An additional local data structure is required to host and manage the batch. 
A batch may contain only a small portion of overall flows, making it
feasible to maintain exact counter-per-flow.

\subsection{Framework API}
\label{sec:framework}
We now introduce a generic CMS framework which, in addition to its normal activity, supports batching, as shown in \autoref{alg:SmartCMS}. 
Then, we examine some batch representations in more detail.

\begin{algorithm}[ht]
    \caption{Abstract CMS with batch, code for data node}
    \label{alg:SmartCMS}
    \footnotesize
    \begin{algorithmic}
        
        \Defs{Locals}{
        $B'$ -- delayed items counter
        }
        
        \Procedure{Init}{$d, w, N, B, mid$}
        \Comment{$mid$ -- maximum identifier, for field width}
            \State $\mathit{Count} \gets new\ \mathit{Matrix}(d, w)$ 
            \Comment{Original CMS, counts up to N}
            
            \State $dt \gets $ \Call{GetBatchRespresentation}{}
            \Comment{Decide on local data type to represent a batch}
            
            \State $ads \gets $ \Call{$dt$.InitBatch}{}
            \Comment{Allocate memory and initiate the representation}
            \State \Call{ResetBatch}{}
        \EndProcedure

        \Event{Arrived($x$)}
            \State \Call{UpdateBatch}{$x$}
        \EndEvent

        \Event{TimeOut}
            \State \Call{IncShare}{}
            \State Reset the $timer$
        \EndEvent

        \Function{Estimate}{$x$}
        \Comment{Original CMS estimation $\hat{c_x}$}
            \State \Return $\min_i{\mathit{Count}[i, h_i(x)]}$ 
            + \Call{$ads$.Estimate}{$x, B'$}
        \EndFunction

        \Procedure{Update}{$x, c_t$}
        \Comment{Update the original CMS}
            \ForAll{$i \in [1..d]$}
                \State $j \gets h_i(x)$ 
                \State $\mathit{Count}[i,j] \gets \mathit{Count}[i,j]+c_t$
                \EndFor
        \EndProcedure

        \Procedure{BulkUpdate}{}
            \State $pair \gets $ \Call{$ads$.GetEnumerator}{}
            \Comment{Walk through all registered pairs}
            \While{\Call{$pair$.MoveNext}{}}
                \State \Call{Update}{$pair.key$, $pair.value$}
                \Comment{By default, $pair.value=1$}
            \EndWhile
        \EndProcedure

        \Procedure{UpdateBatch}{$x$}
            \If{$B' \geq B$}
                \State \Call{IncShare}{}
                \Comment{Share instantly to avoid reallocation}
            \EndIf 
            \State $B' \gets B'+1$
            \Comment{Count the items as they arrive}

                \State \Call{$ads$.Update}{$x, B'$}
                \Comment{Update the additional data structure}
        \EndProcedure

        \Procedure{ResetBatch}{}
            \State $B' \gets 0$
            \Comment{Reset counter of delayed items}
                \State \Call{$ads$.Reset}{}
                \Comment{Reset additional data structure}
        \EndProcedure

        \Procedure{IncShare}{}
        \Comment{Incremental share of batch representation}
            \If{$B' = 0$}
                \State 
                Send minimum message to signal $alive$
            \Else{}
                \State \Call{BulkUpdate}{}
                \Comment{Update the original CMS periodically}
                \State \Call{$ads$.Send}{}
                \Comment{Send items or counters, depending on representation}
                
                \State \Call{ResetBatch}{}
                \Comment{Reset to reuse with next batch}
            \EndIf
        \EndProcedure
        
        \Procedure{FullShare}{}
        \Comment{Full share of $\mathit{Count}$}
            \State \Call{SendCounters}{}
        \EndProcedure
    \end{algorithmic}
\end{algorithm}

Define $B'$ to count the delayed items between the consecutive shares, $B' \leq B$.
Initially, as well as after sharing the batch, a counter $B'$ is zeroed and a batch representation is reset.
When an item $x$ arrives, 
$B'$ is incremented and an item is processed into a batch, where $B'$'s recent value can be used as an index in Buffer~representations. 
As CMS is updated with the batched changes upon sharing, the answer to a point query combines the estimate from a prior cycle and the frequency within the current.

A general scheme for sending procedures is provided in \autoref{alg:SendScheme}. 
Representations that record counters need to implement both $\mathit{EncodeFixed}$ and $\mathit{EncodeVar}$ functions,
while others need to implement $\mathit{EncodeItems}$.
When encoding rows of a fixed length, we can specify a row-width (the actual number of batched elements) into a message's header. 
However, when encoding variable-length rows, which is also the case when the sketch is partitioned by cells, we need to encode these rows so that a redundant node can recognize the sketch row to which each decoded element belongs.
For this purpose, besides the elements associated with a row, we first specify the number of elements to appear on that row.
We call this metadata encoding.

\begin{algorithm}[ht]
    \caption{General scheme for send procedures}
    \label{alg:SendScheme}
    \footnotesize
    \begin{algorithmic}
        \Procedure{SendItems}{}
            \State $msg \gets \Call{EncodeItems}{B'}$
            \ForAll{$r \in \{\text{covering-nodes}\}$}
                \State 
                Send a message $msg$ to a node $r$
                \Comment{Multi-cast to its $r_c$ covering nodes}
            \EndFor
        \EndProcedure

        \Procedure{SendCounters}{}
            \State $\{ Msg[1], \dots, Msg[p] \} \gets \Call{EncodePartitions}{}$
            \Comment{Array of $p$ messages, one per each partition}
            \ForAll{$r \in \{\text{covering-nodes}\}$}
            \Comment{Filter only the covering nodes}
                \State $msg \gets \{\}$
                \Comment{Begin with empty message}
                \For{$part \gets 1$ to $p$}
                \Comment{Construct a message for a node $r$}
                    \If{$r$ covers a partition $part$}
                        \State 
                        \Call{$msg$.Append}{$Msg[part]$}
                        \Comment{Only $f$ nodes cover each partition}
                    \EndIf
                \EndFor
                \State Send a message $msg$ to a node $r$
            \EndFor
        \EndProcedure

        \Function{EncodePartitions}{}
        \Comment{Encode counters divided into partitions}
            \If{$p_{type}=cells$}
                \State \Return \Call{EncodeVar}{$B'$}
                \Comment{Encode variable-length rows with metadata}
            \EndIf
            \State \Return \Call{EncodeFixed}{$p_{type}$, $B'$}
            \Comment{Encode fixed-length rows}
        \EndFunction
    \end{algorithmic}
\end{algorithm}

\subsection{Batch Representations}
\label{sec:batch-rep-intro}
We examine several data structures to economically represent and encode a batch of changes.
We consider two main categories: item-based and counter-based data structures, each of which can be implemented as either a Buffer or a Compact Hash Table.
\autoref{tab:batch-summary} lists the main aspects of their complexity. 
We then compare the representations to each other, as well as to the baseline option of a full share. 

\paragraph{Buffer of Items.}
The na\"ive approach for handling incremental shares is to record the items that arrived in a local buffer $\mathit{ItemBuff}$. 
Upon filling the buffer with at most $B$ items during a cycle, we simply send the content to all the covering nodes, as defined by a coverage mapping.
A receiving redundant node will then extract the identifiers and execute an update procedure against its sum-sketch, recalculating the hash functions for each delivered item. 

Buffer of Items is a simple and easy-to-implement, 
but also has some drawbacks:
\begin{itemize}
\item Extra hash calculations are performed at the redundant nodes upon update.
\item All buffered items must be sent to all its covering nodes, $r_c \geq f$. 
\item There is no room for privacy since items' identifiers are openly shared.
\item Duplicates as different items of a flow are handled independently of each other.
\end{itemize}
To overcome these, we may ($i$) turn to counter-based representations, and/or ($ii$) avoid duplicates through aggregation.

\paragraph{Buffer of Counters.}
Turning to counter-based representation, only a data node performs CMS hash calculations. 
The resulting indices are captured in a buffer of counters $\mathit{CntBuff}_{d \times B}$.
For each counter, a data node knows exactly which $f$ redundant nodes cover it, so it can send the bare minimum of $f$ copies.
A receiving redundant node will then extract the indices and increment the corresponding counts of its sum-sketch.
\begin{note}
As all nodes are identically configured, we cannot guarantee that the privacy is protected with counters.
However, it is clearly better preserved than if sharing items.
\end{note}

\paragraph{Hash Table of Flows.}
Turning to duplicate avoidance, 
$\mathit{FlwHash}$ captures the identifiers and their corresponding frequencies using \emph{key-value} pairs, implying longer elements than with $\mathit{ItemBuff}$. 
Although an extra capacity is required due to the \emph{load threshold} of a hash table itself, recall that with compact hash tables $\alpha \rightarrow 1$, which is a key benefit in space considerations.

\paragraph{Hash Table of Counters.}
Similarly, duplicates in $\mathit{CntBuff}_{d \times B}$ can be resolved using $d$ hash tables, where each $\mathit{Cnt_i{Hash}}$ is associated with a sketch row $i$ and captures the indices and their corresponding frequencies.

\paragraph{Difference Matrix.}
Another approach to track modified counters is with an additional $\mathit{CntDiff}_{d \times w}$ matrix, having the same structure as the original sketch $\mathit{Count}_{d \times w}$, but with shorter entries since difference counters only need to count up to a batch size $B$. 
Although it requires more space than a hash table $\mathit{Cnt{Hash}}$, its computational efficiency is intuitive, as only CMS hashes are calculated. 

\subsubsection{Hosting a Batch}
Recall that within a stream of $N$ items, there are only $n \leq N$ flows (distinct items).
Splitting a stream's timeline by the regular time intervals would result in $q$ cycles, each having an associated batch.
For ease of reference, we assume that the batches are full, i.e., contain $B$ items each.
With incremental share, we zoom in to a given cycle.

Denote $b$ the number of \emph{flows} that arrived during that particular cycle, and $c \approx b \cdot d$ the number of modified sketch \emph{counters}.
Throughout the timeline of the stream, 
$\sum_{i=1}^{q} B_{i} = N$, but 
$\sum_{i=1}^{q} b_{i} \geq n$, since the same flow contributes to every batch in which it appears.
Denote $\beta=\frac{B}{b}$ the average frequency-per-flow within a given batch, then $1 \leq \beta \leq \frac{N}{n}$.

When allocating the memory for the batch, we may initiate the capacity based on the worst-case scenario of $B$ items but potentially end up with a sparse data structure, or use some $\hat{B} \leq B$ as an estimation. 
In the latter case, when the actual $B'$ is about to exceed the estimation, we can either resize the data structure or send the share prematurely before the cycle ends.
Both options imply costly operations associated with dynamic resizing or communication overhead, as a result of extra system calls and power consumption due to additional transmission, which would be critical for devices running on battery.

Although the true values can be discovered in retrospect, we need to decide on batch's representation and allocate the memory in advance.
We could use some estimations of $\hat{\beta}$,  
and once the desired $\hat{\beta}$ is chosen, we can further estimate the number of unique flows within a batch as $\hat{b} \gets \ceil*{\frac{B}{\hat{\beta}}}$, and derive the initial capacity of a hash table for the $\ceil*{\frac{\hat{b}}{\alpha}}$ buckets.
For this purpose, we measure some statistics among different known large Internet traces, where flow frequencies are characterized by skewness. 
We find interest in $\beta_{avg}$ and other percentiles among the batches of each trace, 
and the respective statistics among all the traces we check.

\begin{note}
If such measurements significantly differ from the stream of interest, a learning algorithm can be executed on the fly to tune the hyperparameters and switch to another compact representation. 
\end{note}

\subsubsection{Representations Metrics}
\autoref{tab:batch-summary} lists the main metrics for batch representation analysis, 
e.g., the extra local space they require, the amount of traffic they produce in each cycle, and the extra computation they imply, 
while omitting the notation $O$ for space-saving.
The abbreviations r. and l. stand for remote and local operations, while CMS and HT are the types of hash function to be calculated.
For further evaluation, the metrics are expressed using hyperparameters.

{
\renewcommand{\arraystretch}{1.2}
\begin{table}[ht]
\footnotesize
{
    \centering
    \begin{tabular}
    {|l||l|l|l|}
    \hline
    Type
    & Local space
    & Payload of a share
    & Extra computation
    \\
    \hline
    \hline
    $\mathit{Count}$
    & $d \cdot w \cdot \bs{N}$
    & $f \cdot d \cdot w \cdot \bs{N}$
    & \\
    \hline
    \hline
    $\mathit{ItemBuff}$
    & $B \cdot \bs{mid}$
    & $r_c \cdot B \cdot \bs{mid}$
    & r. CMS: $f \cdot B \cdot d$
    \\
    \hline
    $\mathit{CntBuff}$
    & $d \cdot B \cdot \bs{w}$
    & $f \cdot d \cdot B \cdot \bs{w}$
    & \\
    \hline\hline
    \multirow{2}{*}{$\mathit{FlwHash}$}
    & \multirow{2}{*}
    {$\ceil*{\frac{\hat{b}}{\alpha}} \cdot 
    (\bs{mid} + \bs{B})$}
    & \multirow{2}{0.29\textwidth}
    {$r_c \cdot b \cdot 
    (\bs{mid} + \bs{B})$}
    & \multirow{2}{0.19\textwidth}
    {l. HT: $B$ 
    \newline 
    r. CMS: $f \cdot b \cdot d$}
    \\
    &&&\\
    \hline
    \multirow{2}{*}{$\mathit{CntHash}$}
    & \multirow{2}{*}
    {$d \cdot \ceil*{\frac{\hat{b}}{\alpha}} \cdot 
    (\bs{w} + \bs{B})$}
    & \multirow{2}{0.29\textwidth}
    {$f \cdot d \cdot \bs{B} +\newline
    f \cdot d \cdot b \cdot 
    (\bs{w} + \bs{B})$}
    & \multirow{2}{0.19\textwidth}
    {l. HT: $B \cdot d$ 
    \newline 
    sort: $d \cdot (b \cdot \log_2{b})$}
    \\
    &&&\\
    \hline\hline
    \multirow{2}{*}{$\mathit{CntDiff}$}
    & \multirow{2}{*}
    {$d \cdot w \cdot \bs{B}$}
    & \multirow{2}{0.29\textwidth}
    {$f \cdot d \cdot \bs{B} +\newline
    f \cdot d \cdot b \cdot 
    (\bs{w} + \bs{B})$}
    & \\
    &&&\\
    \hline
    \end{tabular}
    \caption{Batch representations}
    \label{tab:batch-summary}
}
\end{table}
}

\section{Analysis}
\label{chap:eval}
We now examine some batch representations in more detail.

\subsection{Buffer of Items}
\label{sec:ItemBuff}

\begin{algorithm}[ht]
    \caption{ItemBuffCMS - ads (buffer of items) for CMS, code for data node}
    \label{alg:ItemBuffCMS}
    \footnotesize
    \begin{algorithmic}
        \Procedure{Init}{$d, w, N, B, mid$}
            \State $\mathit{ItemBuff} \gets new\ \mathit{Array}(B)$ 
            \Comment{Capacity for $B$ identifiers, same length as $mid$}
        \EndProcedure

        \Procedure{Reset}{}
            \State $\mathit{ItemBuff} \gets 0_{B}$
            \Comment{Resetting index in \autoref{alg:SmartCMS} is enough}
        \EndProcedure

        \Function{Estimate}{$x, cnt$}
        \Comment{Estimation within a batch}
            \State \Return{\Call{Count}{$x, cnt$}}
        \EndFunction

        \Procedure{Update}{$x$, $idx$}
        \Comment{$idx$ is handled by \autoref{alg:SmartCMS}}
            \State $\mathit{ItemBuff}[idx] \leftarrow x$
            \Comment{Each delayed item has its own entry in the buffer, indexed by $idx$}
        \EndProcedure

        \Procedure{Send}{}
            \State \Call{SendItems}{}
            \Comment{Need to implement EncodeItems}
        \EndProcedure
        
        \Function{EncodeItems}{$cnt$}
        \Comment{Send heading $cnt$ items, $cnt$ is accumulated by \autoref{alg:SmartCMS}}
            \State $msg \gets \{\}$
            \ForAll{$idx \in [1..cnt]$}
            \Comment{Construct a message from $\mathit{ItemBuff}$'s actual items}
                \State \Call{$msg$.Append}{$\mathit{ItemBuff}[idx]$}
            \EndFor
            \State \Return{$msg$}
        \EndFunction

        \Function{Count}{$x, cnt$}
        \Comment{Count instances of $x$ within a batch}
            \State $count \gets 0$
            \ForAll{$idx \in [1..cnt]$}
                \If{$\mathit{ItemBuff}[idx] = x$}
                    \State $count\mathrel{++}$
                \EndIf
            \EndFor
            \State \Return{$count$}
        \EndFunction
    \end{algorithmic}
\end{algorithm}

\paragraph*{Maintenance.} 
When an item $x$ arrives, in addition to normal CMS execution, an item is buffered into a batch by setting $\mathit{ItemBuff}[idx] \leftarrow x$,
where $idx$ is the recent value of a counter $B'$, managed by \autoref{alg:SmartCMS} while counting the items as they~arrive. 

\paragraph*{Space.} 
The buffer has a capacity for $B$ entries, each of which is $\bs{mid}$ bits long.

\paragraph*{Communication.}
Upon sharing, there are exactly $B' \leq B$ filled entries in a buffer that need to be sent to each of the $r_c$ covering redundant nodes.
However, with the $single$ partition (see \autoref{fig:part-table}), 
there are only $r_c=f$ covering~nodes.

\paragraph*{Remote computation.}
Sharing items implies that every receiving redundant node must execute an update procedure against its sum-sketch, which covers some counters of a sending data node.
Although each sketch counter (cell) is mapped to one partition only, a redundant node might cover just a few. 
Therefore, sketch partitioning type has a significant impact on the overall extra computation:
\begin{description}
\item[Single.]
With the $single$ partition, there are only $r_c=f$ covering nodes, each of which must recalculate all $d$ hash functions for each delivered item.

\item[Rows.]
With partitioning by $rows$ (see \autoref{fig:part-rows}), 
each covering node is responsible for a disjoint range of rows and their corresponding hash functions of a copy, while all $r_c$ covering nodes together hold $f$ copies of a sketch.

\item[Cells.]
However, with partitioning by $cells$, the counters of the same row could be mapped to different partitions. 
Hence, for each extracted item $x$, every redundant node that covers a row $i$, must recalculate the same column index $h_i(x)$ first, and test the counter $[i,h_i(x)]$ for membership next.
When a redundant node discovers that counter's coordinates are out of its sum-partition's boundaries, such a counter is ignored.
\end{description}

\paragraph*{Metrics.}
We summarize a Buffer of Items representation and an impact of various sketch partitioning types on communication and computation overhead.
{
\renewcommand{\arraystretch}{1.2}
\begin{table}[ht]
\footnotesize
{
    \centering
    \begin{tabular}
    {|l|l|c||l|l|}
    \hline
    Type
    & Local space
    & $p_{type}$
    & Payload of a share
    & Extra computation
    \\
    \hline
    \multirow{3}{*}
    {$\mathit{ItemBuff}$}
    & \multirow{3}{*}
    {$B \cdot \bs{mid}$}
    & single
    & $f \cdot B \cdot \bs{mid}$
    & \multirow{2}{*}
    {r. CMS: 
    $f \cdot B \cdot d$}
    \\
    \cline{3-4}
    & & rows
    & \multirow{2}{*}
    {$r_c \cdot B \cdot \bs{mid}$}
    & \\
    \cline{3-3}\cline{5-5}
    & & {cells}
    & & {r. CMS:
    $r_c \cdot B \cdot d$}
    \\
    \hline
    \end{tabular}
    \caption{Buffer of Items analysis}
    \label{tab:tbl-buff-item}
}
\end{table}
}

\subsubsection{Full Share vs. Buffer of Items}
Recall that {full share} of the original sketch requires neither additional local data structure nor additional hash calculations. 
Thus, it is more space- and computation-efficient than other methods.

\paragraph*{Full share.}
Might be more economical in terms of \underline{communication} than sharing buffer of items when:
{\small
$$f \cdot d \cdot w \cdot \bs{N} 
\leq r_c \cdot B \cdot \bs{mid} \Rightarrow 
{B \geq d \cdot w \cdot
\underbrace{\frac{f}{r_c}}_{\leq 1} 
\cdot \frac{\bs{N}}{\bs{mid}}}$$
}

\paragraph{Buffer of items.}
The communication traffic produced by buffer of items is linear in $r_c \geq f$.
Hence, with the single partition we can achieve the best results for this representation. 
We consider {buffer of items} over full share when $B$ is small enough to compensate for the identifier length:
{\small
$$\boxed{B \leq d \cdot w \cdot \frac{f}{r_c} \cdot \frac{\bs{N}}{\bs{mid}}} 
\leq d \cdot w \cdot \frac{\bs{N}}{\bs{mid}}$$
}

Generally, we consider using the buffer when $B \leq \frac{w}{2}$. 
For example, let $\delta=10^{-2}$ and require a single partition, resulting in $d=5$ and $r_c=f$. 
Hence, even for long 256-bit identifiers and 32-bit sketch counters, 
$B \leq \frac{5}{8} \cdot w$.

\subsection{Buffer of Counters}
\label{sec:CntBuff}

\paragraph*{Hardware friendly.}
With \emph{memory hierarchy} and CPU cache levels, the data is transferred between the main memory and the cache in blocks of fixed size, called cache-lines. 
Hence, it is more efficient to edit data entries that are loaded together into the same cache-line.
In our case, there are $f$ covering nodes for each sketch counter. 
Therefore, a sending data node aims to encode the modified counters in an easy-to-consume manner, helping all its covering nodes to update their sum-sketches row by row.
This might increase the \emph{cache hit ratio} and enhance the performance.

\begin{algorithm}[ht]
    \caption{CntBuffCMS - ads (buffer of counters) for CMS, code for data node}
    \label{alg:CntBuffCMS}
    \footnotesize
    \begin{algorithmic}
        \Procedure{Init}{$d, w, N, B$}
            \State $\mathit{CntBuff} \gets new\ \mathit{Matrix}(d, B)$ 
            \Comment{Capacity for $d \times B$ indices, length like $w$}
        \EndProcedure

        \Procedure{Reset}{}
            \State $\mathit{CntBuff} \gets 0_{d \times B}$
        \EndProcedure

        \Function{Estimate}{$x, cnt$}
        \Comment{Estimation within a batch}
            \State \Return $\min_i{\Call{Count}{i, h_i(x), cnt}}$
        \EndFunction

        \Procedure{Update}{$x$, $idx$}
            \Comment{Each delayed item has its own column in the buffer}
            \ForAll{$i \in [1..d]$}
                \State $\mathit{CntBuff}[i, idx] \gets h_i(x)$
                \Comment{Record counter index in buffer}
            \EndFor
        \EndProcedure

        \Procedure{Send}{}
            \State \Call{SendCounters}{}
            \Comment{Need to implement Encode functions}
        \EndProcedure

        \Function{Count}{$i, j, cnt$}
        \Comment{Count instances of $j$ within a batch}
            \State $count \gets 0$
            \ForAll{$idx \in [1..cnt]$}
                \If{$\mathit{CntBuff}[i, idx] = j$}
                    \State $count\mathrel{++}$
                \EndIf
            \EndFor
            \State \Return{$count$}
        \EndFunction
        
        \end{algorithmic}
\end{algorithm}

\paragraph*{Maintenance.}
When an item $x$ arrives, in addition to normal CMS execution, the item's counters indices are buffered into a batch by setting $\mathit{CntBuff}[i,idx] \leftarrow h_i(x)$.

\paragraph*{Space.}
The buffer has a capacity for 
$d \times B$ entries, each of which is $\bs{w}$ bits long.

\paragraph*{Communication.}
Upon sharing, there are exactly $B' \leq B$ filled columns with $d$ indices each, each of which needs to be sent to only $f$ covering redundant nodes, using the coverage mapping from \autoref{sec:coverage-mapping}.
The buffer consists of fixed-length rows, turning an encoding procedure into a simple sequential concatenation of modified indices, using $\mathit{EncodeFixed}$. 
However, when the sketch is partitioned by $cells$, $\mathit{EncodeVar}$ must be used. 

\paragraph*{Computation.}
While encoding a variable-length row of a buffer, a sending data node tests each element for membership. 
Moreover, a redundant node must be able to recognize the sketch row, each decoded element belongs to.
To this end, for each row of a partition, we additionally specify a metadata stating the {number of elements} to follow in that row.
Such metadata must be sent to all redundant nodes that cover a particular row, i.e., $O(r_c)$.

\paragraph*{Metrics.}
We summarize a Buffer of Counters representation and an impact of various sketch partitioning types on communication and computation overhead. 

{
\renewcommand{\arraystretch}{1.2}
\begin{table}[ht]
\footnotesize
{
    \centering
    \begin{tabular}
    {|l|l|c||l|l|}
    \hline
    Type
    & Local space
    & $p_{type}$
    & Payload of a share
    & Extra computation
    \\
    \hline
    \multirow{3}{*}
    {$\mathit{CntBuff}$}
    & \multirow{3}{*}
    {$d \cdot B \cdot \bs{w}$}
    & single
    & \multirow{2}{*}
    {$f \cdot d \cdot B \cdot \bs{w}$}
    & \\
    \cline{3-3}\cline{5-5}
    & & rows
    & & membership: $d$
    \\
    \cline{3-3}
    \cline{4-5}
    & & cells
    & $+$\hfill$r_c \cdot d \cdot \bs{B}$
    & membership: $d \cdot B$
    \\
    \hline
    \end{tabular}
    \caption{Buffer of Counters analysis}
    \label{tab:tbl-buff-cnt}
}
\end{table}
}

\subsubsection{Full Share vs. Buffer of Counters}
Full share might be more economical in terms of \underline{communication} than sharing a buffer of counters only when: 
{\small
$${w \cdot \bs{N} \leq B \cdot \bs{w}} \Rightarrow 
B \geq w \cdot
\underbrace{\frac{\bs{N}}{\bs{w}}}_{\geq 1} \geq w$$}

Hence, we generally prioritize sharing a buffer of counters over full share.

\subsubsection{Buffer of Counters vs. Buffer of Items}
\paragraph{Computation.}
{Buffer of Counters} avoids the additional remote hash calculations required with a Buffer of Items.

\paragraph{Space and Communication.}
However, {Buffer of Counters} might be more economical than buffer of items, in terms of both extra local space and communication traffic, only when
${\bs{mid} \geq d \cdot \bs{w}}$, which is rare.
Hence, sharing items is generally preferred over counters. 

\subsection{Hash Table of Flows}
\label{sec:FlwHash}
Hash table of flows and their corresponding frequencies is represented by an additional $\mathit{FlwHash}$ hash table with a \emph{load threshold} $0 < \alpha \leq 1$.
It is based on the Buffer of Items, but also maintains flows' true frequencies to avoid duplicates in identifiers. 

\begin{algorithm}[ht]
    \caption{FlwHashCMS - ads (hash of flows) for CMS, code for data node}
    \label{alg:FlwHashCMS}
    \footnotesize
    \begin{algorithmic}
        \Defs{Locals}{
        $\alpha$ - the load threshold of a hash table\\
        $\hat{\beta}$ - the average frequency factor for a batch 
        }
        
        \Procedure{Init}{$d, w, N, B, mid$}
            \State $\hat{b} \gets \ceil*{\frac{B}{\hat{\beta}}}$
            \Comment{Estimated occupation size - number of flows}
            \State $bk \gets \ceil*{\frac{\hat{b}}{\alpha}}$
            \Comment{Initial capacity - number of buckets required for a given size}
            
            \State $\mathit{FlwHash} \gets new\ \mathit{HashTable}(bk)$ 
            \Comment{id-frequency pairs}
        \EndProcedure

        \Procedure{Reset}{}
            \State \Call{$\mathit{FlwHash}$.Clear}{}
            \Comment{Remove all elements at once}
        \EndProcedure

        \Function{Estimate}{$x$}
        \Comment{Estimation within a batch}
            \State \Return $\mathit{FlwHash}[x]$
        \EndFunction

        \Procedure{Update}{$x$}
            \State $\mathit{FlwHash}[x]\mathrel{++}$
        \EndProcedure

        \Procedure{Send}{}
            \State \Call{SendItems}{}
            \Comment{Need to implement Encode functions}
        \EndProcedure
        
        \Function{EncodeItems}{}
            \State $msg \gets \{\}$
            \State $pair \gets $ \Call{$\mathit{FlwHash}$.GetEnumerator}{}
            \Comment{Walk through all registered pairs}
            
            \While{\Call{$pair$.MoveNext}{}}
            \Comment{Construct a message from $\mathit{FlwHash}$'s $b$ actual pairs}
                \State \Call{$msg$.Append}{$pair$}
            \EndWhile
            \State \Return{$msg$}
        \EndFunction
    \end{algorithmic}
\end{algorithm}

\paragraph*{Maintenance.}
Initially, as well as after sharing the hash table, 
all entries in $\mathit{FlwHash}$ are deleted at once.
When an item $x$ arrives, in addition to normal CMS execution, the flow frequency is incremented. 

\paragraph*{Space.}
A hash table with the load threshold $\alpha$ has an initial capacity for the
$\ceil*{\frac{\hat{b}}{\alpha}}$ buckets, where $\hat{b}$ is the estimated number of pairs to occupy this hash table. 
Clearly, none of the frequencies can exceed the batch size $B$, while zero value means that a respective key was not observed yet during the cycle.
Thus, 
$\bs{mid}+\bs{B}$ bits are sufficient to represent the identifier-frequency~pair. 

\paragraph*{Communication.}
Upon sharing, there are exactly $b \leq B$ flows in a hash table that need to be sent to each of the $r_c$ covering redundant nodes.
Similarly to the Buffer of Items representation, but with only $b$ flows, each of which requires additional $\bs{B}$ bits to record its frequency within a batch. 

\paragraph*{Remote computation.}
Similarly to the Buffer of Items, but with only $b \leq B$ flows. 
Every receiving redundant node will extract the received identifiers and execute an update procedure against its sum-sketch, i.e., for each of $b$ flows, received from a particular data node, it will calculate $O(d)$ CMS hash functions it is responsible for, and update the corresponding counters by adding the extracted~frequency.

\paragraph*{Metrics.}
We summarize a Hash Table of Flows representation and an impact of various sketch partitioning types on communication and computation overhead.
There are two types of hash calculations: the sum-sketch $d$ hash functions that need to be computed per each flow, denoted as hash (CMS), and the hash function(s) of $\mathit{FlwHash}$ to calculate the bucket index upon arrival of each item, denoted as hash (HT).

{\renewcommand{\arraystretch}{1.2}
\begin{table}[ht]
\footnotesize
{
    \centering
    \begin{tabular}
    {|l|l|c||l|l|}
    \hline
    Type
    & Local space
    & $p_{type}$
    & Payload of a share
    & Extra computation
    \\
    \hline
    \multirow{5}{*}
    {$\mathit{FlwHash}$}
    & \multirow{5}{*}
    {$\ceil*{\frac{\hat{b}}{\alpha}} \cdot 
    (\bs{mid} + \bs{B})$}
    & \multirow{2}{*}{single}
    & \multirow{2}{0.22\textwidth}
    {$f \cdot b \cdot \newline 
    (\bs{mid} + \bs{B})$}
    & \multirow{3}{0.19\textwidth}
    {l. HT: $B$ \newline 
    r. CMS: $f \cdot b \cdot d$}
    \\
    &&&&\\
    \cline{3-4}
    & & rows
    & \multirow{3}{0.22\textwidth}
    {$r_c \cdot b \cdot \newline
    (\bs{mid} + \bs{B})$}
    & \\
    \cline{3-3}\cline{5-5}
    & & \multirow{2}{*}{cells}
    & & \multirow{2}{0.19\textwidth}
    {l. HT: $B$ 
    \newline 
    r. CMS: $r_c \cdot b \cdot d$}
    \\
    &&&&\\
    
    \hline
    \end{tabular}
    \caption{Hash Table of Flows analysis}
    \label{tab:rep-flow-hash}
}
\end{table}
}

\subsubsection{Hash Table of Flows vs. Buffer of Items}
Hash table of flows tries to improve a buffer of items through duplicate avoidance, but requires longer elements to additionally store the corresponding frequencies. 
Refer to \autoref{sec:eval-hash} for more details. 

\paragraph{Computation.}
The extra remote CMS hash calculations, associated with these representations, are linear in $b \leq B$ respectively. 
However, a hash table of flows requires local hash computations (HT) on each item arrival.
Further evaluation is needed, using different compact hash tables.

\paragraph{Communication.}
To evaluate the communication traffic produced by these representations, we need to compare 
$b \cdot (\bs{mid} + \bs{B})$ vs $B \cdot \bs{mid}$, or alternatively 
$\frac{\bs{mid} + \bs{B}}{\bs{mid}}$ vs $\frac{B}{b}$.
To do so, we compare $\theta$ vs $\beta$.

We consider a {hash table of flows} over a buffer of items when 
$\boxed{\beta_{avg} \geq \theta} = 1+
\frac{\bs{B}}{\bs{mid}}$.
This property holds for the Internet traces that we checked. 
Therefore, we generally prefer a hash table over a buffer.
However, for applications holding the property ${\beta_{avg} < \theta}$, a buffer of items might be preferred.

\paragraph{Space.}
Comparing space requirements is a bit trickier, since although a hash table of flows implies longer elements, it might still occupy less memory than a buffer of items.
To this end we compare 
$\ceil*{\frac{1}{\alpha} \cdot \hat{b}} \cdot \theta$ vs $B$, where 
$\hat{b} \leftarrow \ceil*{\frac{B}{\hat{\beta}}}$.
Therefore, when 
$\boxed{\hat{\beta} > \frac{1}{\alpha} \cdot \theta}$, a hash table of flows is preferred.

\paragraph{Space $\Rightarrow$ Communication.}
For any $\hat{\beta} \leq \beta_{avg}$: 
if $\hat{\beta}$ holds space efficiency property for a hash table, it also holds its communication efficiency, as 
$\beta_{avg} \geq \hat{\beta} > \frac{1}{\alpha} \cdot \theta \geq \theta$.

\subsection{Hash Table of Counters}
\label{sec:CntHash}
Hash table of counters (indices and their corresponding frequencies) is represented by additional $d$ hash tables $\mathit{Cnt_{i}Hash}$, each responsible for the corresponding $h_i$ CMS hash function and the sketch row $i$ accordingly.

\begin{algorithm}[ht]
    \caption{CntHashCMS - ads (hash of counters) for CMS, code for data node}
    \label{alg:CntHashCMS}
    \footnotesize
    \begin{algorithmic}
        \Defs{Locals}{
        $\alpha$ - the load threshold of a hash table\\
        $\hat{\beta}$ - the average frequency factor for a batch 
        }
        
        \Procedure{Init}{$d, w, N, B, mid$}
            \State $\hat{b} \gets \ceil*{\frac{B}{\hat{\beta}}}$
            \Comment{Estimated occupation size - number of flows}
            \State $bk \gets \ceil*{\frac{\hat{b}}{\alpha}}$
            \Comment{Initial capacity - number of buckets required for a given size}
            
            \ForAll{$i \in [1..d]$}
                \State $\mathit{Cnt_{i}Hash} \gets new\ \mathit{HashTable}(bk)$ 
                \Comment{index-change pairs}
            \EndFor
        \EndProcedure

        \Procedure{Reset}{}
            \ForAll{$i \in [1..d]$}
                \State \Call{$\mathit{Cnt_{i}Hash}$.Clear}{}
                \Comment{Remove all elements at once}
            \EndFor
        \EndProcedure

        \Function{Estimate}{$x$}
        \Comment{Estimation within a batch}
            \State \Return $\min_i{\mathit{Cnt_{i}Hash}[h_i(x)]}$
        \EndFunction

        \Procedure{Update}{$x$}
            \ForAll{$i \in [1..d]$}
                \State $\mathit{Cnt_{i}Hash}[h_i(x)]\mathrel{++}$
            \EndFor
        \EndProcedure

        \Procedure{Send}{}
            \ForAll{$i \in [1..d]$}
                \State \Call{$\mathit{Cnt_{i}Hash}$.Sort}{}
                \Comment{Sort the hashed elements by keys, cache friendly for redundant nodes}
            \EndFor
            
            \State \Call{SendCounters}{}
            \Comment{Need to implement Encode functions}
        \EndProcedure
    \end{algorithmic}
\end{algorithm}

\paragraph*{Maintenance.}
Initially, as well as after sharing the hash tables, all entries in $\mathit{Cnt_{i}Hash}$ hash tables are deleted at once.
When an item $x$ arrives, 
normal CMS execution obtains a counter index 
per each sketch hash function, 
and its frequency is incremented in the hash tables.

\paragraph*{Space.}
With $d$ hash tables, each $\mathit{Cnt_{i}Hash}$ hosts up to $\hat{b}$ counters of a corresponding sketch row $i$, while the actual number of modified counters may vary from table to table due to hash collisions. 
The hash tables have the same load threshold $\alpha$. 
Thus, each hash table has an initial capacity for the $\ceil*{\frac{\hat{b}}{\alpha}}$ buckets, while $\bs{w}+\bs{B}$ bits are sufficient to represent the index-change~pair.

\paragraph*{Communication.}
Upon sharing, there are exactly $c \leq b \cdot d$ modified sketch counters that need to be sent to covering nodes.
Given a coverage mapping, a data node knows the partition each counter belongs to and which $f$ redundant nodes cover it, so it can map each hashed pair while encoding, and send only $f$ required copies of each pair. 

We apply the metadata encoding method for variable-length rows, specifying the number of elements in each hash table, and encode the tables in sequence as if encoding the sketch itself row by row.
This technique was introduced with the Buffer of Counters representation when the sketch is partitioned by cells. 
Every covering node needs to receive the metadata that is relevant to it. 
However, when the sketch is partitioned by rows (or single), no two partitions own the elements of the same hash table. 
Thus, only $f$ copies of table's metadata need to be sent, instead of overall $r_c$.

\paragraph*{Computation.}
Similarly to the Buffer of Counters, the elements are tested for membership while encoding. 
With the $single$ partition, no membership testing is required, since all the elements belong to the same one and only partition.
With partitioning by $rows$, each hash table as a whole is a member of some sketch partition, resulting in $O(d)$ membership tests while encoding, one per~table.
However, with partitioning by $cells$, 
all the elements of such a table might need to be tested while encoding.

\begin{algorithm}[ht]
    \caption{CntHashCMS (continued) - Implement the encoding functionality}
    \label{alg:CntHashCMS2}
    \footnotesize
    \begin{algorithmic}
        \Function{EncodeFixed}{$p_{type}$}
            \State $\{Msg[1],\dots,Msg[p]\} \gets \{\{\},\dots,\{\}\}$
            \For{$i \gets 1$ to $d$}
            \Comment{Encode the tables in sequence to match sketch rows}
                \State $part \gets (p_{type}=single)$ ? 1 : \Call{GetPartitionNumber}{$i$}
                \Comment{No membership test with $single$}
                
                \State $count \gets $ \Call{$\mathit{Cnt_{i}Hash}$.Count}{}
                \Comment{Count the registered pairs}
                
                \State \Call{$Msg[part]$.Append}{$count$}
                \Comment{Specify the number of pairs as metadata}
                
                \State $pair \gets $ \Call{$\mathit{Cnt_{i}Hash}$.GetEnumerator}{}
                \Comment{Walk through all registered pairs}
                
                \While{\Call{$pair$.MoveNext}{}}
                \Comment{Construct a message from $\mathit{Cnt_{i}Hash}$'s actual pairs}
                    \State \Call{$Msg[part]$.Append}{$pair$}
                \EndWhile
            \EndFor
            \State \Return{\{$Msg[1],\dots,Msg[p]\}$}
            \State
        \EndFunction

        \Function{EncodeVar}{}
            \State $\{Msg[1],\dots,Msg[p]\} \gets \{\{\},\dots,\{\}\}$
            \State $\{Bin[1],\dots,Bin[p]\} \gets \{\{\},\dots,\{\}\}$
            \Comment{Reuse within a table, bin for each partition}
            
            \For{$i \gets 1$ to $d$}
            \Comment{Encode the tables in sequence to match sketch rows}
                \State \LeftComment{0}{Get the range of partitions a (partial) table $i$ appears on}
                \State $part_{min} \gets$ \Call{GetPartitionNumber}{$i,1$}
                \Comment{Partition of the first sketch counter in row $i$}
                
                \State $part_{max} \gets$ \Call{GetPartitionNumber}{$i,w$}
                \Comment{Partition of the last sketch counter in row $i$}
                
                \State $partial \gets (part_{min} \neq part_{max})$
                \Comment{Is it a partial table with multiple partitions?}

                \For{$part \gets part_{min}$ to $part_{max}$}
                    \State \Call{$Bin[part]$.Clear}{}
                    \Comment{Empty the bins required for $\mathit{Cnt_{i}Hash}$}
                \EndFor

                \State
                \State \LeftComment{0}{Divide pairs into bins associated with partitions}
                \State $pair \gets $ \Call{$\mathit{Cnt_{i}Hash}$.GetEnumerator}{}
                
                \While{\Call{$pair$.MoveNext}{}}
                    \State $part \gets partial$ ? \Call{GetPartitionNumber}{$i,pair.key$} : $part_{max}$
                    \State \Call{$Bin[part]$.Insert}{$pair$}
                \EndWhile

                \State
                \State \LeftComment{0}{Encode the messages, one per partition}
                \For{$part \gets part_{min}$ to $part_{max}$}
                    \State $count \gets $ \Call{$Bin[part]$.Count}{}
                    \Comment{Bin contains index-change pairs}
                        
                    \State \Call{$Msg[part]$.Append}{$count$}
                    \Comment{Specify the number of pairs as metadata}
                    
                    
                    \ForAll{$pair \in$ \Call{$Bin[part]$.Values}{}}
                        \State \Call{$Msg[part]$.Append}{$pair$}
                        \Comment{List the pairs into a message}
                    \EndFor
                \EndFor
            \EndFor
            \State \Return{$\{Msg[1],\dots,Msg[p]\}$}
        \EndFunction
        
    \end{algorithmic}
\end{algorithm}

\paragraph*{Metrics.}
We summarize a Hash Table of Counters representation and an impact of various implementations and sketch partitioning types on communication and computation overhead.

{\renewcommand{\arraystretch}{1.2}
\begin{table}[H]
\footnotesize
{
    \centering
    \begin{tabular}
    {|l|l|c||l|l|}
    \hline
    Type
    & Local space
    & $p_{type}$
    & Payload of a share
    & Extra computation
    \\
    \hline
    \multirow{5}{*}
    {$\mathit{CntHash}$}
    & \multirow{5}{*}{$d \cdot \ceil*{\frac{\hat{b}}{\alpha}} \cdot 
    (\bs{w} + \bs{B})$}
    & {single}
    & \multirow{3}{0.22\textwidth}
    {$f \cdot d \cdot \bs{B} +
    f \cdot d \cdot b \cdot \newline
    (\bs{w} + \bs{B})$}
    & {l. HT: $B \cdot d$
    }
    \\
    \cline{3-3}\cline{5-5}
    & & \multirow{2}{*}
    {rows}
    & & \multirow{2}{0.19\textwidth}
    {l. HT: $B \cdot d$
    \newline
    membership: $d$}
    \\
    &&&&
    \\
    \cline{3-5}
    & & \multirow{2}{*}
    {cells}
    & \multirow{2}{0.22\textwidth}
    {$r_c \cdot d \cdot \bs{B} +
    f \cdot d \cdot b \cdot \newline
    (\bs{w} + \bs{B})$}
    & \multirow{2}{0.19\textwidth}
    {l. HT: $B \cdot d$
    \newline
    membership: $d \cdot b$}
    \\
    &&&&
    \\
    \hline
    \end{tabular}
    \caption{Hash Table of Counters analysis}
    \label{tab:rep-count-hash}
}
\end{table}
}

\subsubsection{Hash Table of Counters vs. Buffer of Counters}
Similar to a comparison of flows vs items, but instead of $\theta$ we define $\theta' = 1 + \frac{\bs{B}}{\bs{w}}$.

\subsection{Difference Matrix}
\label{sec:CntDiff}
The difference matrix is represented by an additional CMS having the same structure as the original CMS but with shorter entries due to a batch size $B$.

\begin{algorithm}[ht]
    \caption{CntDiffCMS - ads (matrix of counters) for CMS, code for data node}
    \label{alg:CntDiffCMS}
    \footnotesize
    \begin{algorithmic}
        \Procedure{Init}{$d, w, N, B, mid$}
            \State $\mathit{CntDiff} \gets new\ \mathit{Matrix}(d, w)$
            \Comment{Difference matrix with $d \cdot w$ counts, counting up to $B$}
        \EndProcedure

        \Procedure{Reset}{}
            \State $\mathit{CntDiff} \gets 0_{d \times w}$
        \EndProcedure

        \Function{Estimate}{$x$}
            \State \Return $\min_i{\mathit{CntDiff}[i, h_i(x)]}$
        \EndFunction

        \Procedure{Update}{$x$}
            \ForAll{$i \in [1..d]$}
                \State $\mathit{CntDiff}[i,h_i(x)]\mathrel{++}$
                \Comment{Update difference matrix}
            \EndFor
        \EndProcedure

        \Procedure{Send}{}
            \State \Call{SendCounters}{}
            \Comment{Need to implement Encode functions}
        \EndProcedure
    \end{algorithmic}
\end{algorithm}

\paragraph*{Communication.} 
Upon sharing, there are $c \approx b \cdot d$ non-zero counters in $\mathit{CntDiff}$, while each row contains up to $b$. 
To encode the share, we might use the same technique as with $CntHash$, or send the entire matrix as with full share.

\paragraph*{Computation.}
The CMS hash functions are calculated once per each item, while upon sharing, the difference matrix is merged into the original sketch as a plus operation of matrices. 
The only extra computation is associated with membership testing due to sketch partitioning. 
However, when the sketch is partitioned by cells, only non-zero counters are tested for membership.

\paragraph*{Metrics:}
We summarize a Difference Matrix representation and an impact of various sketch partitioning types on communication and computation overhead. 
This representation is very simple and is quite intuitive solution, with negligible additional local computation and no extra computational overhead on the redundant nodes, but it requires almost doubling the local space.
We note that when $c << d \cdot w$, leading to a sparse matrix (many zero entries), this representation is too space-consuming.

{\renewcommand{\arraystretch}{1.2}
\begin{table}[H]
\footnotesize
{
    \centering
    \begin{tabular}
    {|l|l|c||l|l|}
    \hline
    Type
    & Local space
    & $p_{type}$
    & Payload of a share
    & Extra computation
    \\
    \hline
    \multirow{4}{*}
    {$\mathit{CntDiff}$}
    & \multirow{4}{*}
    {$d \cdot w \cdot \bs{B}$}
    & {single}
    & \multirow{2}{0.3\textwidth}
    {$f \cdot d \cdot \bs{B} +\newline
    f \cdot d \cdot b \cdot 
    (\bs{w} + \bs{B})$}
    & \\
    \cline{3-3}\cline{5-5}
    & & {rows}
    & & {membership: $d$}
    \\
    \cline{3-5}
    & & \multirow{2}{*}
    {cells}
    & \multirow{2}{0.3\textwidth}
    {$r_c \cdot d \cdot \bs{B} +\newline
    f \cdot d \cdot b \cdot 
    (\bs{w} + \bs{B})$}
    & {membership: $d \cdot b$}
    \\
    &&&&\\
    \hline
    \end{tabular}
    \caption{Difference Matrix analysis}
    \label{tab:rep-cms-diff}
}
\end{table}
}

\subsection{Summary}
With every representation, we face some challenges.

\paragraph*{Item-based representations.} 
When sharing \underline{items}, all the buffered items must be sent to every covering redundant node, 
which in turn must execute CMS update procedure upon each delivered item, i.e., repeating the same hash calculations as already performed by a sending data node.
Communication traffic is linear in $r_c \geq f$, where the minimum can be achieved using the $single$ partition, but results in extra traffic otherwise.
Moreover, it is not recommended to partition the sketch by $cells$, since it also implies extra computations associated with membership testing.

\paragraph*{Counter-based representations.}
When sharing \underline{counters} that do not match the exact structure (or order), we need to encode the share, providing a redundant node with the information essential for successful decoding.

\paragraph*{Extra local space.}
Additional local space, required for buffering a batch, might exceed the space of the original sketch.
In such a case, we try to reduce the communication overhead associated with sending {sparse} data structure, using compact encoding.
Otherwise, it would be more economical to send the original sketch as with full~share. 
\section{Evaluation}
\label{chap:experiment}

\subsection{Usage of Compact Hash Tables}
\label{sec:eval-hash}
We evaluate the usage of compact hash tables by comparing the item-based representations, $\mathit{FlwHash}$ and $\mathit{ItemBuff}$.
We generally prioritize reducing communication traffic and avoiding early transmissions,
but space efficiency is also~desired.

To evaluate the maintenance traffic produced by these representations, we need to compare 
$\frac{\bs{mid} + \bs{B}}{\bs{mid}}=\theta$ vs $\frac{B}{b}=\beta$. 
We note that $\theta$ depends on pre-defined settings, while $\beta$ depends on actual data. 
We find these parameters very useful in further evaluation of compact hash tables, and in the over-allocation vs early-transmission trade-off in~particular.

\subsubsection{Implementation}
In our experiment, we measured CAIDA Internet traces using various batch sizes, and hence, we consider the results as the range for any smaller Internet trace (e.g., within a campus). 
The source code for measuring $\beta$ and generating the plots is available on GitHub~\cite{Cohen_Measuring_average_frequency_2024}. 
The traces are processed using the \verb|Go| programming language and its built-in \verb|map| data structure, while the plots are generated using \verb|Python| with its \verb|pandas| and \verb|matplotlib| libraries.

{
\renewcommand{\arraystretch}{1.2}
\begin{table}[ht]
\footnotesize
{
    \centering
    \begin{tabular}
    {|l||l||l|l|l|l|l|l|l|}
    \hline
    {Trace}, $N/n$
    & $B$ & 50 & 100 & 250 & 500 & 1000 & 2000 & 4000\\
    \hline
    \hline
    {80-bit identifiers:} 
    & $\theta$ & 1.075 & 1.09 & 1.1 & 1.11 & 1.125 & 1.14 & 1.15 \\
    \hline
    Chicago16Small 
    & $\beta_{avg}$ 
    & 1.51& 1.61& 1.74& 1.85& 1.98& 2.12& 2.27 \\
    \cline{2-9}
    \multirow{2}{*}{$\frac{\num{1000001}}{\num{165509}}=6.04$}
    & $\beta_{5\%}$ 
    & 1.14& 1.23& 1.37& 1.5& 1.64& 1.78& 1.94\\
    \cline{2-9}
    & $\beta_{25\%}$ 
    & 1.32& 1.43& 1.55& 1.68& 1.81& 1.97& 2.13\\
    \cline{2-9}
    & $\beta_{50\%}$ 
    & 1.43& 1.56& 1.69& 1.82& 1.96& 2.1& 2.27\\
    \cline{2-9}
    & $\beta_{75\%}$ 
    & 1.61& 1.72& 1.87& 1.97& 2.1& 2.24& 2.4\\
    \cline{2-9}
    & $\beta_{95\%}$ 
    & 2.08& 2.13& 2.19& 2.27& 2.39& 2.52& 2.61\\
    \hline
    Chicago1610Mil 
    & $\beta_{avg}$ 
    & 1.56& 1.67& 1.82& 1.94& 2.07& 2.22& 2.38\\
    \cline{2-9}
    \multirow{2}{*}{$\frac{\num{10000000}}{\num{988835}}=10.11$}
    & $\beta_{5\%}$ 
    & 1.22& 1.33& 1.47& 1.60& 1.76& 1.94& 2.13\\
    \cline{2-9}
    & $\beta_{25\%}$ 
    & 1.39& 1.49& 1.66& 1.78& 1.93& 2.09& 2.28\\
    \cline{2-9}
    & $\beta_{50\%}$ 
    & 1.52& 1.61& 1.77& 1.9& 2.045& 2.2& 2.37\\
    \cline{2-9}
    & $\beta_{75\%}$ 
    & 1.67& 1.79& 1.94& 2.05& 2.18& 2.31& 2.47\\
    \cline{2-9}
    & $\beta_{95\%}$ 
    & 2.08& 2.17& 2.27& 2.35& 2.46& 2.57& 2.68\\
    
    \hline\hline
    {64-bit identifiers:}
    & $\theta$ 
    & 1.09 & 1.11 & 1.125 & 1.14 & 1.16 & 1.17 & 1.19\\
    \hline
    ny19A 
    & $\beta_{avg}$ 
    & 2.07& 2.26& 2.48& 2.64& 2.86& 3.11& 3.41\\
    \cline{2-9}
    \multirow{2}{*}{$\frac{\num{30098745}}{\num{1522382}}=19.77$}
    & $\beta_{5\%}$ 
    & 1.47& 1.64& 1.88& 2.09& 2.34& 2.61& 2.93\\
    \cline{2-9}
    & $\beta_{25\%}$ 
    & 1.72& 1.89& 2.12& 2.31& 2.56& 2.84& 3.15\\
    \cline{2-9}
    & $\beta_{50\%}$ 
    & 1.92& 2.13& 2.36& 2.53& 2.76& 3.03& 3.33\\
    \cline{2-9}
    & $\beta_{75\%}$ 
    & 2.27& 2.44& 2.66& 2.81& 3.02& 3.27& 3.56\\
    \cline{2-9}
    & $\beta_{95\%}$ 
    & 3.125& 3.33& 3.47& 3.6& 3.75& 3.93& 4.17\\
    \hline
    ny19B 
    & $\beta_{avg}$ 
    & 1.34& 1.41& 1.5& 1.58& 1.67& 1.79& 1.95\\
    \cline{2-9}
    \multirow{2}{*}{$\frac{\num{63284829}}{\num{2968038}}=21.32$}
    & $\beta_{5\%}$ 
    & 1.11& 1.2& 1.32& 1.41& 1.52& 1.67& 1.83\\
    \cline{2-9}
    & $\beta_{25\%}$ 
    & 1.19& 1.28& 1.4& 1.49& 1.59& 1.73& 1.89\\
    \cline{2-9}
    & $\beta_{50\%}$ 
    & 1.28& 1.37& 1.47& 1.56& 1.65& 1.78& 1.94\\
    \cline{2-9}
    & $\beta_{75\%}$ 
    & 1.43& 1.49& 1.57& 1.64& 1.72& 1.84& 1.99\\
    \cline{2-9} 
    & $\beta_{95\%}$ 
    & 1.72& 1.75& 1.79& 1.82& 1.86& 1.955& 2.09\\
    \hline
    SJ14.small 
    & $\beta_{avg}$ 
    & 1.54& 1.655& 1.78& 1.87& 1.97& 2.12& 2.32\\
    \cline{2-9}
    \multirow{2}{*}{$\frac{\num{188511031}}{\num{2922904}}=64.49$}
    & $\beta_{5\%}$ 
    & 1.22& 1.32& 1.47& 1.59& 1.72& 1.89& 2.11\\
    \cline{2-9} 
    & $\beta_{25\%}$ 
    & 1.35& 1.47& 1.61& 1.72& 1.85& 2.01& 2.22\\
    \cline{2-9}
    & $\beta_{50\%}$ 
    & 1.52& 1.61& 1.75& 1.845& 1.95& 2.1& 2.3\\
    \cline{2-9}
    & $\beta_{75\%}$ 
    & 1.67& 1.79& 1.91& 1.98& 2.08& 2.21& 2.4\\
    \cline{2-9}
    & $\beta_{95\%}$ 
    & 2& 2.13& 2.21& 2.24& 2.3& 2.4& 2.56\\
    \hline
    \end{tabular}
    \caption{Measuring $\hat{\beta}$}
    \label{tab:beta-scale}
}
\end{table}
}

\subsubsection{Results}
The plots visualize how $\beta$ percentiles evolve as a function of batch size. 
For each Internet trace evaluated, we also mark the baselines for the traffic and space efficiency of a hash table over a buffer. 
\autoref{fig:trace-80bit} and \autoref{fig:trace-64bit} simulate a hash table with $\alpha=0.8$, while \autoref{fig:trace-64bit-09} visualizes the impact of the higher load threshold.

From \autoref{tab:beta-scale} we learn that even for small batches, all these traces hold $\beta_{5\%} > \theta$, meaning that any hash table is more communication efficient than a buffer.
Choosing $\hat{\beta} = \beta_{5\%}$ implies that only up to 5\% of the batches must be sent earlier, making it the best candidate in terms of early-transmission reduction.
However, it is much harder to remain efficient in space, due to the multiplication factor $\frac{1}{\alpha} \geq 1$, which is dictated by the implementation of a hash table. 
In fact, for small batches and $\alpha=0.8$, most of these traces hold (purple line)
$\beta_{5\%} \leq \frac{1}{\alpha} \cdot \theta$, requiring more space than with a buffer. 
When looking for a space-efficient candidate for $\hat{\beta}$, we recognize that all these traces hold (green line) $\beta_{50\%} \leq \beta_{avg}$.
Unfortunately, this means that half of batches will encounter early transmission, which we aim to avoid. 

{
\begin{figure}[ht]
    \begin{subfigure}[b]{0.32\textwidth}
        \centering
        \includegraphics[width=\textwidth]{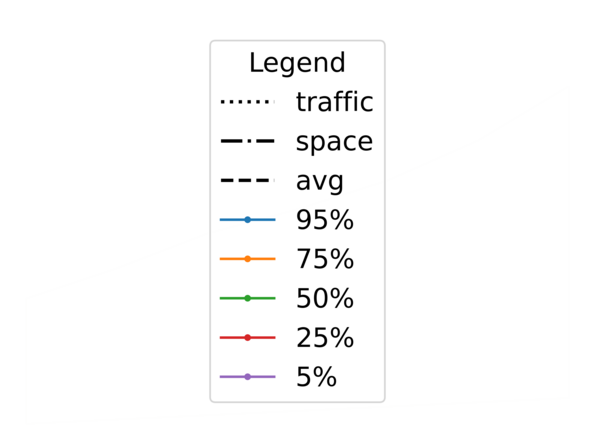}
        \label{fig:beta-legend}
    \end{subfigure}
    \hfill
    \begin{subfigure}[b]{0.32\textwidth}
        \centering
        \includegraphics[width=\textwidth]{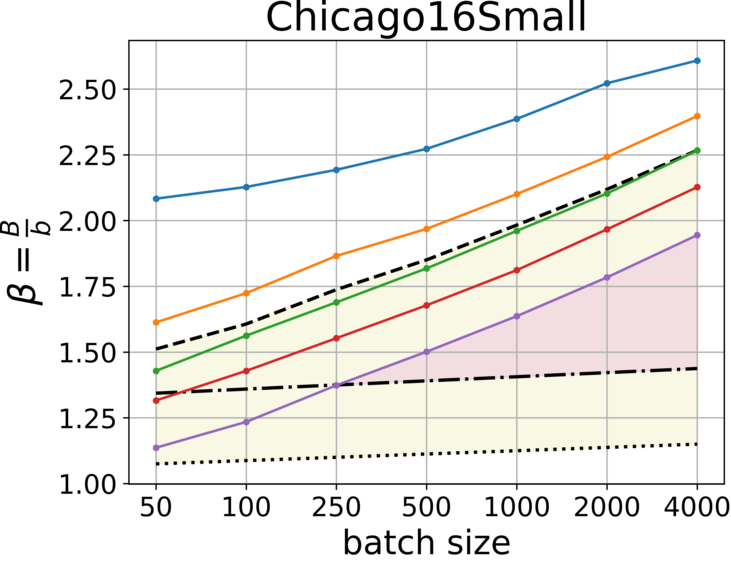}
        \label{fig:beta-Chicago16Small}
    \end{subfigure}
    \hfill
    \begin{subfigure}[b]{0.32\textwidth}
        \centering
        \includegraphics[width=\textwidth]{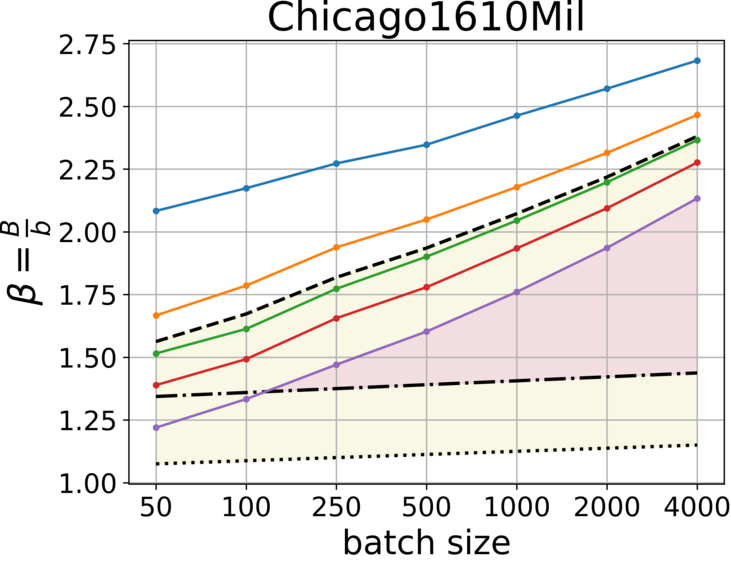}
        \label{fig:beta-Chicago1610Mil}
    \end{subfigure}
    \caption{Internet traces with 80-bit identifiers, $\alpha=0.8$}
    \label{fig:trace-80bit}
\end{figure}
\begin{figure}[ht]
    \begin{subfigure}[b]{0.32\textwidth}
        \centering
        \includegraphics[width=\textwidth]{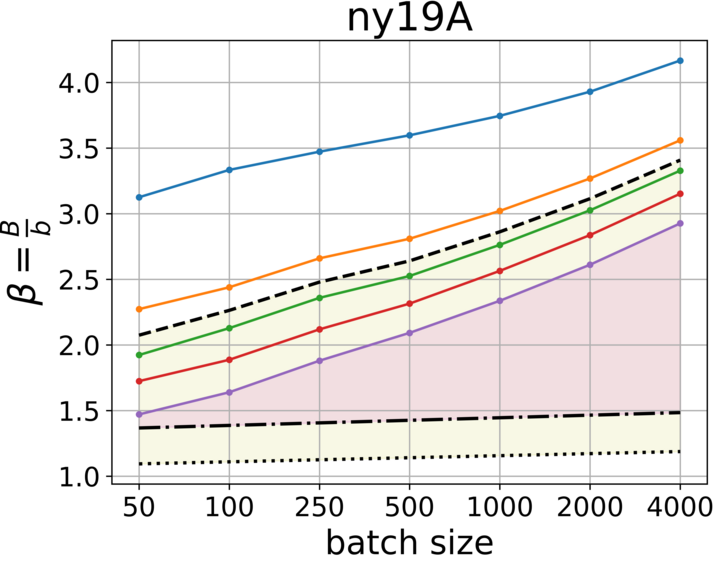}
        \label{fig:beta-ny19A}
    \end{subfigure}
    \hfill
    \begin{subfigure}[b]{0.32\textwidth}
        \centering
        \includegraphics[width=\textwidth]{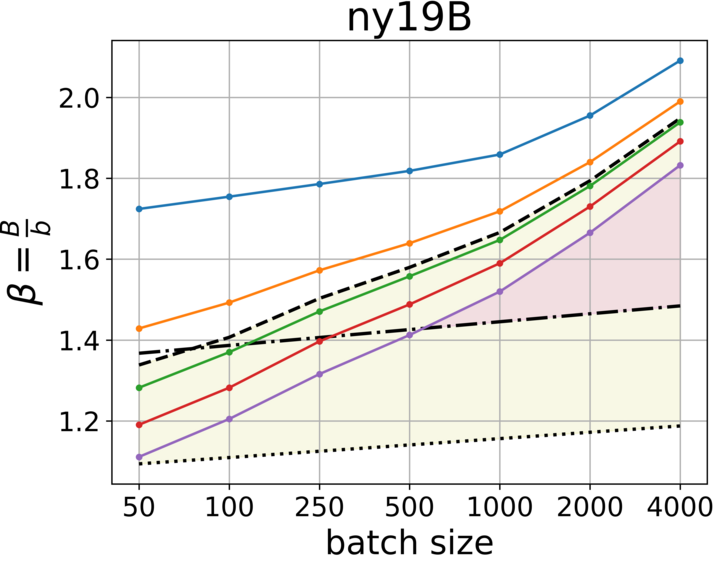}
        \label{fig:beta-ny19B}
    \end{subfigure}
    \hfill
    \begin{subfigure}[b]{0.32\textwidth}
        \centering
        \includegraphics[width=\textwidth]{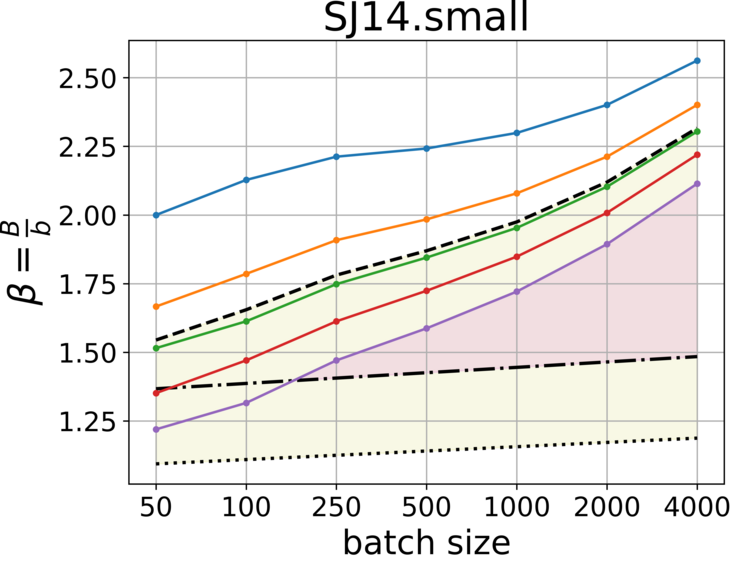}
        \label{fig:beta-SJ14.small}
    \end{subfigure}
    \caption{Internet traces with 64-bit identifiers, $\alpha=0.8$}
    \label{fig:trace-64bit}
\end{figure}
\begin{figure}[h!]
    \begin{subfigure}[b]{0.32\textwidth}
        \centering
        \includegraphics[width=\textwidth]{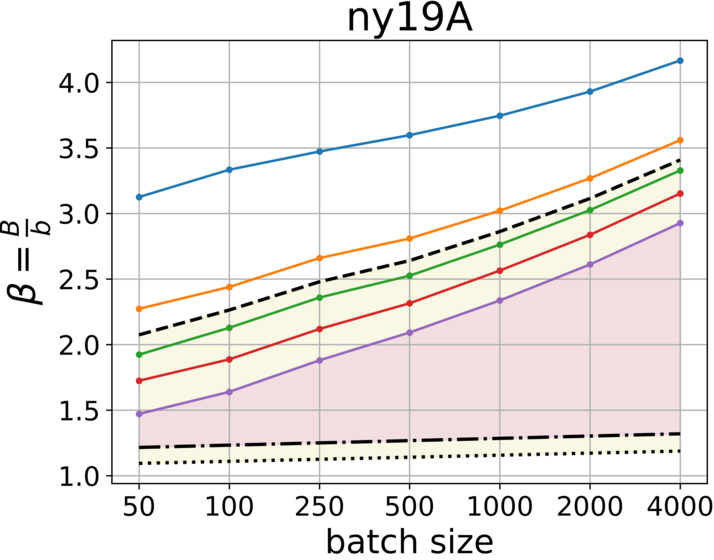}
        \label{fig:beta-ny19A-09}
    \end{subfigure}
    \hfill
    \begin{subfigure}[b]{0.32\textwidth}
        \centering
        \includegraphics[width=\textwidth]{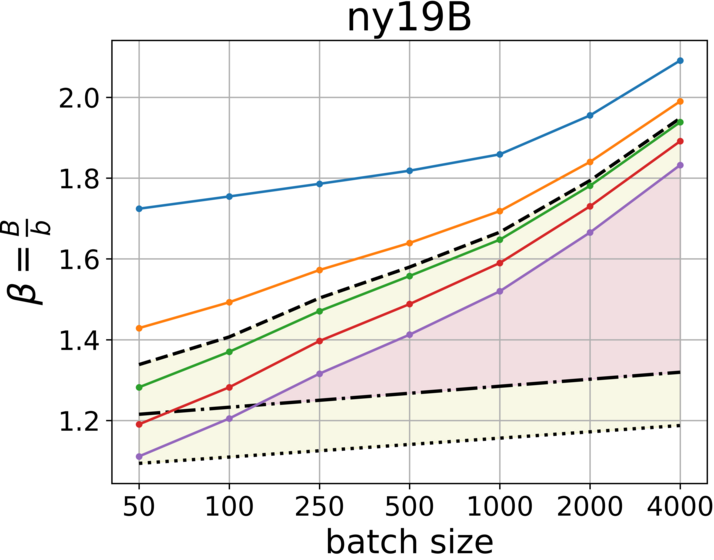}
        \label{fig:beta-ny19B-09}
    \end{subfigure}
    \hfill
    \begin{subfigure}[b]{0.32\textwidth}
        \centering
        \includegraphics[width=\textwidth]{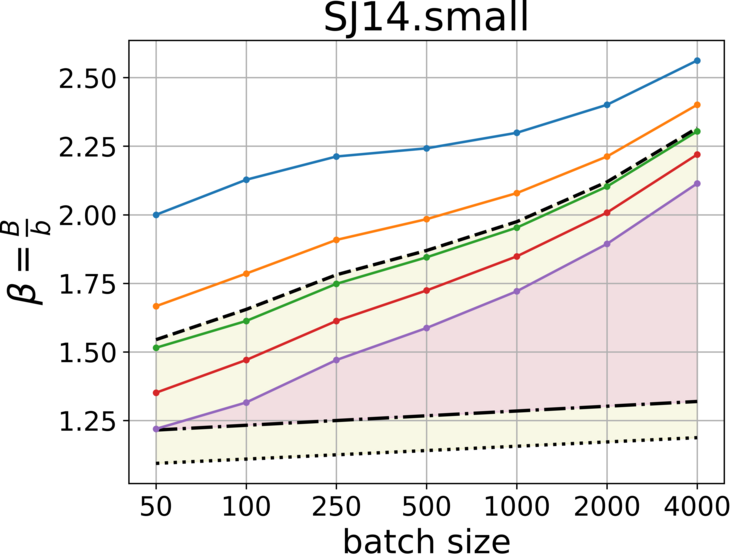}
        \label{fig:beta-SJ14.small-09}
    \end{subfigure}
    \caption{Internet traces with 64-bit identifiers, $\alpha=0.9$}
    \label{fig:trace-64bit-09}
\end{figure}
}

\subsubsection{Recommendations}
With low $\beta$ percentiles, we reduce the probability of early transmission at the cost of space.
To choose the right $\hat{\beta}$ for a given batch size $B$, we look for the lowest percentile that holds:
($i$) $\hat{\beta} \leq \beta_{avg}$ and $\hat{\beta} \geq \theta$ for communication \emph{traffic} efficiency (yellow zone), and 
($ii$) $\hat{\beta} > \frac{1}{\alpha}\cdot\theta$ for \emph{space} efficiency.
Recall that any space-efficient 
$\hat{\beta}>\beta_{5\%}$ 
implies early transmission of up to the same percentage of batches, which we aim to avoid.
Also, we get noisy $\beta$ for small batch sizes.
Therefore, for $B \leq 100$, perhaps it is better to use a standard hash table and allocate memory twice a batch size in advance to avoid dynamic reallocation. 

\paragraph*{Next Experiment.}
The plots reveal an approach to gain both space and traffic efficiency simultaneously, by increasing the batch size (magenta zone).
However, we need to remember that batch size has a significant impact on the estimation error of a recovered sketch. 

\subsection{Estimation Error after Recovery}
\label{sec:eval-error}
Denote $t$ a cycle of the latest backup, and $\hat{c_x}^{(t)}$ the estimation answered by a history CMS upon a point query $Q(x)$.
We assume that all the nodes read the same state of a system, i.e., each and every non-failed node is aware of all the failed nodes at the beginning of the $(t+1)^{\mathrm{th}}$ cycle.
By this time, a failed data node could lose up to $B$ items of a batch. 
For simplicity, we assume a single failure, $f=1$.

The recovered CMS might answer $Q(x)$ with $\hat{c_x} \gets \hat{c_x}^{(t)} + B$, ensuring one-sided error even for the worst case scenario. 
With probability at least $(1-\delta)$, 
$\hat{c_x}^{(t)} + B \leq {c_x}^{(t)} + (\epsilon + \frac{B}{N_t}) N_t$.
Another option is to answer with $\hat{c_x} \gets \hat{c_x}^{(t)}$, implying two-sided error due to out of sync counts.
We now empirically evaluate the mean relative error (MRE) on various traces and batch sizes.

\subsubsection{Implementation}
In this experiment, we implement CMS in \verb|Go| using built-in \verb|maphash| for hashing string keys with \verb|seed|, and configure CMS with $(\epsilon, \delta) = (10^{-6}, 10^{-2})$.
Similar to our prior experiment, we process the traces using \verb|Go|, and visualize the results using \verb|Python|. 

We split the stream's timeline into batches of size $B$, having values 100, 500, or 2000. 
Then we sample a batch to emulate its failure, after $1/3$, $1/2$ or $2/3$ of overall batches.
For a chosen failed batch, we sample a point of failure within a batch, e.g., after 10\%, 50\% or 90\% of delayed items.

At the point of failure, we aim to measure the impact of +B on mean relative error. 
For this purpose, we manage the map of all seen flows from the beginning of the stream until the point of failure. 
We also manage the CMS until the latest backup at point $t$, as well as a difference matrix of a failed batch. 
To measure the error range, we process every flow seen so far in a stream and measure the distance between $(\hat{c_x}^{(t)} + B)$ and both non-failed CMS estimator $\hat{c_x}$ and the true frequency $c_x$.

\subsubsection{Results}
{
\begin{figure}[ht]
    \begin{subfigure}[b]{0.32\textwidth}
        \centering
        \includegraphics[width=\textwidth]{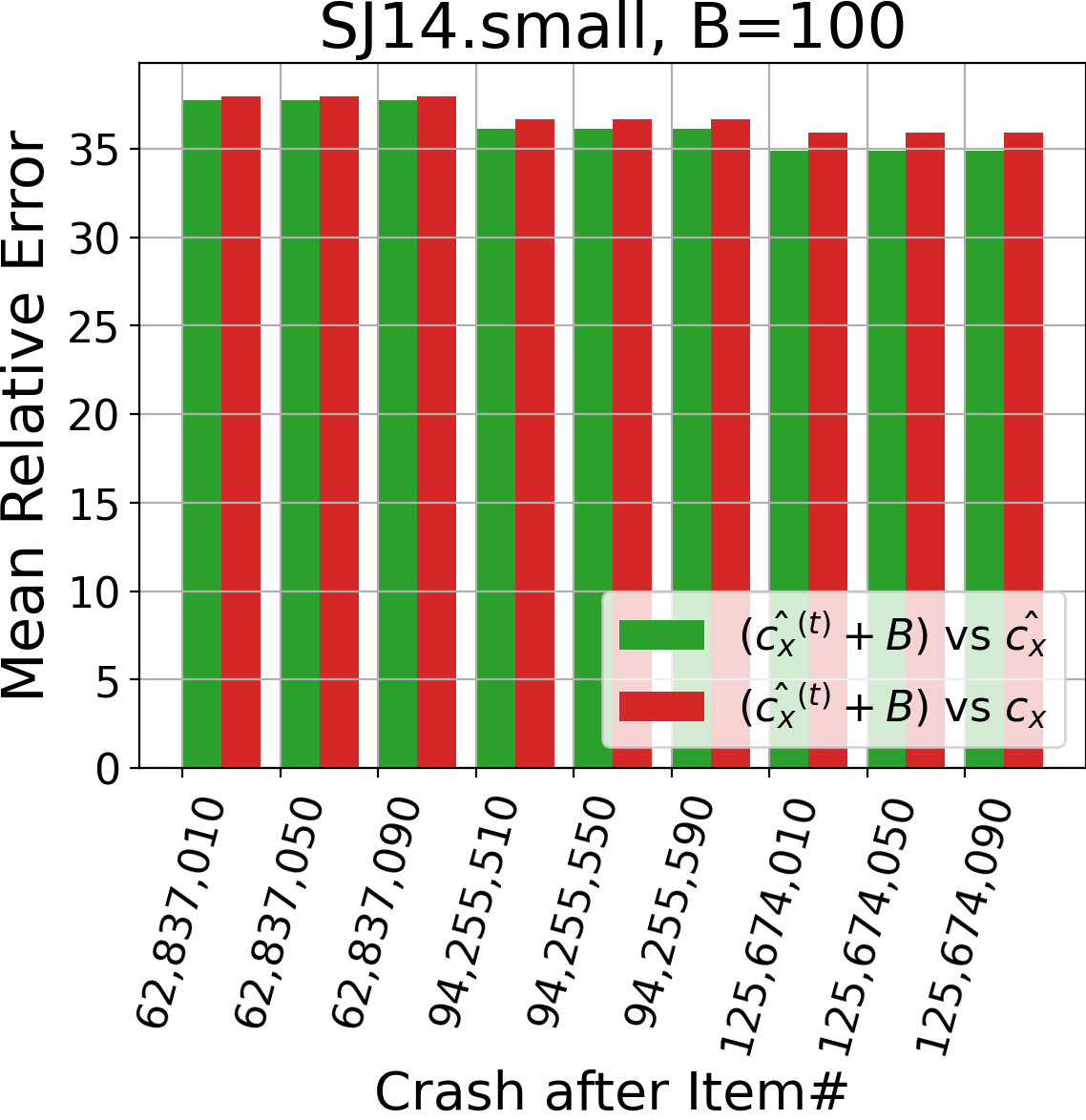}
        \label{fig:MRE-recovery-SanJose-100}
    \end{subfigure}
    \hfill
    \begin{subfigure}[b]{0.32\textwidth}
        \centering
        \includegraphics[width=\textwidth]{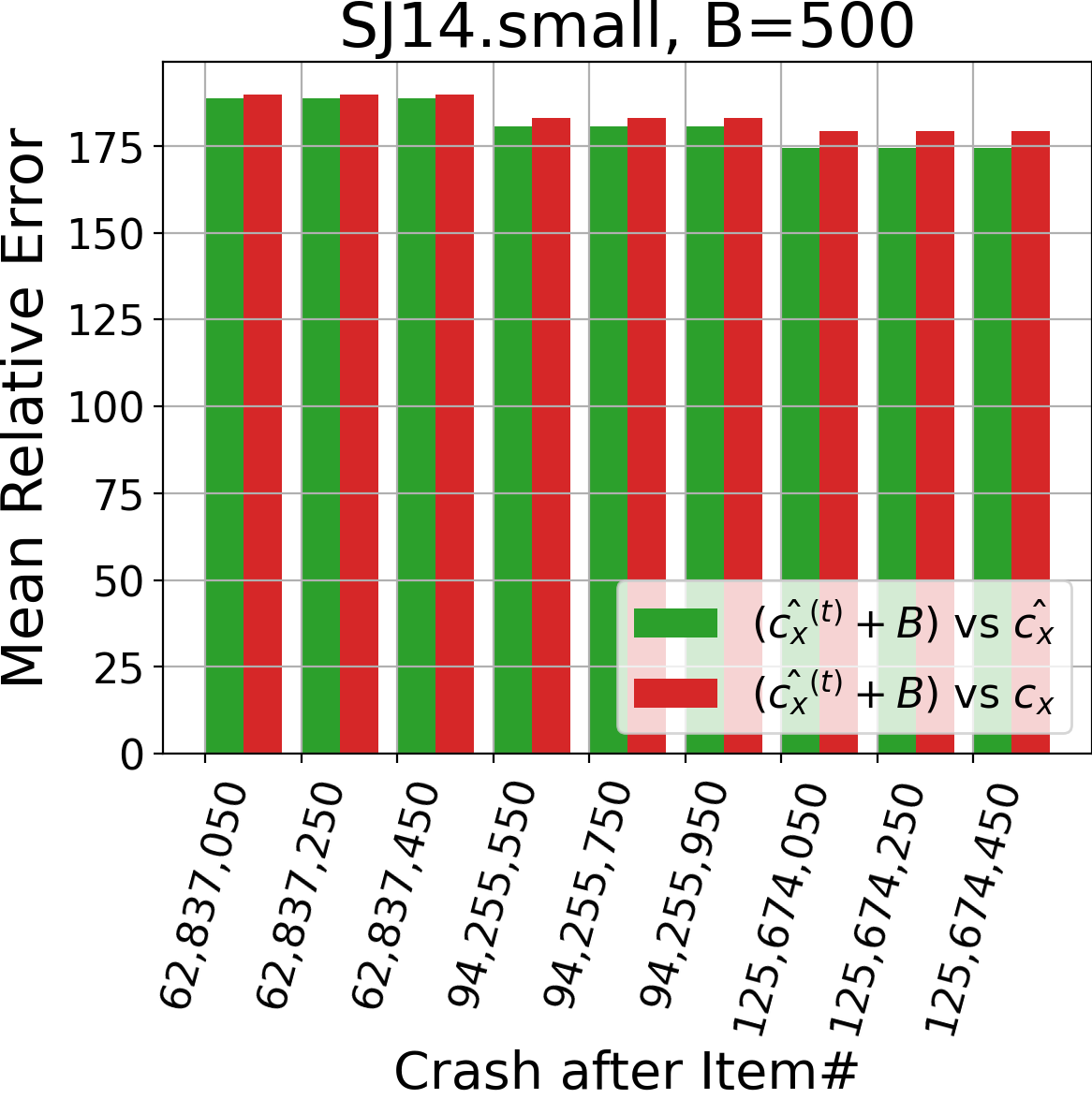}
        \label{fig:MRE-recovery-SanJose-500}
    \end{subfigure}
    \hfill
    \begin{subfigure}[b]{0.32\textwidth}
        \centering
        \includegraphics[width=\textwidth]{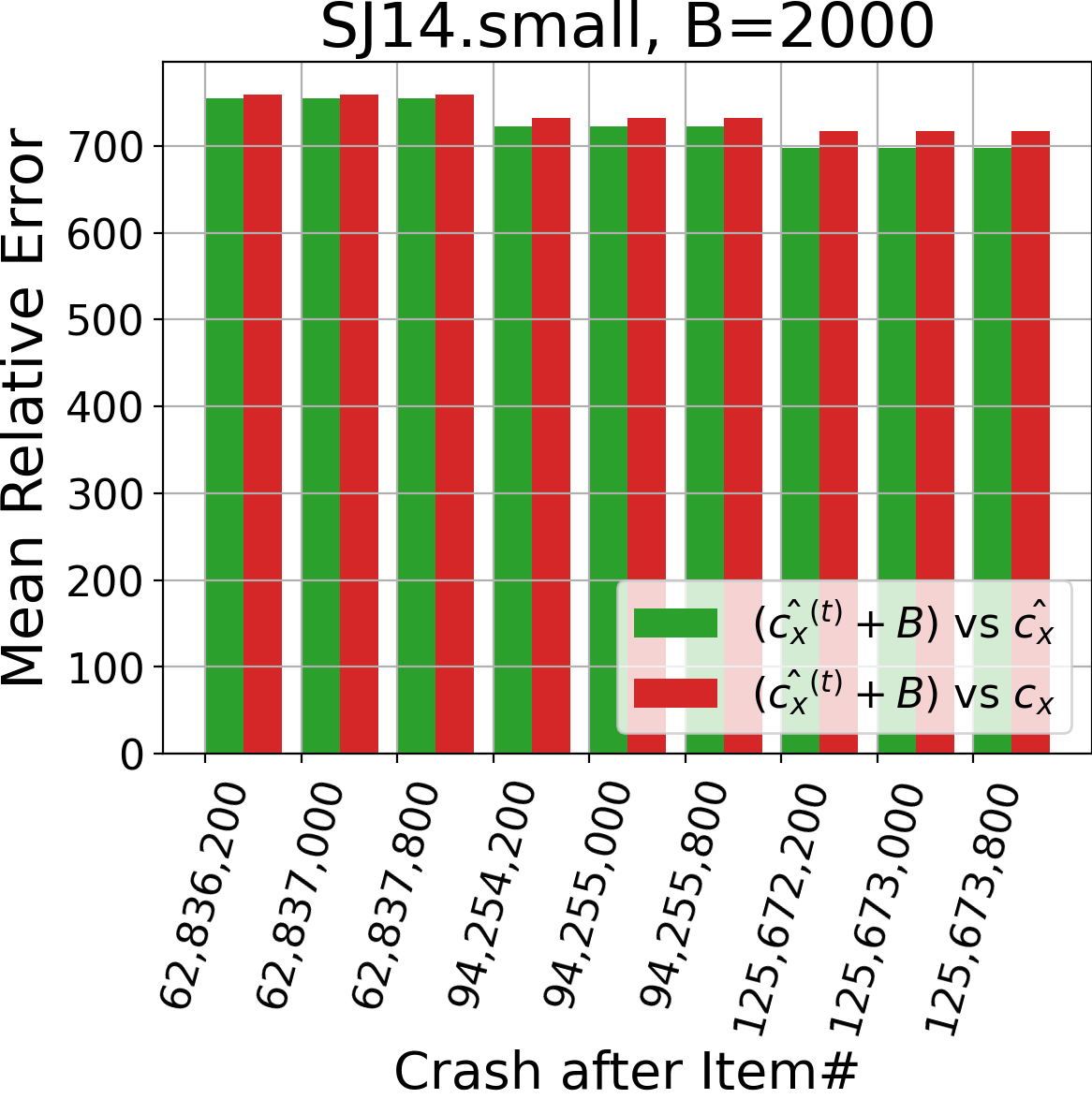}
        \label{fig:MRE-recovery-SanJose-2000}
    \end{subfigure}
    \caption{Impact of +B on Mean Relative Error in San Jose}
    \label{fig:MRE-recovery-SanJose}
\end{figure}
}
From the above we learn that adding $B$ to ensure one-sided error drifts the estimator far away. 
Our next experiment is to measure the distance between the true frequency $c_x$ and both the latest backup $\hat{c_x}^{(t)}$ and non-failed CMS $\hat{c_x}$.

{
\begin{figure}[ht]
    \begin{subfigure}[b]{0.32\textwidth}
        \centering
        \includegraphics[width=\textwidth]{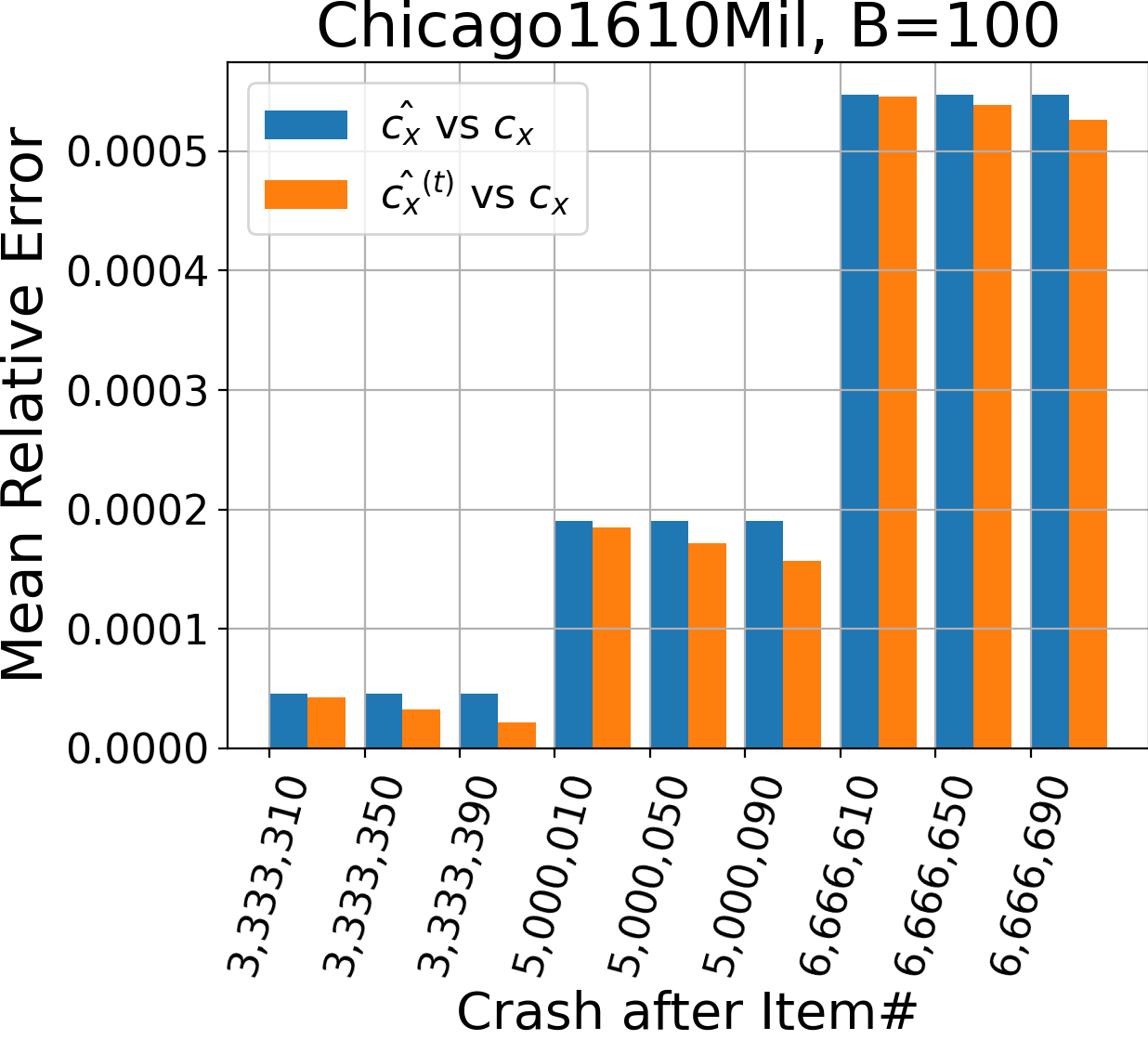}
        \label{fig:MRE-backup-Chicago-100}
    \end{subfigure}
    \hfill
    \begin{subfigure}[b]{0.32\textwidth}
        \centering
        \includegraphics[width=\textwidth]{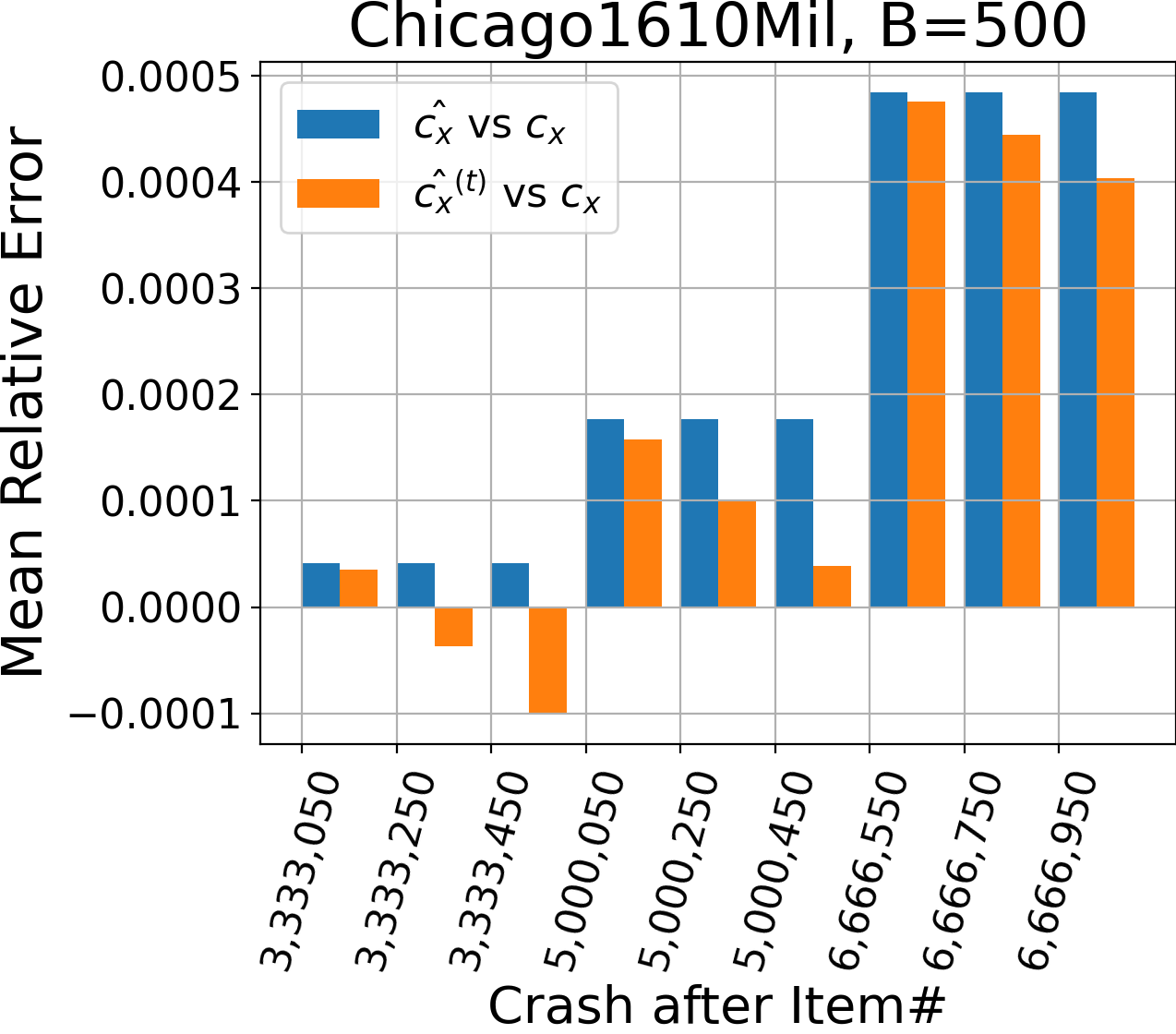}
        \label{fig:MRE-backup-Chicago-500}
    \end{subfigure}
    \hfill
    \begin{subfigure}[b]{0.32\textwidth}
        \centering
        \includegraphics[width=\textwidth]{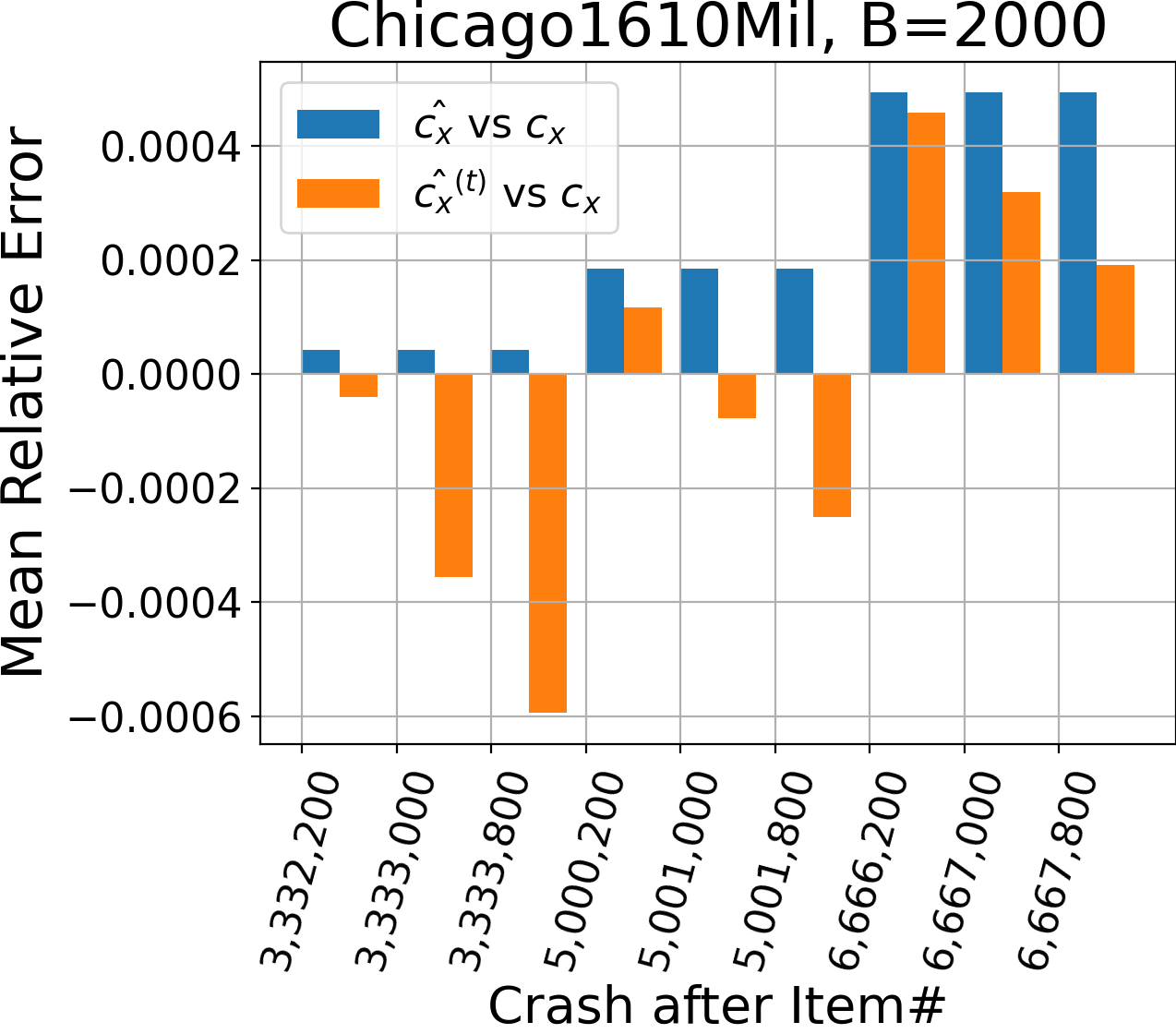}
        \label{fig:MRE-backup-Chicago-2000}
    \end{subfigure}
    \caption{Impact of batch loss on Mean Relative Error in Chicago}
    \label{fig:MRE-backup-Chicago}
\end{figure}
\begin{figure}[ht]
    \begin{subfigure}[b]{0.32\textwidth}
        \centering
        \includegraphics[width=\textwidth]{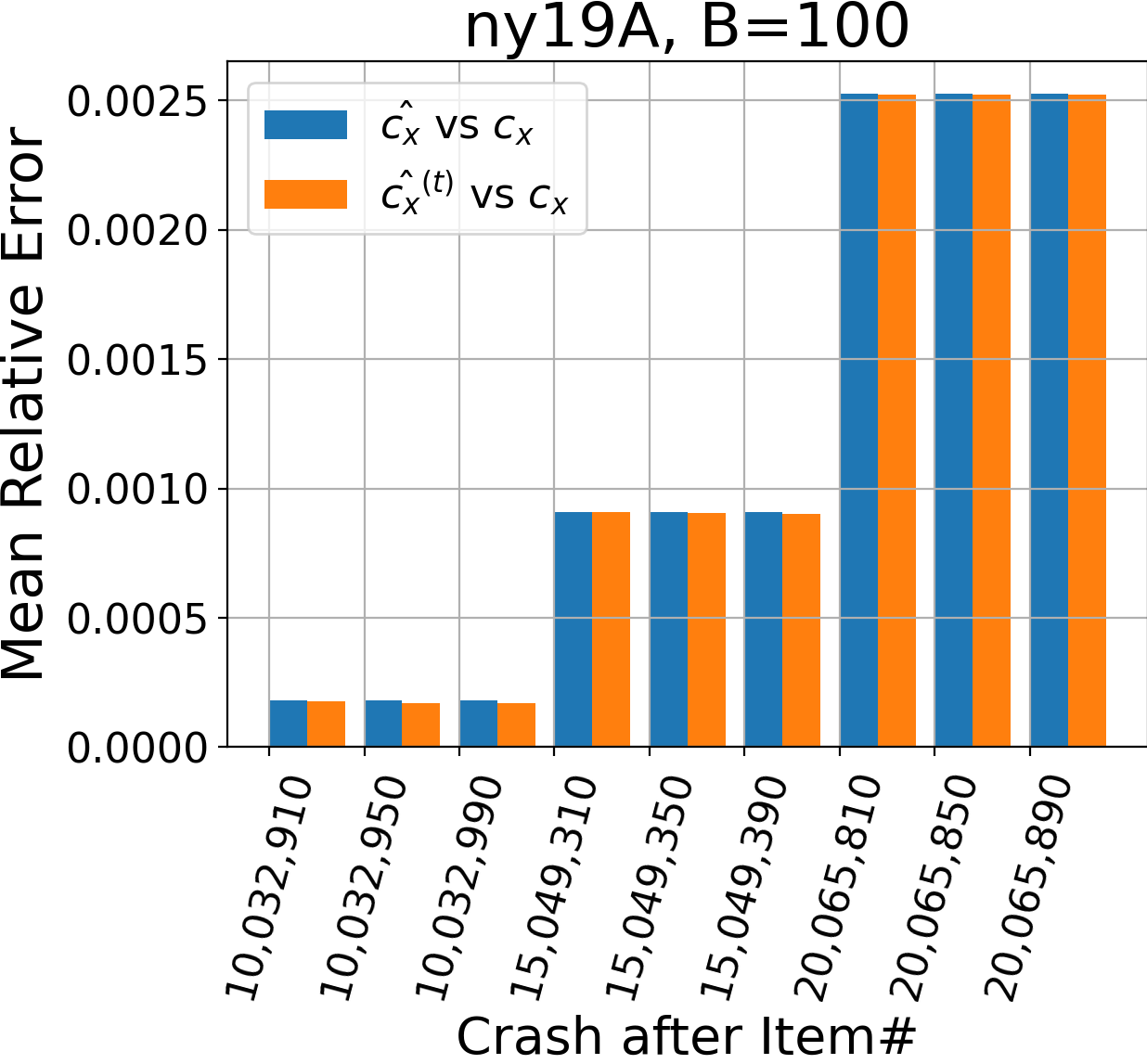}
        \label{fig:MRE-backup-NewYork-100}
    \end{subfigure}
    \hfill
    \begin{subfigure}[b]{0.32\textwidth}
        \centering
        \includegraphics[width=\textwidth]{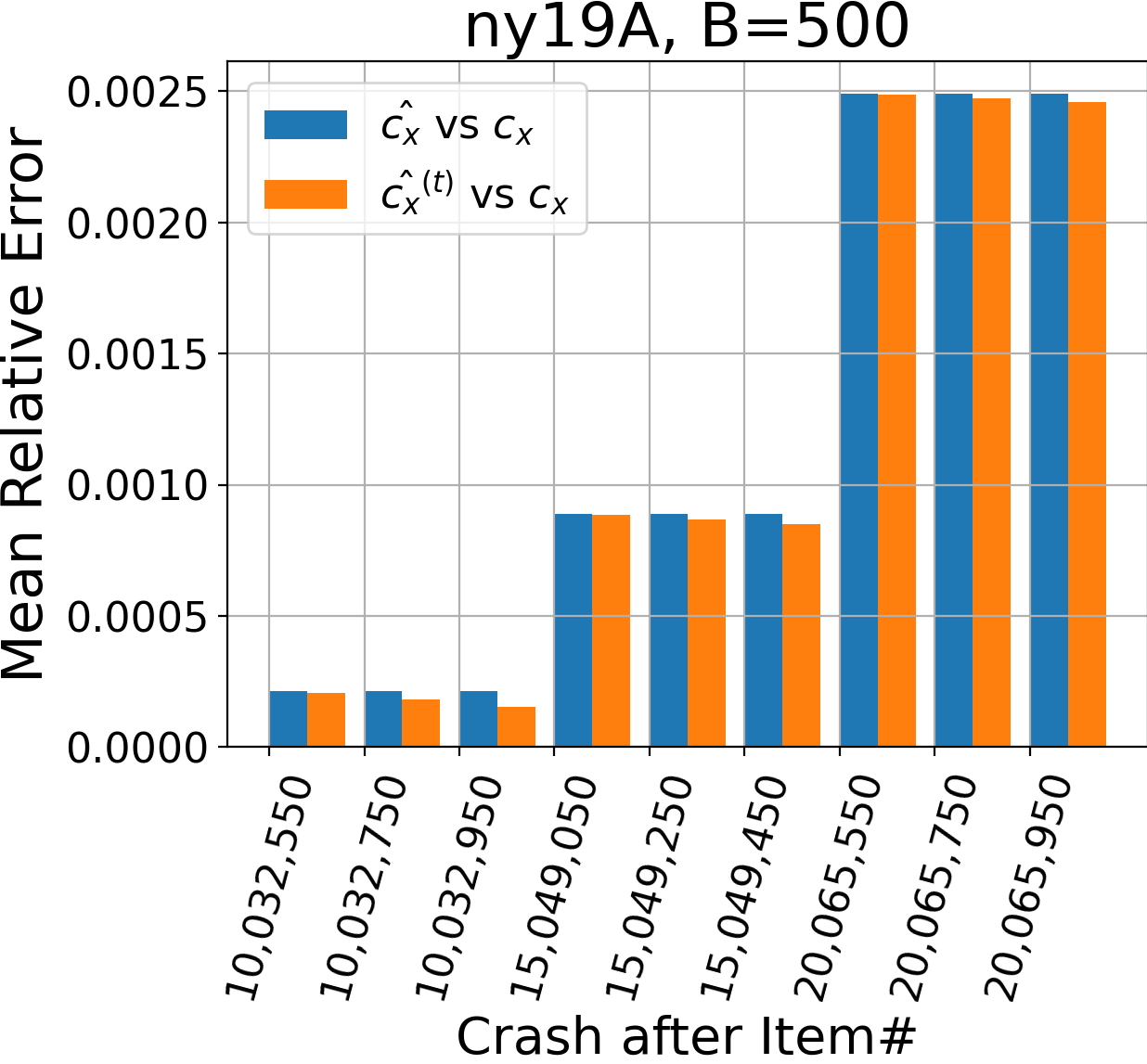}
        \label{fig:MRE-backup-NewYork-500}
    \end{subfigure}
    \hfill
    \begin{subfigure}[b]{0.32\textwidth}
        \centering
        \includegraphics[width=\textwidth]{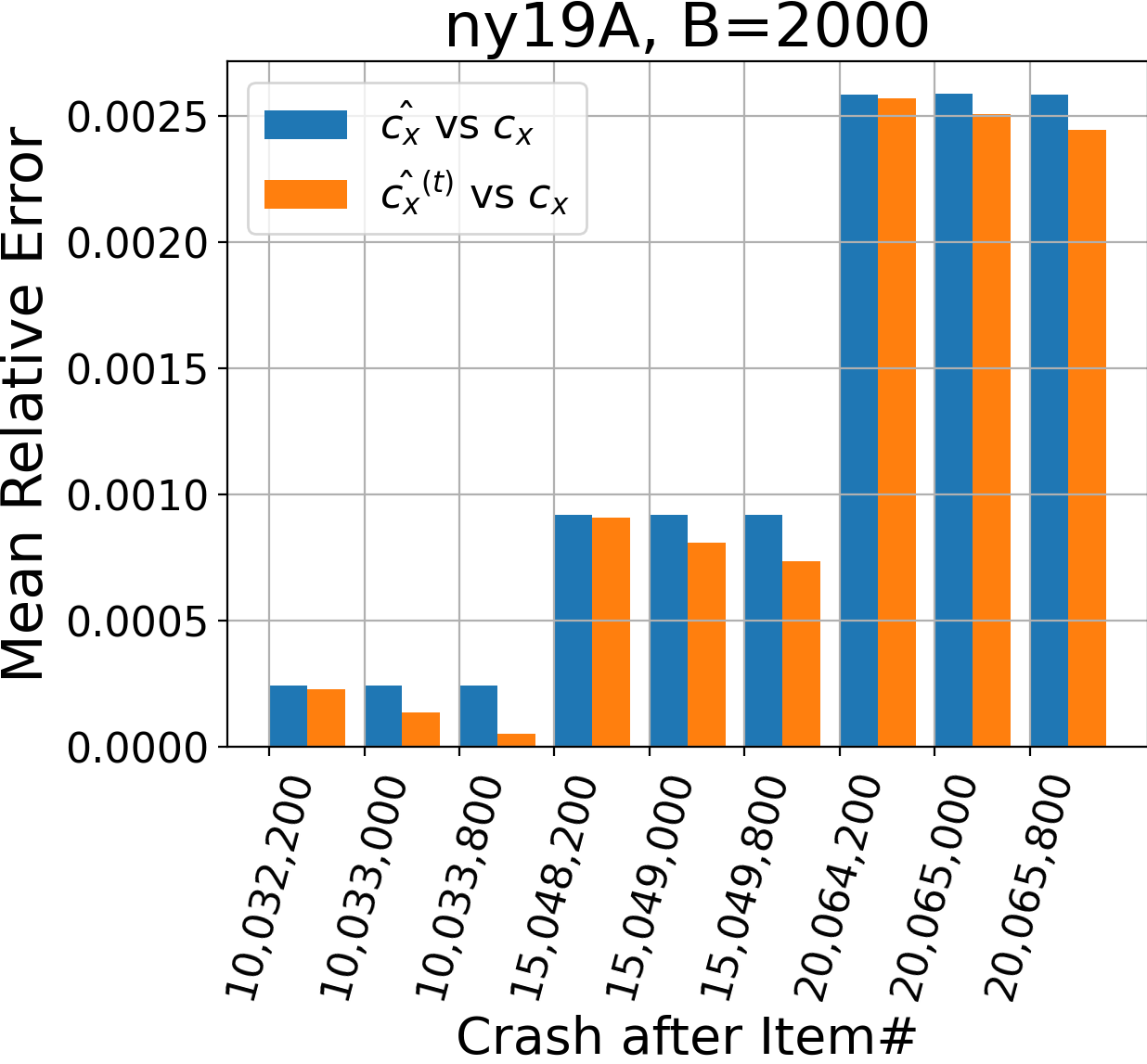}
        \label{fig:MRE-backup-NewYork-2000}
    \end{subfigure}
    \caption{Impact of batch loss on Mean Relative Error in New York}
    \label{fig:MRE-backup-NewYork}
\end{figure}
\begin{figure}[h!]
    \begin{subfigure}[b]{0.32\textwidth}
        \centering
        \includegraphics[width=\textwidth]{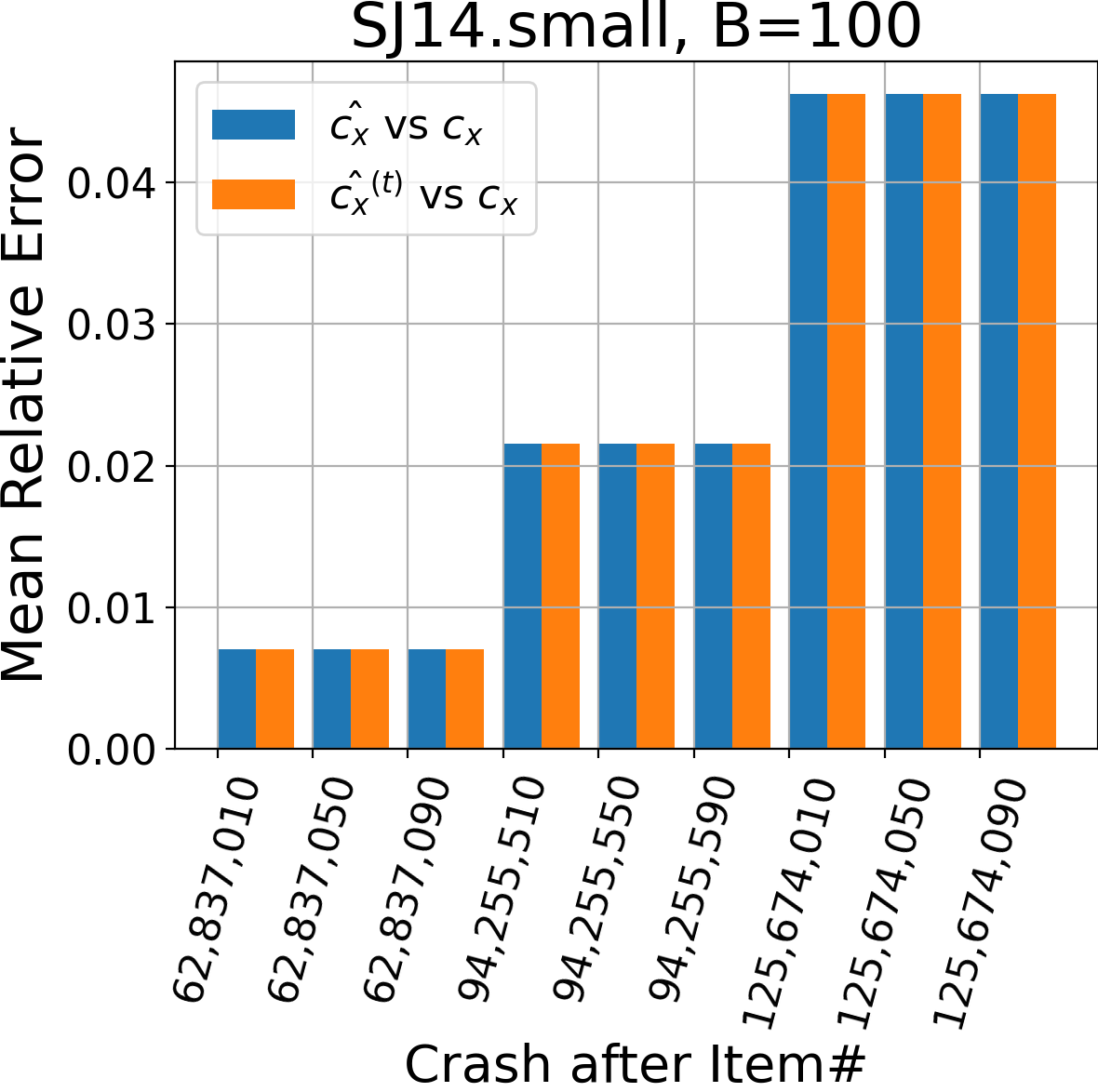}
        \label{fig:MRE-backup-SanJose-100}
    \end{subfigure}
    \hfill
    \begin{subfigure}[b]{0.32\textwidth}
        \centering
        \includegraphics[width=\textwidth]{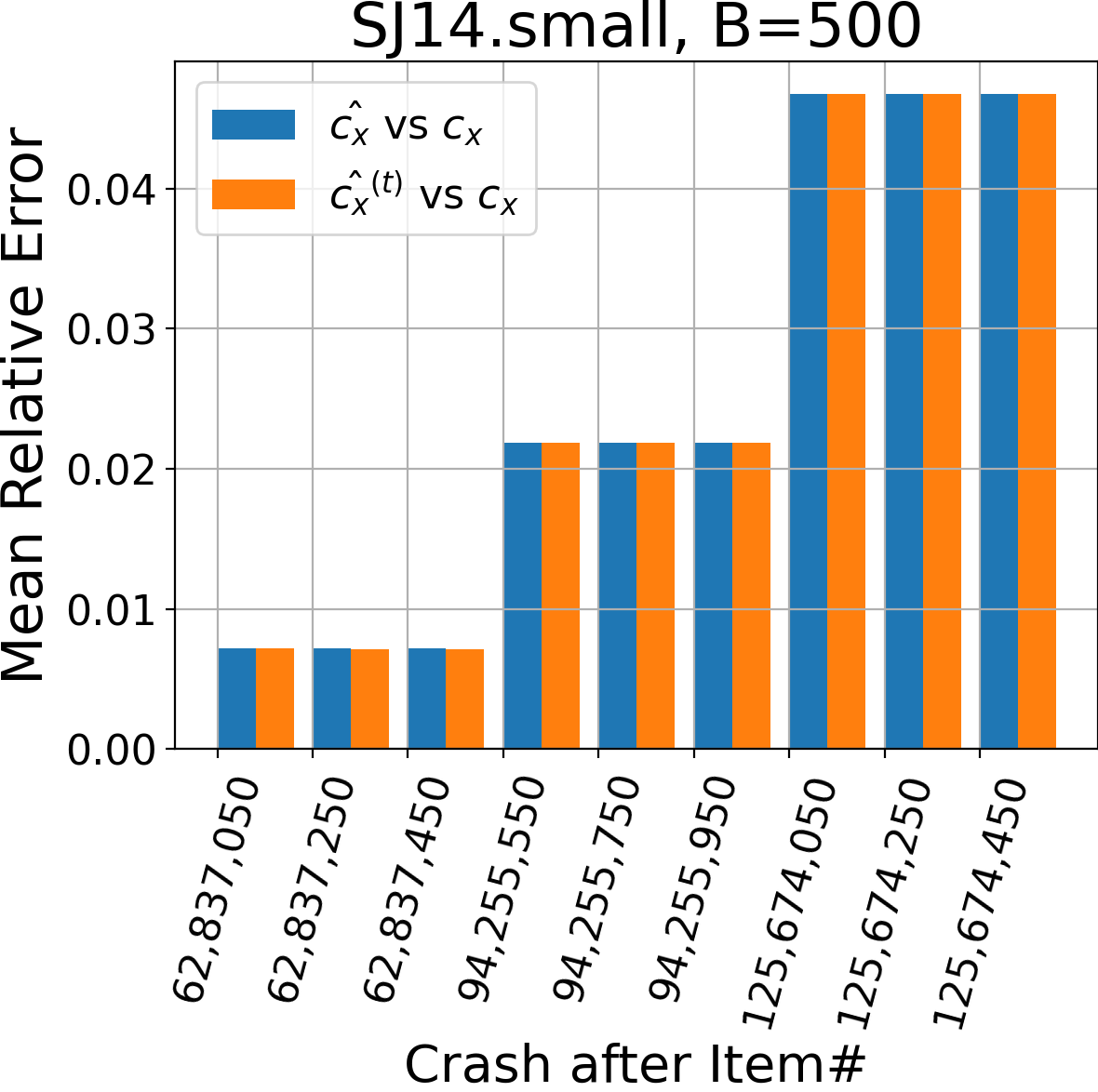}
        \label{fig:MRE-backup-SanJose-500}
    \end{subfigure}
    \hfill
    \begin{subfigure}[b]{0.32\textwidth}
        \centering
        \includegraphics[width=\textwidth]{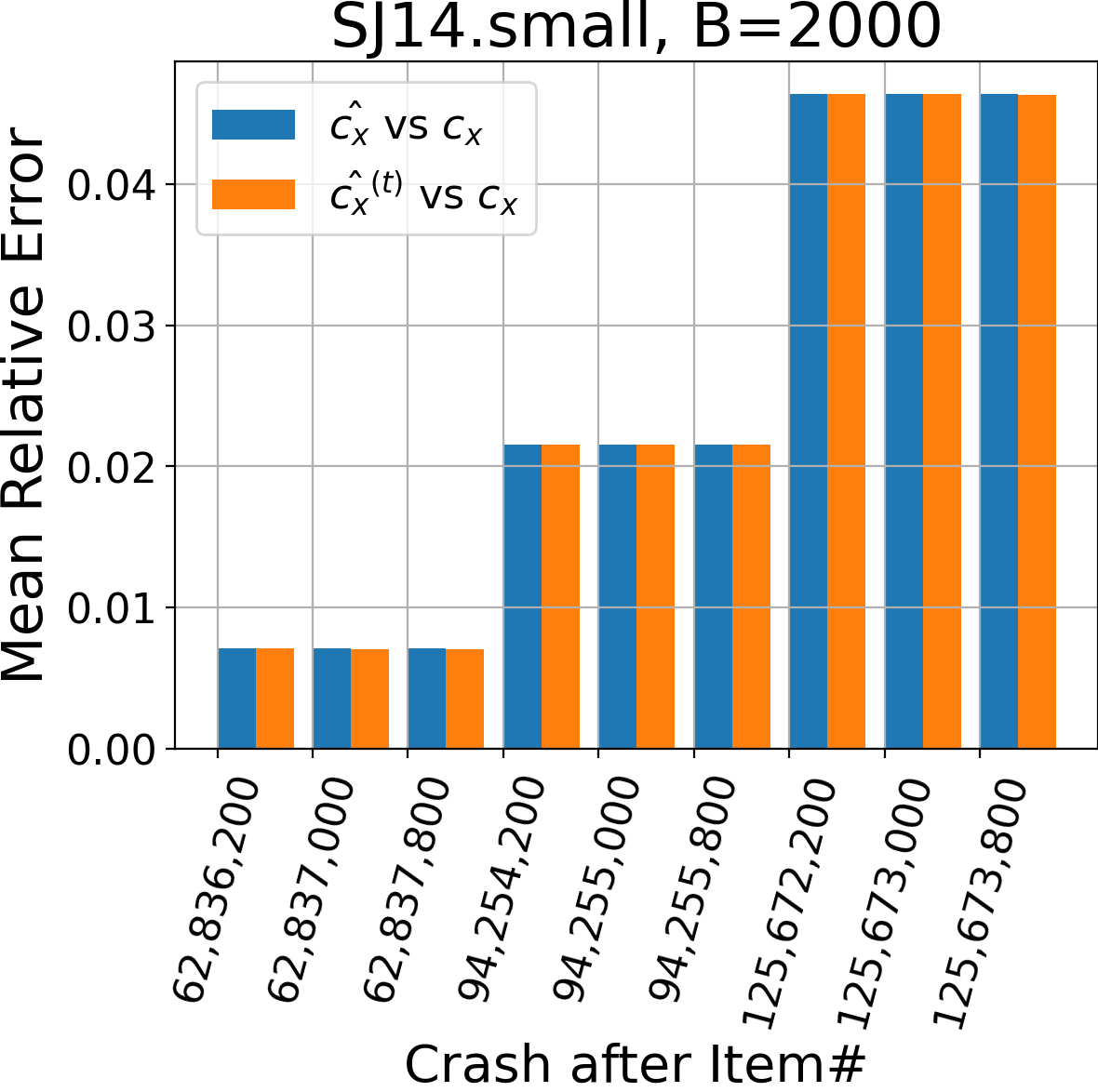}
        \label{fig:MRE-backup-SanJose-2000}
    \end{subfigure}
    \caption{Impact of batch loss on Mean Relative Error in San Jose}
    \label{fig:MRE-backup-SanJose}
\end{figure}
}

As we can see, for large traces, we get one-sided empirical error from true frequency. 
This reveals an interesting direction, to periodically skip batches in large traces, to control an additive error due to overestimation.

\subsubsection{Recommendations}
No need to add $B$ when answering a point query after recovery, as the backup estimation still overestimates in most of the~time.

\section{Conclusions}
\label{chap:conclusions}
In this work, we have presented the motivation for recoverable sketches in a network environment.
Using the role-based approach, the original sketches are held by data nodes as replicas, while the redundant nodes provide them with backup services using erasure-like redundancy. 
Our distributed redundancy strategy is based on the combination of these two roles, which can be implemented as different services running on the same node. 
We introduced a concept of sketch partitioning for load balancing and described different coverage mappings that define the relations between the nodes. 

We discovered an interesting imbalanced mapping, which can be useful in a heterogeneous setting and suits fairness design. 
For future work, we would like to implement our design and test the variety of settings in real-world datasets, exploring the potential of industry use. 
An interesting experiment is to empirically evaluate the usage of compact hash tables vs. the difference matrix. 
A promising direction is to integrate our design into the RDMA model, benefitting from its zero-copy, while performing matrix operations using the difference matrix for batching.

\appendix
\section{}
\label{chap:appendix}

\subsection{Core Parameters}

\paragraph*{CMS accuracy guarantees.}
Denote $\epsilon \in (0,1]$ and $\delta \in (0,1)$ accuracy and confidence guarantees of CMS, meaning that the error in answering a query is within an additive factor of $\epsilon$ with probability $(1-\delta)$.
In other words, 
$(1-\delta)$ is the confidence for not overestimating too much, and 
the probability to exceed the error is bounded by $\delta$.
The sketch table consists of $d$ rows and $w$ columns, such that $d=\ceil*{\ln{\frac{1}{\delta}}}$ and $w=\ceil*{\frac{e}{\epsilon}}$, and $d$ hash functions 
$h_1 \dots h_d: \{1 \dots n\} \rightarrow \{1 \dots w\}$ 
are chosen uniformly at random from a pairwise independent family. 
It is reasonable that $d \leq 12$.
In particular, in our examples and evaluations we refer to $d=5$ hash functions, fitting $\delta=10^{-2}$ that ensures 99\% of~confidence.

{\renewcommand{\arraystretch}{1.4}
\begin{table}[ht]
\footnotesize
{
    \centering
    \begin{tabular}
    {|p{1cm}||l|c|l|c|c|}
    \hline
    $\delta$
    & $d=\ceil*{\ln{\frac{1}{\delta}}}$
    & $\ceil*{\log_2{d}}$
    & Probably correct
    & Number of nines
    & Max confidence
    \\
    \hline
    \hline
    $10^{-1}$
    & $\ceil*{\ln{10^1}}=3$
    & 2
    & 90\%
    & 1
    & 95.0212\%
    \\
    \hline
    $10^{-2}$
    & $\ceil*{\ln{10^2}}=5$
    & 3
    & 99\%
    & 2
    & 99.3262\%
    \\
    \hline
    $10^{-3}$
    & $\ceil*{\ln{10^3}}=7$
    & 3
    & 99.9\%
    & 3
    & 99.9088\%
    \\
    \hline
    $10^{-4}$
    & $\ceil*{\ln{10^4}}=10$
    & 4
    & 99.99\% 
    & 4
    & 99.9954\%
    \\
    \hline
    $10^{-5}$
    & $\ceil*{\ln{10^5}}=12$
    & 4
    & 99.999\% 
    & 5
    & 99.9993\%
    \\
    \hline
    \end{tabular}
    \caption{Choosing $\delta$}
    \label{tab:delta-scale}
}
\end{table}
}

More formally, for a stream of $N$ items, the estimate $\hat{c_x}=\min_{i} \mathit{Count}[i, h_i(x)]$ answered by CMS has the following guarantees: $c_x \leq \hat{c_x}$, where $c_x$ is the actual frequency of $x$; 
and, with probability at least $(1-\delta)$, $\hat{c_x} \leq c_x + \epsilon N$.
In our examples, we consider $\epsilon \leq 10^{-3}$.
Hence, $w> \ceil*{\log_2{w}} \geq 12 \geq d$.

{\renewcommand{\arraystretch}{1.4}
\begin{table}[ht]
\footnotesize
{
    \centering
    \begin{tabular}
    {|p{1cm}||l|c|c|}
    \hline
    $\epsilon$ 
    & $w=\ceil*{\frac{e}{\epsilon}}$
    & $\ceil*{\log_2{w}}$
    & Min error for $w$: $\frac{e}{w}$
    \\
    \hline
    \hline
    $1$
    & $\ceil*{e \cdot {10^0}}=3$
    & 2
    & 90.6093\%
    \\
    \hline
    $10^{-1}$
    & $\ceil*{e \cdot {10^1}}=28$
    & 5
    & 9.7081\%
    \\
    \hline
    $10^{-2}$
    & $\ceil*{e \cdot {10^2}}=272$
    & 9 
    & 0.9993\%
    \\
    \hline
    $10^{-3}$ 
    & $\ceil*{e \cdot {10^3}}=\num{2719}$
    & 12 
    & 0.0999\%
    \\
    \hline
    $10^{-4}$
    & $\ceil*{e \cdot {10^4}}=\num{27183}$
    & 15
    & 0.0099\%
    \\
    \hline
    $10^{-5}$
    & $\ceil*{e \cdot {10^5}}=\num{271829}$
    & 19
    & 0.0009\%
    \\
    \hline
    $10^{-6}$
    & $\ceil*{e \cdot {10^6}}=\num{2718282}$
    & 22
    & 0.0001\%
    \\
    \hline
    $10^{-7}$
    & $\ceil*{e \cdot {10^7}}=\num{27182819}$
    & 25
    & $\frac{0.0001}{10}\%$
    \\
    \hline
    $10^{-8}$
    & $\ceil*{e \cdot {10^8}}=\num{271828183}$
    & 29
    & $\frac{0.0001}{10^2}\%$
    \\
    \hline
    $10^{-9}$
    & $\ceil*{e \cdot {10^9}}=\num{2718281829}$
    & 32
    & $\frac{0.0001}{10^3}\%$
    \\
    \hline
    \end{tabular}
    \caption{Choosing $\epsilon$}
    \label{tab:epsilon-scale}
}
\end{table}
}

\paragraph*{Stream counter.}
We find it reasonable to bound counter length by a size of processor register, e.g., 
in 32-bit architecture we can count up to $N \leq 2^{32}-1$ items, which is more than $4 \cdot 10^{9}$.
CMS hash functions map $\{1 \dots n\}$ to $\{1 \dots w\}$, supporting $N \geq n>w$.
CMS table consists of $d \cdot w$ counters, thus we could maintain exact counters for $n \leq d \cdot w$ flows.
When a sketch becomes saturated, we archive it and turn to use a new instance.
That is, once in awhile we reset the counters of CMS, starting a new epoch.

\paragraph*{Identifier.}
A Universally Unique Identifier (UUID) is a 128-bit label used for information in computer systems.
Internet flows can be identified with 148 bits for IPv6, and with up to 104 bits for IPv4.
However, some applications might use much longer identifiers, e.g., email address, Uniform Resource Identifier (URI). 
In such a case, a cryptographic hash function can be applied to produce shorter fixed-length fingerprints, e.g., cryptocurrency transactions on the blockchain are recognized by a transaction ID (TXID) of 256 bits.

For network applications, when identifier contains IPv4 address, 
$\ceil*{\log_2{mid}} \geq 32$, 
which is an upper bound for stream counter length in 32-bit architecture, and with IPv4 2-tuple, we need 64 bits.
Yet, it's hard to think of an application having short 10-bit identifiers.
Therefore, we generally assume that 
$\ceil*{\log_2{N}} \leq \ceil*{\log_2{mid}} \leq 256$,
and in particular, 
$\ceil*{\log_2{B}} \leq \ceil*{\log_2{mid}}$.

\paragraph*{Batch counter.}
A batch counter is required only when $B \geq 2$; 
otherwise, we update the redundant nodes instantly on each item arrival, and no additional data structure is required to represent a batch.
We note that even with maximum TCP and IPv4 headers (60 bytes each), and 256-bit identifiers (32 bytes), we still have enough room for at least 40 identifiers to fill the MTU of 1500 bytes.
On the other hand, with minimum TCP and IPv4 headers (20 bytes each), and 32-bit identifiers (single IP address), even a jumbo frame of 9000 bytes cannot fit more than 2240 identifiers.
For a large batch of $B=4000$ items we need 12-bit counters, and generally we assume that 16 bits are more than enough.
By default, we consider $B=1000$ as a sufficient size for large batches, requiring 10-bit counters, while with a single byte we can count up to 255 delayed items.

\subsection{Missing Proofs}
\label{sec:proofs}
Some statements in this work were left without proof.
We now provide the proofs through concrete examples.

\begin{lemma}
\label{lemma:f-span}
\autoref{alg:RedundantMatrix} generates a redundant matrix ${MR}_{f} \in \mathcal{M}_{f \times k}(\mathbb{R})$, where any subset of $f$ out of its $k$ columns spans $\mathbb{R}^{f}$.
\end{lemma}
\begin{proof}
Alternatively, we show for $MR_f$ that the determinants of the corresponding $f \times f$ sub-matrices are~non-zero.
To that end, we calculate the determinants for all $\binom{k}{f}$ permutations for $k=5$ and $f \in [1..5]$, starting with the edge~cases:

\paragraph*{Case \emph{f} = \emph{k}.}
This means that all $k$ data vectors were erased, and we need to recover them from redundancy.
For the full matrix, there is only one option, since $\binom{k}{k}=1$.
\begin{center}
\footnotesize
$\det(MR_{5}) = 
\begin{vmatrix}
1 & 1 & 1 & 1 & 1\\
1 & 2 & 3 & 4 & 5\\
1 & 2 & 4 & 7 & 11\\
1 & 2 & 4 & 8 & 15\\
1 & 2 & 4 & 8 & 16
\end{vmatrix} = 1 \neq 0$.
\end{center}

\paragraph*{Case \emph{f} = 1.}
This means that any single data vector was erased. 
There are $\binom{k}{1}=k$ options to form $1 \times 1$ sub-matrix from 
$MR_1=
\begin{bsmallmatrix}
1 & 1 & 1 & 1 & 1
\end{bsmallmatrix}$, each of which consists of the only number $1$, having 
$\det(\begin{bsmallmatrix}1\end{bsmallmatrix}) = 1 \neq 0$.

\paragraph*{Case \emph{f} = 2.}
In case of two data erasures, we need to show non-zero determinants of $\binom{5}{2}=10$ possible $2 \times 2$ sub-matrices of $MR_2$.

{
\footnotesize
$\begin{vmatrix}
1 & 1\\
1 & 2
\end{vmatrix}=1$,
$\begin{vmatrix}
1 & 1\\
1 & 3
\end{vmatrix}=2$,
$\begin{vmatrix}
1 & 1\\
1 & 4
\end{vmatrix}=3$,
$\begin{vmatrix}
1 & 1\\
1 & 5
\end{vmatrix}=4$,

$\begin{vmatrix}
1 & 1\\
2 & 3
\end{vmatrix}=1$,
$\begin{vmatrix}
1 & 1\\
2 & 4
\end{vmatrix}=2$,
$\begin{vmatrix}
1 & 1\\
2 & 5
\end{vmatrix}=3$,

$\begin{vmatrix}
1 & 1\\
3 & 4
\end{vmatrix}=1$,
$\begin{vmatrix}
1 & 1\\
3 & 5
\end{vmatrix}=2$,
$\begin{vmatrix}
1 & 1\\
4 & 5
\end{vmatrix}=1$.
}

\paragraph*{Case \emph{f} = 3.}
There are $\binom{5}{3}=10$ possible $3 \times 3$ sub-matrices of $MR_3$.

{
\footnotesize
$\begin{vmatrix}
1 & 1 & 1\\
1 & 2 & 3\\
1 & 2 & 4
\end{vmatrix}=1$,
$\begin{vmatrix}
1 & 1 & 1\\
1 & 2 & 4\\
1 & 2 & 7
\end{vmatrix}=3$,
$\begin{vmatrix}
1 & 1 & 1\\
1 & 2 & 5\\
1 & 2 & 11
\end{vmatrix}=6$,

$\begin{vmatrix}
1 & 1 & 1\\
1 & 3 & 4\\
1 & 4 & 7
\end{vmatrix}=3$,
$\begin{vmatrix}
1 & 1 & 1\\
1 & 3 & 5\\
1 & 4 & 11
\end{vmatrix}=8$,
$\begin{vmatrix}
1 & 1 & 1\\
1 & 4 & 5\\
1 & 7 & 11
\end{vmatrix}=6$,

$\begin{vmatrix}
1 & 1 & 1\\
2 & 3 & 4\\
2 & 4 & 7
\end{vmatrix}=1$,
$\begin{vmatrix}
1 & 1 & 1\\
2 & 3 & 5\\
2 & 4 & 11
\end{vmatrix}=3$,
$\begin{vmatrix}
1 & 1 & 1\\
2 & 4 & 5\\
2 & 7 & 11
\end{vmatrix}=3$,
$\begin{vmatrix}
1 & 1 & 1\\
3 & 4 & 5\\
4 & 7 & 11
\end{vmatrix}=1$.
}

\paragraph*{Case \emph{f} = 4.}
There are $\binom{5}{4}=5$ possible $4 \times 4$ sub-matrices of $MR_4$.

{
\footnotesize
$\begin{vmatrix}
1 & 1 & 1 & 1\\
1 & 2 & 3 & 4\\
1 & 2 & 4 & 7\\
1 & 2 & 4 & 8
\end{vmatrix}=1$,
$\begin{vmatrix}
1 & 1 & 1 & 1\\
1 & 2 & 3 & 5\\
1 & 2 & 4 & 11\\
1 & 2 & 4 & 15
\end{vmatrix}=4$,
$\begin{vmatrix}
1 & 1 & 1 & 1\\
1 & 2 & 4 & 5\\
1 & 2 & 7 & 11\\
1 & 2 & 8 & 15
\end{vmatrix}=6$,

$\begin{vmatrix}
1 & 1 & 1 & 1\\
1 & 3 & 4 & 5\\
1 & 4 & 7 & 11\\
1 & 4 & 8 & 15
\end{vmatrix}=4$,
$\begin{vmatrix}
1 & 1 & 1 & 1\\
2 & 3 & 4 & 5\\
2 & 4 & 7 & 11\\
2 & 4 & 8 & 15
\end{vmatrix}=1$.
}
\end{proof}

\begin{corollary}
\label{lemma:f-tolerance}
From the above, for $k \leq 5$ and $f \leq k$, once $MR_f$ is available, we are able to fully recover up to \underline{$f$ data} erasures.
Similarly, once all data vectors are available, we are able to reconstruct any $f$ redundant erasures.
\end{corollary}

\begin{lemma}
\label{lemma:2-tolerance-dedicated}
$MR_f$ tolerates at least \underline{any two} erasures, for $f \geq 2$.
\end{lemma}
\begin{proof}
From \autoref{lemma:f-tolerance} we learn that the lemma holds for any two data or redundant erasures.
We now show that it also holds for a mixed case, i.e., one erased data in addition to one redundant erasure.
In such a setting, we are left with $(k-1) \geq 1$ data and $(f-1) \geq 1$ redundant vectors available. 
Let us note that all the coefficients of $MR_f$ are non-zero, meaning that every $1 \times 1$ sub-matrix has a non-zero determinant.
Hence, \emph{any} of the redundant vectors can be used together with the remaining $(k-1)$ data vectors to recover a single data erasure. 
We recover the erased data first and reconstruct the redundancy next.
\end{proof}

\begin{example}
The following example is fully recoverable for $f=k=5$. 
Suppose that $D_1, D_3$ and $D_5$, as well as $R_1$ and $R_3$ failed.
\begin{center}
\footnotesize
$
{G} \vec{D}^\intercal =
\begin{bmatrix}
{I}_5 \\
{MR}_5
\end{bmatrix}
\vec{D}^\intercal=
\begin{matrix}
\begin{pmatrix}
\msout{D_1}\\ 
D_2\\ 
\msout{D_3}\\ 
D_4\\ 
\msout{D_5}
\end{pmatrix}\\*
\begin{pmatrix}
\msout{R_1}\\ 
R_2\\ 
\msout{R_3}\\ 
R_4\\ 
R_5
\end{pmatrix}
\end{matrix}
=
\begin{matrix}
\begin{bmatrix}
\msout{1} & \msout{0} & \msout{0} & \msout{0} & \ \msout{0}\\
0 & 1 & 0 & 0 & \ 0\\
\msout{0} & \msout{0} & \msout{1} & \msout{0} & \ \msout{0}\\
0 & 0 & 0 & 1 & \ 0\\
\msout{0} & \msout{0} & \msout{0} & \msout{0} & \ \msout{1}\\
\end{bmatrix}\\*
\begin{bmatrix}
\msout{1} & \msout{1} & \msout{1} & \msout{1} & \msout{1}\\
1 & 2 & 3 & 4 & 5\\
\msout{1} & \msout{2} & \msout{4} & \msout{7} & \msout{11}\\
1 & 2 & 4 & 8 & 15\\
1 & 2 & 4 & 8 & 16
\end{bmatrix}
\end{matrix}
\begin{pmatrix}
D_1\\ 
D_2\\ 
D_3\\ 
D_4\\ 
D_5
\end{pmatrix}
$
\end{center}
After replacing each failed data vector with a redundant vector that is still available, we are left with $(R_2, D_2, R_4, D_4, R_5)$. 
As before, we recover the data first and reconstruct the redundancy next.

\begin{center}
\footnotesize
$
\begin{pmatrix}
R_2\\ 
D_2\\ 
R_4\\ 
D_4\\ 
R_5
\end{pmatrix} 
=
\begin{bmatrix}
1 & 2 & 3 & 4 & 5\\
0 & 1 & 0 & 0 & 0\\
1 & 2 & 4 & 8 & 15\\
0 & 0 & 0 & 1 & 0\\
1 & 2 & 4 & 8 & 16
\end{bmatrix}
\begin{pmatrix}
D_1\\ 
D_2\\ 
D_3\\ 
D_4\\ 
D_5
\end{pmatrix}
\Rightarrow
\begin{pmatrix}
D_1\\ 
D_2\\ 
D_3\\ 
D_4\\ 
D_5
\end{pmatrix}=
\begin{bmatrix}
\ \ 4 & -2 & -28 & \ \ 8 & \ \ 25\\
\ \ 0 & \ \ 1 & \ \ 0 & \ \ 0 & \ \ 0\\
-1 & \ \ 0 & \ \ 11 & -4 & -10\\
\ \ 0 & \ \ 0 & \ \ 0 & \ \ 1 & \ \ 0\\
\ \ 0 & \ \ 0 & -1 & \ \ 0 & \ \ 1
\end{bmatrix}
\begin{pmatrix}
R_2\\ 
D_2\\ 
R_4\\ 
D_4\\ 
R_5
\end{pmatrix}$
\end{center}
\end{example}

\begin{example}
\label{ex:dedicated-d1-d2-r1}
In contrast, the following example clarifies that $MR_f$ cannot guarantee full recovery of more than just any two erasures, not even for $f=3$. 
When $D_1, D_2$ and $R_1$ fail, we are left with a non-invertible matrix.

\begin{center}
\footnotesize
$
{G} \vec{D}^\intercal =
\begin{bmatrix}
{I}_5 \\
{MR}_3
\end{bmatrix}
\vec{D}^\intercal=
\begin{matrix}
\begin{pmatrix}
\msout{D_1}\\ 
\msout{D_2}\\ 
{D_3}\\ 
D_4\\ 
D_5
\end{pmatrix}\\*
\begin{pmatrix}
\msout{R_1}\\ 
R_2\\ 
R_3
\end{pmatrix}
\end{matrix}
=
\begin{matrix}
\begin{bmatrix}
\msout{1} & \msout{0} & \msout{0} & \msout{0} &\ \msout{0}\\
\msout{0} & \msout{1} & \msout{0} & \msout{0} &\ \msout{0}\\
0 & 0 & 1 & 0 & \ 0\\
0 & 0 & 0 & 1 & \ 0\\
0 & 0 & 0 & 0 &\ 1
\end{bmatrix}\\*
\begin{bmatrix}
\msout{1} & \msout{1} & \msout{1} & \msout{1} & \msout{1}\\
1 & 2 & 3 & 4 & 5\\
1 & 2 & 4 & 7 & 11
\end{bmatrix}
\end{matrix}
\begin{pmatrix}
D_1\\ 
D_2\\ 
D_3\\ 
D_4\\ 
D_5
\end{pmatrix}
$
\end{center}
However, the fact that we focus on CMS reveals an additional point of view on recovery.
Denote $X=(D_1+2 D_2)=(R_2-3 D_3-4 D_4-5 D_5)$, which can be extracted from the non-failed nodes.
Clearly, $0 \leq D_1 \leq X$ and $0 \leq D_2 \leq \floor*{\frac{X}{2}}$.
Therefore, the recovery procedure can return semi-CMSs $D_1' \gets X$ and $D_2' \gets \floor*{\frac{X}{2}}$, which never underestimate the counts of the original $D_1$ and $D_2$, but might have a greater error.
\end{example}

\begin{corollary}
As we focus on CMS, we are able to semi-recover any $f$ erasures, while returning semi-CMSs for cases when exact recovery is impossible.
\end{corollary}

\begin{conjecture}
To tolerate any $f$ erasures, a symmetric Pascal matrix $P_k$ can be used instead of $MR_k$, in expense of larger coefficients.
\end{conjecture}
The construction method of $P$ is quite common to that of $MR$.
The first row and column are filled with
1’s, and all other entries are calculated inductively.
This time, such an entry depends on $up$ and $left$ entries, i.e., in terms of \autoref{alg:RedundantMatrix}, the last line would look like this: $P[row][col] \gets P[row-1][col]+P[row][col-1]$.
\begin{center}
\footnotesize
$MR_5 = 
\begin{bmatrix}
1 & 1 & 1 & 1 & 1\\
1 & 2 & 3 & 4 & 5\\
1 & 2 & 4 & 7 & 11\\
1 & 2 & 4 & 8 & 15\\
1 & 2 & 4 & 8 & 16
\end{bmatrix}$ 
vs 
$P_5 = 
\begin{bmatrix}
1 & 1 & 1 & 1 & 1\\
1 & 2 & 3 & 4 & 5\\
1 & 3 & 6 & 10 & 15\\
1 & 4 & 10 & 20 & 35\\
1 & 5 & 15 & 35 & 70
\end{bmatrix}$ 
vs
${V}=\begin{bmatrix}
1^0 & \ 1^1 & \ 1^2 & \ 1^3 & \ 1^4\\
2^0 & 2 & 4 & 8 & 16\\
3^0 & 3 & 9 & 27 & 81\\
4^0 & 4 & 16 & 64 & 256\\
5^0 & 5 & 25 & 125 & 625
\end{bmatrix}$
\end{center}

\begin{lemma}
\label{lemma:2-tolerance-distributed}
With distributed redundancy, $G_4'$ tolerates up to two node failures.
\end{lemma}
\begin{proof}
In \autoref{sec:distributed-recovery} we described the construction of $MR_k'$ by circular displacement of rows from $MR_k$.
Our goal is to show the recoverability of $G_4'$ for each of $\binom{4}{1}=4$ possible failures of a single node, as well as for each of $\binom{4}{2}=6$ possible failures of two nodes.
According to the construction, $G_4'$ consists of pairs $(D_i, R_i)$ as follows:
\begin{center}
\footnotesize
$
\begin{pmatrix}
D_1\\ 
D_2\\ 
D_3\\ 
D_4
\end{pmatrix},
\begin{pmatrix}
R_1\\ 
R_2\\ 
R_3\\ 
R_4
\end{pmatrix}=
\begin{matrix}
\begin{bmatrix}
1 & 0 & 0 & 0\\
0 & 1 & 0 & 0\\
0 & 0 & 1 & 0\\
0 & 0 & 0 & 1
\end{bmatrix}, 
\begin{bmatrix}
1 & 2 & 4 & 7\\
1 & 2 & 4 & 8\\
1 & 1 & 1 & 1\\
1 & 2 & 3 & 4
\end{bmatrix}
\end{matrix}
$
\end{center}

\paragraph*{Case node 1.}
$D_1 = (R_2-2D_2-4D_3-8D_4)$

\paragraph*{Case node 2.}
$D_2 = \frac{1}{2} \cdot (R_1-D_1-4D_3-7D_4)$

\paragraph*{Case node 3.}
$D_3 = \frac{1}{4} \cdot (R_1-D_1-2D_2-7D_4)$

\paragraph*{Case node 4.}
$D_4 = \frac{1}{7} \cdot (R_1-D_1-2D_2-4D_3)$

\paragraph*{Case nodes 1 and 2.}
\begin{center}
\footnotesize
$
\begin{pmatrix}
R_3\\ 
R_4\\ 
D_3\\ 
D_4
\end{pmatrix} 
=
\begin{bmatrix}
1 & 1 & 1 & 1\\
1 & 2 & 3 & 4\\
0 & 0 & 1 & 0\\
0 & 0 & 0 & 1
\end{bmatrix}
\begin{pmatrix}
D_1\\ 
D_2\\ 
D_3\\ 
D_4
\end{pmatrix}
\Rightarrow
\begin{pmatrix}
D_1\\ 
D_2\\ 
D_3\\ 
D_4
\end{pmatrix}=
\begin{bmatrix}
\ \ 2 & -1 & \ \ 1 & \ \ 2\\
-1 & \ \ 1 & -2 & -3\\
\ \ 0 & \ \ 0 & \ \ 1 & \ \ 0\\
\ \ 0 & \ \ 0 & \ \ 0 & \ \ 1
\end{bmatrix}
\begin{pmatrix}
R_3\\ 
R_4\\ 
D_3\\ 
D_4
\end{pmatrix}$
\end{center}

\paragraph*{Case nodes 1 and 3.}
\begin{center}
\footnotesize
$
\begin{pmatrix}
R_2\\ 
D_2\\ 
R_4\\ 
D_4
\end{pmatrix} 
=
\begin{bmatrix}
1 & 2 & 4 & 8\\
0 & 1 & 0 & 0\\
1 & 2 & 3 & 4\\
0 & 0 & 0 & 1
\end{bmatrix}
\begin{pmatrix}
D_1\\ 
D_2\\ 
D_3\\ 
D_4
\end{pmatrix}
\Rightarrow
\begin{pmatrix}
D_1\\ 
D_2\\ 
D_3\\ 
D_4
\end{pmatrix}=
\begin{bmatrix}
-3 & -2 & \ \ 4 & \ \ 8\\
\ \ 0 & \ \ 1 & \ \ 0 & \ \ 0\\
\ \ 1 & \ \ 0 & -1 & -4\\
\ \ 0 & \ \ 0 & \ \ 0 & \ \ 1
\end{bmatrix}
\begin{pmatrix}
R_2\\ 
D_2\\ 
R_4\\ 
D_4
\end{pmatrix}$
\end{center}

\paragraph*{Case nodes 1 and 4.}
\begin{center}
\footnotesize
$
\begin{pmatrix}
R_2\\ 
D_2\\ 
D_3\\ 
R_3
\end{pmatrix} 
=
\begin{bmatrix}
1 & 2 & 4 & 8\\
0 & 1 & 0 & 0\\
0 & 0 & 1 & 0\\
1 & 1 & 1 & 1
\end{bmatrix}
\begin{pmatrix}
D_1\\ 
D_2\\ 
D_3\\ 
D_4
\end{pmatrix}
\Rightarrow
\begin{pmatrix}
7 D_1\\ 
D_2\\ 
D_3\\ 
7 D_4
\end{pmatrix}=
\begin{bmatrix}
-1 & -6 & -4 & \ \ 8\\
\ \ 0 & \ \ 1 & \ \ 0 & \ \ 0\\
\ \ 0 & \ \ 0 & \ \ 1 & \ \ 0\\
\ \ 1 & -1 & -3 & -1
\end{bmatrix}
\begin{pmatrix}
R_2\\ 
D_2\\ 
D_3\\ 
R_3
\end{pmatrix}$
\end{center}

\paragraph*{Case nodes 2 and 3.}
\begin{center}
\footnotesize
$
\begin{pmatrix}
D_1\\ 
R_1\\ 
R_4\\ 
D_4
\end{pmatrix} 
=
\begin{bmatrix}
1 & 0 & 0 & 0\\
1 & 2 & 4 & 7\\
1 & 2 & 3 & 4\\
0 & 0 & 0 & 1
\end{bmatrix}
\begin{pmatrix}
D_1\\ 
D_2\\ 
D_3\\ 
D_4
\end{pmatrix}
\Rightarrow
\begin{pmatrix}
D_1\\ 
2 D_2\\ 
D_3\\ 
D_4
\end{pmatrix}=
\begin{bmatrix}
\ \ 1 & \ \ 0 & \ \ 0 & \ \ 0\\
-1 & -3 & \ \ 4 & \ \ 5\\
\ \ 0 & \ \ 1 & -1 & -3\\
\ \ 0 & \ \ 0 & \ \ 0 & \ \ 1
\end{bmatrix}
\begin{pmatrix}
D_1\\ 
R_1\\ 
R_4\\ 
D_4
\end{pmatrix}$
\end{center}

\paragraph*{Case nodes 2 and 4.}
\begin{center}
\footnotesize
$
\begin{pmatrix}
D_1\\ 
R_1\\ 
D_3\\ 
R_3
\end{pmatrix} 
=
\begin{bmatrix}
1 & 0 & 0 & 0\\
1 & 2 & 4 & 7\\
0 & 0 & 1 & 0\\
1 & 1 & 1 & 1
\end{bmatrix}
\begin{pmatrix}
D_1\\ 
D_2\\ 
D_3\\ 
D_4
\end{pmatrix}
\Rightarrow
\begin{pmatrix}
D_1\\ 
5 D_2\\ 
D_3\\ 
5 D_4
\end{pmatrix}=
\begin{bmatrix}
\ \ 1 & \ \ 0 & \ \ 0 & \ \ 0\\
-6 & -1 & -3 & \ \ 7\\
\ \ 0 & \ \ 0 & \ \ 1 & \ \ 0\\
\ \ 1 & \ \ 1 & -2 & -2
\end{bmatrix}
\begin{pmatrix}
D_1\\ 
R_1\\ 
D_3\\ 
R_3
\end{pmatrix}$
\end{center}

\paragraph*{Case nodes 3 and 4.}
\begin{center}
\footnotesize
$
\begin{pmatrix}
D_1\\ 
D_2\\ 
R_1\\ 
R_2
\end{pmatrix} 
=
\begin{bmatrix}
1 & 0 & 0 & 0\\
0 & 1 & 0 & 0\\
1 & 2 & 4 & 7\\
1 & 2 & 4 & 8
\end{bmatrix}
\begin{pmatrix}
D_1\\ 
D_2\\ 
D_3\\ 
D_4
\end{pmatrix}
\Rightarrow
\begin{pmatrix}
D_1\\ 
D_2\\ 
4 D_3\\ 
D_4
\end{pmatrix}=
\begin{bmatrix}
\ \ 1 & \ \ 0 & \ \ 0 & \ \ 0\\
\ \ 0 & \ \ 1 & \ \ 0 & \ \ 0\\
-1 & -2 & \ \ 8 & -7\\
\ \ 0 & \ \ 0 & -1 & \ \ 1
\end{bmatrix}
\begin{pmatrix}
D_1\\ 
D_2\\ 
R_1\\ 
R_2
\end{pmatrix}$
\end{center}
\end{proof}


\printbibliography

\end{document}